\documentclass[%
 reprint,
 amsmath,amssymb,
 aps,
]{revtex4-2}

\usepackage[labelformat=simple,font=small]{caption}
\usepackage{graphicx}% Include figure files
\usepackage{dcolumn}% Align table columns on decimal point
\usepackage{bm}% bold math
\usepackage[colorlinks=true,bookmarks=true]{hyperref}
\usepackage[labelformat=simple,font=small]{subcaption}

\begin{document}

%\preprint{}

\title{RTP Pockels Cell with Nanometer-Level Position Control}


\author{Caryn Palatchi}
%\thanks{}
 \email{palatchi@jlab.org}
\author{Kent Paschke}%
 \email{paschke@virginia.edu}
\affiliation{%
  Department of Physics, University of Virginia, Charlottesville, VA 22904
}%


\date{\today}

\begin{abstract}
MOLLER is a future experiment designed to measure parity violation in Moller scattering to extremely high precision. MOLLER will measure the right-left scattering differential cross-section parity-violating asymmetry $A_{PV}$, in the elastic scattering of polarized electrons off an unpolarized $LH_2$ target to extreme ppb precision. To make this measurement, the polarized electron source, generated with a circularly polarized laser beam, must have the ability to switch quickly between right and left helicity polarization states. The polarized source must also maintain minimal right-left helicity correlated beam asymmetries, including energy changes, position changes, intensity changes, or spot-size changes. These requirements can be met with appropriate choice and design of the Pockels cell used to generate the circularly polarized light. Rubidium Titanyl Phosphate (RTP) has been used in recent years for ultra-fast Pockels cell switches due to its lack of piezo-electric resonances at frequencies up to several hundred MHz.  However, crystal non-uniformity in this material leads to poorer extinction ratios than in commonly used KD*P Pockels cells when used in $\lambda/2$-wave configuration. It leads to voltage dependent beam steering when used in $\lambda/4$-wave configuration. Here we present an innovative RTP Pockels cell design which uses electric field gradients to counteract crystal non-uniformities and control beam steering down to the nm-level. We demonstrate this RTP Pockels cell design is capable of producing precisely controlled polarized electron beam at Jefferson Laboratory, a national accelerator facility, for current experiments, including the recent PREX II measurement, as well as the future MOLLER experiment. 

%\begin{description}
%Accelerators 29.20.-c ,Pockels effect 78.20.Jq, electro-optical effects, 78.20.Jq, Optical switches 42.79.Ta,
%\end{description}
\end{abstract}

%\keywords{}

\maketitle

\section{Introduction to Parity Violation Experiments}

In electron beam accelerator facilities like Jefferson Lab, parity-violation experiments are performed which measure asymmetries between reactions with positive(right) and negative(left) helicity states. These experiments measure a parity-violating asymmetry in the differential electron scattering cross section off a target at an angle corresponding to a known energy transfer, and the experimental asymmetry is defined as

\begin{equation}
A_{exp} = \frac{d\sigma^+-d\sigma^-}{d\sigma^++d\sigma^-} 
\end{equation}

where $d\sigma^{+(-)}$ refers to the differential cross-section, proportional to the detected rates, for positive and negative electron beam helicity states respectively. The right and left handed longitudinally polarized electrons for such experiments come from right and left circularly polarized light. A Pockels cell controls the spin of the electron beam by switching the polarization state of the laser beam generating it. The Pockels cell is fed a randomized helicity signal which applies either positive or negative high voltages, producing either right or left circularly polarized light, which is incident on a photocathode, producing consecutive windows of right-handed and left-handed electrons. The electrons are accelerated and then sent into the experimental hall where the differential cross-section asymmetries are measured.

To achieve high precision measurements on $A_{exp}$, the Pockels cell must satisfy both statistical and systematic requirements as regards the electron beam produced. Regarding statistical experimental requirements, in helicity switching, time windows are generated in the electron bunch train at a selected flip rate, with the sign of the beam's longitudinal polarization in each window assigned on a pseudo random basis. Frequency selection for helicity flipping affects the noise, measurement widths, and statistical errors significantly. The future MOLLER experiment \cite{MOLLERcd1} is a high data rate experiment, with 10X higher data rate than the recent PREX II \cite{PREX2paper} experiment in which the Pockels cell presented here was used.  The Pockels cell which controls the electron beam must switch helicity states very quickly, with minimal dead-time, to obtain sufficient statistical precision at high data rates.

In parity experiments the differential cross-section asymmetries are extremely small. The symmetry between incident right and left helicity beams is of importance in achieving systematic experimental requirements. Because this measurement compares right and left handed, opposite helicity, electrons and looks for changes in scattering rates, any change in the polarized beam, correlated with the helicity reversal, can be a potential source for systematic error, or a false asymmetry on $A_{exp}$: this includes energy changes, position changes, intensity changes, or spot-size changes. To first order, this can be written as:
\begin{equation} \label{eq:falseAsym}
A_{raw}= A_{det} - A_Q + \alpha \Delta E + \sum_{}^{} \beta_i \Delta x_i
\end{equation}
where $A_{raw}$ is the beam current normalized detector asymmetry, $A_Q$ is the beam charge asymmetry, $\Delta E$ is the helicity correlated energy difference, $\Delta x_i$ are the  helicity correlated position differences, and $\alpha, \beta_i$ are the coupling constants, both calculated and measured through cross correlations and linear regression in data analysis. Additionally, helicity correlated changes to the spot size of the beam can also give rise to systematic errors.  

\begin{figure*}%[H]
    %\centering
    \captionsetup{justification=raggedright,singlelinecheck=false}
      \begin{subfigure}{0.26\textwidth}
               \caption{Intensity Asymmetry}
  \includegraphics[width=\textwidth]{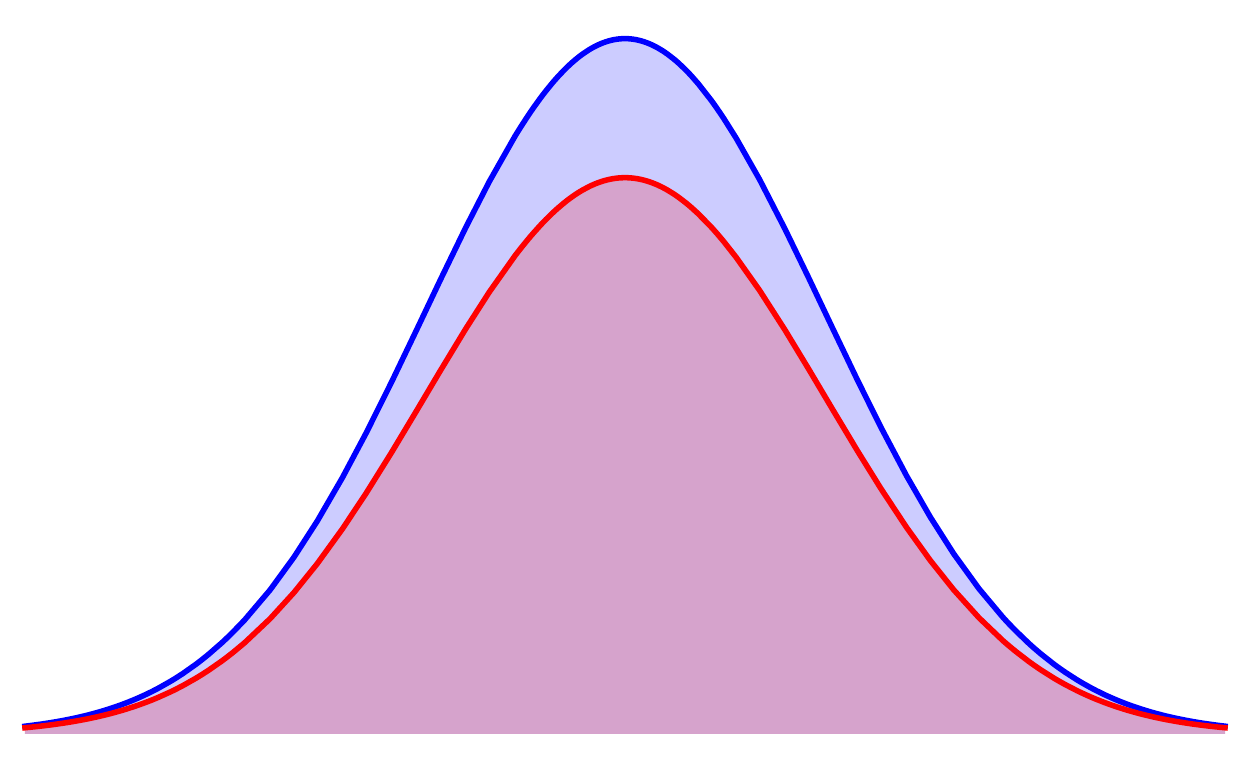}
      \end{subfigure}      
        \begin{subfigure}{0.2\textwidth}
                 \caption{Position Difference}
  \includegraphics[width=\textwidth]{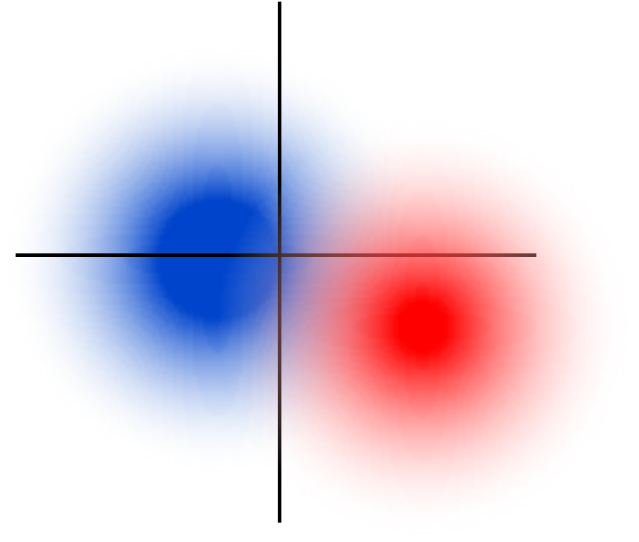}
      \end{subfigure}                  
        \begin{subfigure}{0.26\textwidth}
                 \caption{Spot-Size Asymmetry}
  \includegraphics[width=\textwidth]{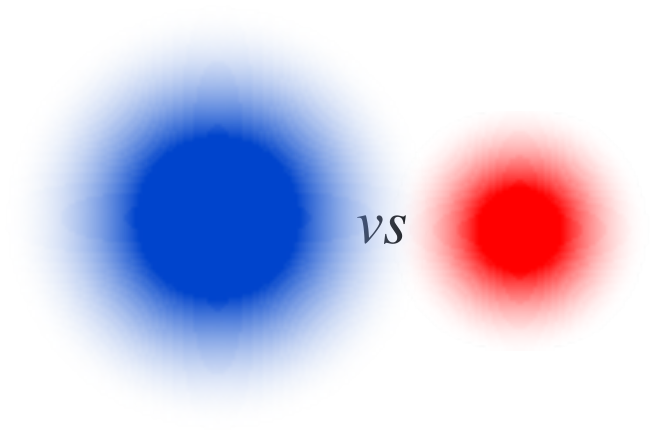}  %{EgradCartoon0.pdf}   
      \end{subfigure}
    \begin{subfigure}{0.27\textwidth}
 \caption{Intensity Asymmetry from Laser Polarization Asymmetry}    
   \includegraphics[width=\textwidth]{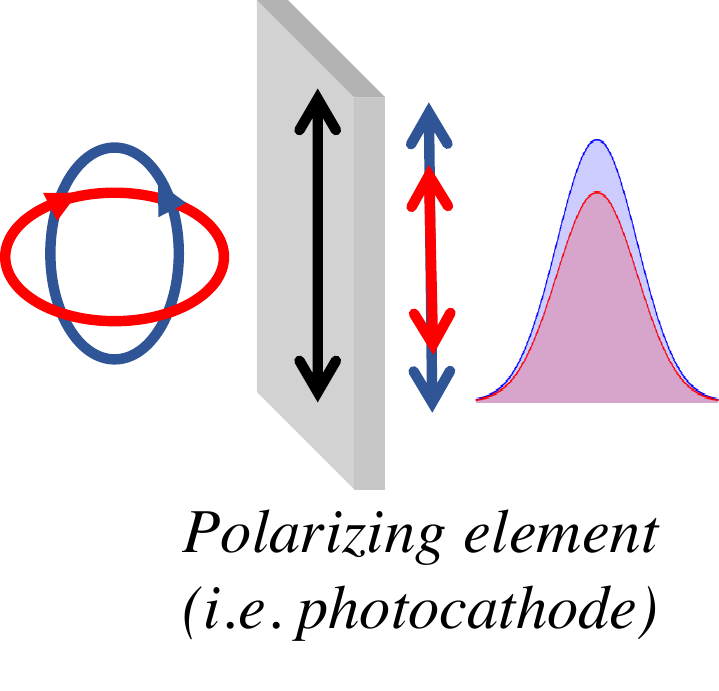}
      \end{subfigure}                          
     \begin{subfigure}{0.25\textwidth}
       \caption{Position Difference from Polarization Gradient}
  \includegraphics[width=\textwidth]{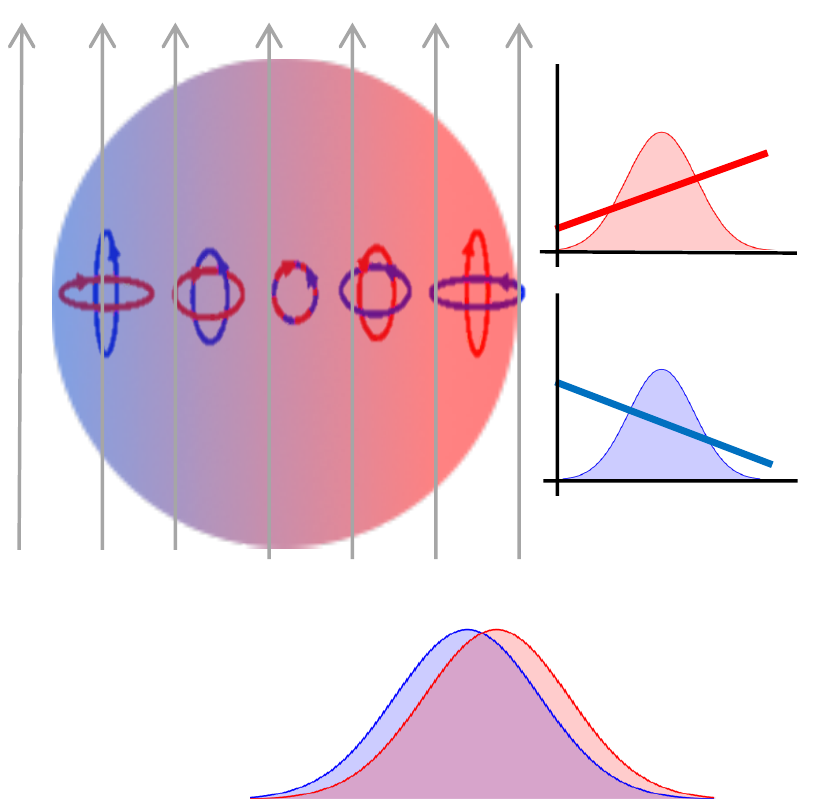}  
      \end{subfigure}      
    \begin{subfigure}{0.25\textwidth}
                            \caption{Spot-Size Asymmetries from Polarization 2nd moment}
  \includegraphics[width=\textwidth]{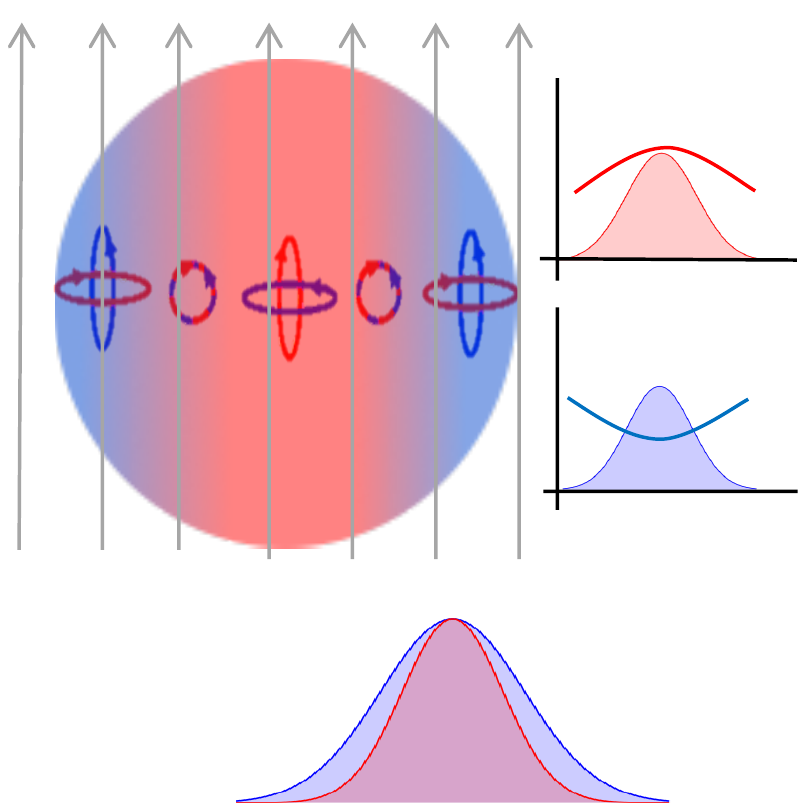}
         %\caption{}
      \end{subfigure}                
     % \hfill
\caption[ Origin of analyzing power dependent beam asymmetries ]{\label{fig:asymtypes}  Origin of analyzing power dependent beam asymmetries: the red and blue ellipses represent polarization ellipses for the opposing right and left circularly polarized states, with residual linear polarization in opposite horizontal and vertical directions.  (a)Intensity asymmetry in gaussian beam where the red and blue curves represent right and left polarization states (b) Illustration of transverse position difference between right and left helicity beams (c) Illustration of spot-size asymmetry (d) When an asymmetry in polarization state is coupled to an analyzing power (a partial polarizer), this can produce an intensity asymmetry (e) If there is a polarization asymmetry gradient transversely across the gaussian beam spot this can produce a position difference between helicity states (f) If there is a 2nd moment in the polarization gradient across the beam spot this can produce a spot-size asymmetry}
\end{figure*}

 For precise comparisons to be made, the two helicity state beams must be extremely symmetric: their intensity, position, and spot-size must be very nearly identical. As illustrated in Fig. \ref{fig:asymtypes} , an intensity asymmetry in the electron beam can arise from a polarization asymmetry in the laser beam when incident on a polarizing element such as the photocathode. A position difference in the electron beam can arise from a polarization gradient in the Pockels cell, a 1st moment effect producing a shift in central laser beam position. A spot size asymmetry can arise from a 2nd moment in polarization gradient, which can broaden or narrow the beam distribution. 

Next generation experiments such as MOLLER \cite{MOLLERcd1} require electron beam off the cathode with position differences of $<$20nm and transverse spot-size asymmetries of $\sim10^{-5}$. These requirements motivated the design of a new RTP (Rubidium Titanyl Phosphate) Pockels cell, with many degrees of freedom, to ensure both fast transition times and small helicity-correlated asymmetries.  This RTP Pockels cell was tested prior to and utilized during the PREX II experiment at Jefferson Laboratory \cite{PREX2paper}.

\section{Pockels Cells}

Pockels cells can be used in a variety of applications including cavity locking in regenerative amplification as well as in electron beam particle accelerators. In regenerative amplifier laser systems, Pockels cell are often used in a $\lambda/2$-configuration for Q-switching. Non-uniformities in the crystal or electric field lead to poor extinction ratios, cavity leaking and poor amplification. In electron accelerators, Pockels cells are used to control the spin of the electron beam by switching the polarization state of the source laser. At the Jefferson Laboratory (JLab) accelerator, the source Pockels cell is used in a $\lambda/4$-wave configuration, switching between right and left circular polarizations, controlling the polarization state of light used to produce polarized electron beam from a GaAs photocathode. Pockels cell non-uniformities, when used in in the $\lambda/4$-wave mode, lead to asymmetries between the generated right and left circular polarized light states and consequently asymmetries in the positive and negative helicity states of the electron beam. In particular, non-uniformities produce helicity-dependent laser beam motion and thus electron-beam position differences.  These helicity-correlated position differences can be detrimental to electron beam accelerator parity experiments which make precise comparisons between positive and negative helicity states. Thus, Pockels cell uniformity is of critical importance for both laser amplification and electron beam accelerator applications.

Fast switching capabilities for high repetition rates is also desirable in these Pockels cell applications. Commonly used commercial KD*P Pockels cells suffer from piezo-electric ringing from acoustic modes when high voltage is suddenly applied to switch polarization states, resulting in a prolonged transition and settle time on the order of 10's of $\mu s$ before the polarization state fully transitions. In regenerative amplifiers for high rep. rate laser systems, a fast-switching Pockels cell with short settle time is desirable. In electron-beam accelerators, when data is taken at a high helicity flip rate, it is also desirable to have a short settle time, to prevent downtime data losses.  In KD*P cells previously used at the Jefferson Lab(JLab) e-beam accelerator this piezo-electric ringing can result in 70-100$\mu$s of settle time during which period the beam quality is poor and data cannot be taken. The statistical losses due to down time goes as $\sim T_{settle} f$, where f is the helicity flip rate. Very high precision parity experiments, like the future MOLLER experiment \cite{MOLLERcd1}, requiring high $\sim$ 2kHz flip rates, depend on the ability to switch helicity state faster and take data at a higher rate than has previously been feasible. %\footnote{Because parity experiments run by flipping helicity states frequently, for the high data rates in the $\sim$2kHz frequency range required high precision experiments, such as the future MOLLER experiment, this $\sim 100\mu s$ settle time results data losses on the 10-20\% level, compromising the statistical precision of experiments. } 
The obvious solution to reducing the settle time of the JLab Pockels cell is to choose a material which suffers less from piezo-electric ringing, even for fast transitions. 

RTP(Rubidium Titanyl Phosphate) is a promising material for Pockels cells due to its fast-switching, high repetition rate capabilities. There is no Pockels cell material which can operate as well at high repetition rates as RTP. Unlike commonly used KD*P cells, RTP suffers minimally from piezo-electric ringing artifacts, the resonances being at much higher frequencies, which in high repetition rate pulsed systems can reduce contrast in half-wave mode laser amplification systems and can cost precious transition time in $\lambda/4$-wave mode electron beam applications. RTP is extremely advantageous in this regard. However, compared with KD*P,  RTP's uniformity is not as good, making for poorer extinction ratios, and in the case of operation in $\lambda/4$ configuration, producing helicity-dependent laser beam motion. In addition RTP is highly birefringent which makes its uniformity in Pockels cells extremely dependent on the precision of face-cut angles; face cut angles as small as 0.1 mrad can have significant impact on extinction ratios and helicity-correlated position differences. 

 Here we present a solution which gives us the best of both worlds: fast transition and improved effective uniformity.
 
 We have demonstrated $\sim10\mu$s transitions in $\lambda/4$-wave configuration with a large aperture (12x12mm$^2$), transverse, RTP cell. Furthermore, we present a new RTP Pockels cell design \cite{patent2} in which crystal intrinsic non-uniformity effects are counteracted with controlled electric-field gradients so that in $\lambda/4$-wave mode, laser beam helicity correlated position motion is controllable and kept at the $\sim$10nrad, 10nm level, while the transition time is kept $<$10$\mu s$.

%Caryn edit this figure so it has better resolution and doesn't have text in it
%\begin{figure*}
%\includegraphics[width=0.7\textwidth]{AsymmetryTypes.png}% Here is how to import EPS art
%\caption[ Origin of analyzing power dependent beam asymmetries ]{\label{fig:asymtypes}  Origin of analyzing power dependent beam asymmetries: Here the red and blue ellipses represent polarization ellipses for the opposing right and left circularly polarized states, with residual linear polarization in opposite directions.  (a) Coupled to an analyzing power, this can produce an intensity asymmetry (b) If there is a polarization gradient across the beam spot this can produce a position difference (c) If there is a 2nd moment in the polarization gradient across the beam spot this can produce a spot-size asymmetry}
%\end{figure*}

\section{Helicity Correlated Beam Asymmetries}
\subsection{Intensity Asymmetry}
A Pockels cell control the polarization state of light passing through it with a voltage induced birefringence via the electro-optic effect. When operated in $\lambda/2$ configuration, the Pockels cell alternates between acting as a $\lambda/2$-wave plate, i.e. rotating incident horizontal polarization into vertical polarization, and having no-birefringence, leaving the incident polarization state unchanged. When operated in $\lambda/4$ wave configuration, the Pockels cell alternates between acting as a quarter-wave plate with its fast axis along $+45^o$ and a quarter wave plate with its fast axis along $-45^o$, i.e. switching incident linearly polarized light into alternating right and left circular polarization states.  Incident horizontally(or vertically) polarized light is exposed to the crystal's fast and slow axes along $\pm45^o$ and splits into two beams which propagate with different phases. The phase shifts for the $\pm45^o$ components of the incident polarization states along primary fast/slow axes of the crystal are
\begin{equation}
\phi^{R(L)}_i= 2 \pi n^{R(L)}_i L^{R(L)} /\lambda
\end{equation}
where R(L) signify right and left circular polarization states of the outgoing light, i refers to the primary fast and slow axes (=y,z for RTP crystals), n is the voltage controlled refractive index, and  $L^{R(L)}=L_0$ is the crystal length which remains fixed when there is no voltage applied and no piezoelectric effect at play.  

When operating in $\lambda/4$-mode, the resultant right and left circular polarization states may not be perfectly circular, having a slight ellipticity, and may not be perfectly symmetric, having different ellipticity for right and left polarization states. This deviation from perfect circular polarization is characterized by the birefringence induced phase shift $\delta$, which each helicity state undergoes.  

\begin{figure}
%\centering
\captionsetup{justification=raggedright,singlelinecheck=false}
\includegraphics[width=0.5\textwidth]{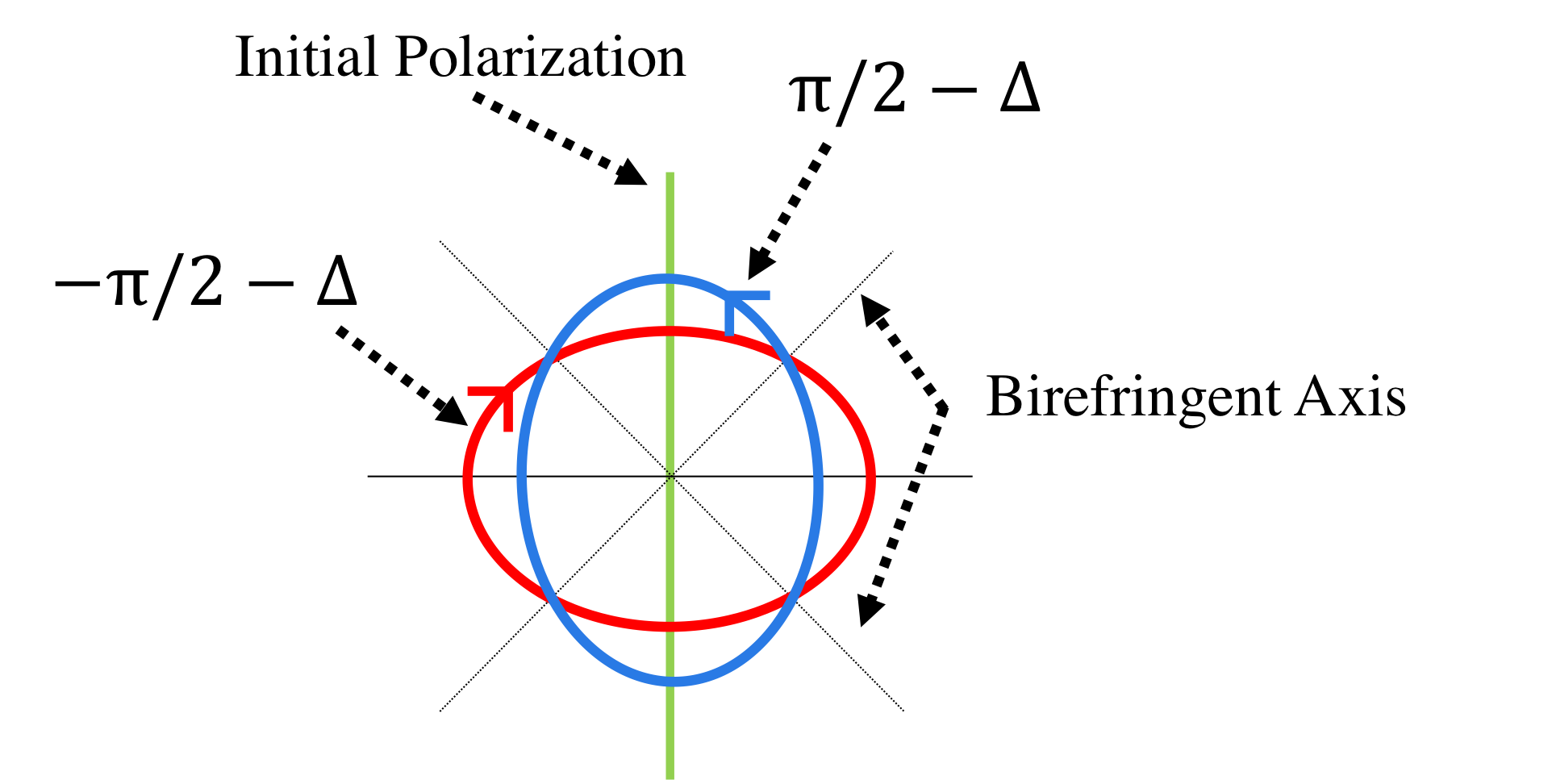}%{DeltaPhase.png}% Here is how to import EPS art
\caption[$\Delta$-phase in S1]{\label{fig:DeltaPhase} Red and blue ellipses illustrate the polarization state of the right and left helicity states after initially vertically polarized light passess through the Pockels cell. The birefringence axis of the Pockels cell is along the diagonal, which controls the degree of linear polarization along the horizontal and vertical dimensions. $\Delta$-phase is defined as the anti-symmetric polarization component, resulting in residual linear polarization along complimentary axes between the two helicity states of light \cite{SilwalThesis}.}
\end{figure}

Each helicity state is nearly circular ($\delta \approx \pm \pi/2$) with small deviations from an $\alpha$-phase and a $\Delta$-phase \cite{NIM}, giving rise to residual linear polarization components as shown in Fig. \ref{fig:DeltaPhase}. The $\alpha$-phase signifies a component of linear polarization which is symmetric in both polarization states, whereas the $\Delta$-phase signifies an asymmetric component between the polarization states. 
\begin{equation}
\delta^{R(L)} = \mp (\pi/2 +\alpha) - \Delta %= \phi^{R(L)}_{u} - \phi^{R(L)}_{v}\]
\end{equation}

In $\lambda/4$-configuration, it is critical to minimize the asymmetric component of linear polarization. The Pockels cell voltages can control the asymmetric component of linear polarization along the horizontal/vertical axes, defined by ``S1" in Stokes parameter terminology.  The Stokes parameters $S_0,S_1,S_2,$ and $S_3$, respectively define the degree of polarization (DoP), the degree of linear polarization (DoLP) along horizontal/vertical axes, the DoLP along the diagonal $\pm45^o$ axes, and the degree of circular polarization (DoCP) \cite{NIM}.  We define a $\Delta$-voltage, also called a PITA (Phase Induced Transmission Asymmetry) voltage, which controls the $\Delta$-phase and an $\alpha$-voltage which controls the $\alpha$-phase as:
\begin{equation}
V^{R(L)} = \pm (|V_{\lambda/4}| + V_\alpha) + V_{\Delta} = \pm V_0  + V_{\Delta} 
\end{equation}
where 
we have defined $V_0 = |V_{\lambda/4}| + V_\alpha$ .
The corresponding electric field in each crystal is $E_z = -V/d$
\begin{equation}
E_z^{R(L)} = \mp (|E_{\lambda/4}| + E_\alpha) - E_\Delta =   \mp E_0 - E_\Delta
\end{equation}

\begin{figure}
%\centering
\captionsetup{justification=raggedright,singlelinecheck=false}
\includegraphics[width=0.5\textwidth]{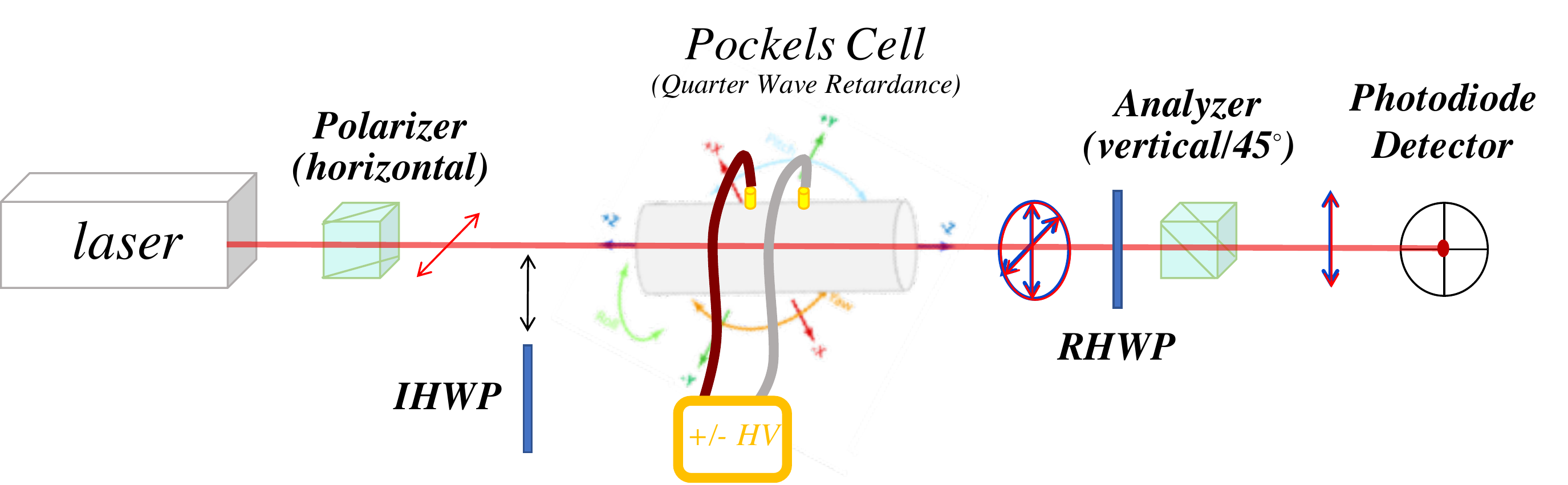}% Here is how to import EPS art
\caption[Laser table in the injector at JLab]{\label{fig:LayoutLaserTable}  Pockels cell alignment laser table configuration. 780nm horizontally polarized light passes through the Pockels cell which produces L or R circularly polarized light. A second rotatable polarizer (analyzer) is inserted which detects any polarization asymmetry on a quad-photodiode detector, which is sensitive to both beam intensity and position.  Angle, translation and voltage adjustments to the Pockels Cell minimize any asymmetry in the degree of linear polarization along both S1 and S2. An insertable half wave plate (IHWP) can be inserted upstream of the Pockels cell to flip the input polarization from horizontal to vertical. A rotating half wave plate (RHWP) is located downstream of the Pockels cell for diagnostic scans. }
\end{figure}

 Such polarization asymmetries lead to intensity asymmetries when exposed to a polarizing element in the beamline. Conversely, intensity asymmetries can be controlled with voltage induced polarization changes when there is a polarizing element. When performing diagnostic tests on the laser table, we use a polarizer with 100\% analyzing power as the polarizing element, and measure transmission and intensity asymmetry $A_I$ with a photodiode, as shown in Fig \ref{fig:LayoutLaserTable} %\footnote{Additionally there is an insertable half-wave-plate IWHP upstream of the PC to convert the H-polarizaton to V-polarization to perform a sign flip to control systematics. There is also a rotating half-wave-plate RHWP downstream of the PC so that the beam polarization can be rotated in accordance with the direction of the analyzing power of the cathode that generates the electron beam}.

  An analyzer is inserted after the Pockels cell with transmission coefficients $T_x, T_y$ along an axis x,y where $\psi$ is the angle subtended between the analyzing direction x and the initial polarization axis (along S1, here we assume the horizontal axis). The transmission through a polarizing element for each polarization state is described by:
\begin{equation}  
T^{R(L)} = T \frac{1}{2} (1+ \epsilon/T \sin(2(\eta-\psi))  \cos \delta^{R(L)})
\end{equation}
where $\epsilon = T_{x}-T_{y}$, $T= (T_{x}-T_{y})/2$ defines the analyzing power of the polarizer(or polarizing element) and $\eta$ is the effective fast-axis of the Pockels cell crystal relative to the horizontal axis. 
 
 In the polarized electron source, the left and right circularly polarized light is incident on a photocathode, which acts as a partial polarizer with slight ($<$6\%) analyzing power. What starts as an asymmetry in laser polarization and would become an asymmetry in light transmission (if the polarizer were optical), instead becomes an asymmetry in the charge of the electron beam. 
 %,
  The charge asymmetry $A_q$ (also referred to as intensity asymmetry $A_I$ for a laser beam), is controlled with the Pockels Cell PITA voltage by inducing $\Delta$-phase polarization changes along S1 (as shown in Fig. \ref{fig:DeltaPhase}) and the analyzing photocathode.
  
The intensity asymmetry $A_I$ is given by
\begin{eqnarray}  
A_I = &&\frac{T^R-T^L}{T^R+T^L} \approx \frac{\epsilon}{T} \sin(2(\eta-\psi)) \frac{1}{2}(\cos \delta^{R} - \cos \delta^{L}) \nonumber\\
&&\approx  -  \frac{\epsilon}{T} \sin(2(\eta-\psi)) \Delta 
\end{eqnarray}
where we have used the approximation $\cos \delta^{R} - \cos \delta^{L}  \approx \delta^R + \delta^L = -2 \Delta$. 

 The $\Delta$-phase can have multiple contributions besides electric fields from various sources, including vacuum windows, angular misalignment of the cell, a rotating HWP downstream of the cell, giving rise to $\Delta_0$. Hence, the electron beam charge asymmetry $A_q$ is given by,
 \begin{equation}
A_q \approx   \frac{\epsilon}{T} (\sin(2(\eta-\psi)) \frac{\pi}{2 |V_{\lambda/4}|} V_\Delta - \Delta_0)
\end{equation}
which when the slow axis of the crystal is along $\eta=45^o$, reduces to 
\begin{equation}
A_q \approx   \frac{\epsilon}{T} (\cos(2\psi) \frac{\pi}{2 |V_{\lambda/4}|} V_\Delta - \Delta_0)
\end{equation}
For 100$\%$ analyzer on the laser table (along S1 where $\psi = 0^o, 90^o = \eta \pm 45^o$) this reduces to 
%Caryn, check the sign is right when plugin voltage here, cause maybe -\Delta -> -Ez -> +V (what you have now) OR -\Delta -> + Ez -> -V. 
\begin{equation} \label{equ:PITAvoltageequtation}
A_I  \approx  -  \Delta  \approx \frac{\pi}{2 |V_{\lambda/4}|} V_\Delta -\Delta_0
\end{equation}
The sensitivity of $A_q$ to $V_\Delta$ is called the PITA-slope $\frac{d A_I}{d V_\Delta}$, and for our RTP system at 780nm is approximately $1053$ppm/V %\footnote{this depends on the definition of PITA running with the 8HV system, but we've empirically observed the PITA-individual slope to be $\sim850ppm/V_{\Delta, indv}$ where each individual plate is at +-800V, which is equivalent to a PITA slope of $\sim1700ppm/V_{\Delta, tot}$ for QWV $\sim1600V$} 
for a $\epsilon/T=100\%$ polarizer along S1. While the $\Delta$-phase has multiple contributions from various sources giving rise to $\Delta_0$, it can be controlled and zeroed out with voltage. Hence, the Pockels cell PITA voltage can control polarization asymmetries along the S1 direction, and can control intensity asymmetries when the polarizing element has a component of analyzing power along the S1 direction (i.e $\epsilon/T \sin(2(\eta-\psi))\neq 0$). 

\subsubsection{Intensity Asymmetry Derivation in RTP Pockels Cells}

%-------
In RTP crystals, the voltage dependent refractive indices (along the primary y and z axes) are given by  \cite{ligonotes}
\begin{eqnarray}
 n^{R(L)}_y = n_{0,y} - \frac{1}{2} n^3_{0,y} r_{23} E_z^{R(L)} \\
   n^{R(L)}_z = n_{0,z} - \frac{1}{2} n^3_{0,z} r_{33} E_z^{R(L)} 
\end{eqnarray}
where typically the electrical field $E_z =-V/d$ is switched to have opposite sign for right and left polarization and is approximately symmetrically flipped such that $E_z^R \approx - E_z^L$ and $V^R \approx - V^L$ the voltages are nearly equal and opposite. 

\begin{figure}
%\centering
\captionsetup{justification=raggedright,singlelinecheck=false}
\includegraphics[width=0.45\textwidth]{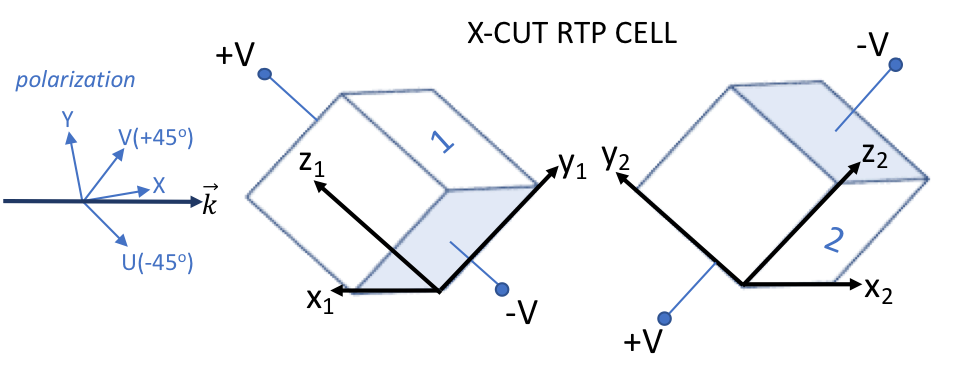}% Here is how to import 
\caption[RTP thermal compensation design]{\label{fig:CristalLaserXcut} RTP thermal compensation design. RTP crystals are highly birefringent and the above two-crystal design is used in Pockels cells to reduce thermal and wavelength dependence. The primary axes of each crystal (x,y,z) are illustrated, as well as the sign of the high voltage applied along $\hat{z}$. Beam propagation direction $\hat{k}$ is along the $\hat{x}$ axis in each crystal in this `X-cut' design. }
\end{figure}

RTP crystals have a high intrinsic birefringence $\Delta n_0=n_{0,z}-n_{0,y}\sim 0.1$ and a single 1cm long RTP crystal, standing alone, functions as a $\sim$1000th order waveplate at 780nm. To avoid severe wavelength dependent and temperature dependent effects of a using such a high order waveplate, two crystals are used in RTP Pockels cells with their fast and slow axes in opposite orientations in a so-called 'thermal compensating' design as shown in Fig. \ref{fig:CristalLaserXcut}. Such a design causes temperature and wavelength shifts of the first RTP crystal to be canceled by the second RTP crystal. %\footnote{analogous to 0th order waveplate designs}. 
The crystals are cut very precisely to be of equal length such that $L_1 \approx L_2 \equiv L_0$ (within 2$\mu$m) so that the net birefringence is near zero when no voltage is applied. Each of the RTP crystals induces equal and opposite phase shifts such that the Pockels cell acts as a zero-order waveplate when inactive:
\begin{eqnarray*}
&& \delta^{R(L)}_1 =   2 \pi \Delta n^{R(L)}_1 L_0 /\lambda =  2 \pi (n^{R(L)}_z-n^{R(L)}_y) L_1 /\lambda 
 \end{eqnarray*}
 \begin{eqnarray*}
 \delta^{R(L)}_2 =  2 \pi \Delta n^{R(L)}_2 L_0 /\lambda =  2 \pi (n^{R(L)}_y-n^{R(L)}_z) L_2 /\lambda 
 \end{eqnarray*}
  \begin{eqnarray}
&& \delta^{R(L)}_{tot} = \delta^{R(L)}_1 + \delta^{R(L)}_2 \nonumber\\
 &&=2 \pi (n^{R(L)}_z-n^{R(L)}_y)(L_1-L_2)/\lambda \approx 0 
\end{eqnarray}

For two crystals of equal length in succession, the phase shift is $\delta = 2 \pi (\Delta n_1 + \Delta n_2) L_0/\lambda $, where $\Delta n_1 =n_z -n_y$ is the birefringence of the 1st crystal and $\Delta n_2 =n_y -n_z$ is the birefringence of the 2nd, and so $\delta \approx 0$. 

When active in $\lambda/4$-mode, the two crystals (1,2) are subjected to opposite sign electric fields $E_{z1}\approx -E_{z2} \equiv E_z$ so that the voltage induced birefringence in both crystals combine additively as follows:
\begin{eqnarray}
n^{R(L)}_{y1} = n_{0,y} - \frac{1}{2} n^3_{0,y} r_{23} E_{z1}^{R(L)} \nonumber \\
 n^{R(L)}_{z1} = n_{0,z} - \frac{1}{2} n^3_{0,z} r_{33} E_{z1}^{R(L)}  \nonumber\\
 n^{R(L)}_{y2} = n_{0,y} - \frac{1}{2} n^3_{0,y} r_{23} E_{z2}^{R(L)}  \nonumber\\
   n^{R(L)}_{z2} = n_{0,z} - \frac{1}{2} n^3_{0,z} r_{33} E_{z2}^{R(L)} 
 \end{eqnarray}
\begin{eqnarray} 
 \delta^{R(L)}_{tot} = &&2 \pi /\lambda (\frac{1}{2} (n^3_{0,y} r_{23} - n^3_{0,z} r_{33})(E_{z1}^{R(L)}L_1 - E_{z2}^{R(L)}L_2)  \nonumber\\
 &&+ (n_{0,z}-n_{0,y})(L_1-L_2))  \nonumber\\
 &&\approx 2 \pi L_0 /\lambda  (n^3_{0,y} r_{23} - n^3_{0,z} r_{33})E_{z}^{R(L)}   
\end{eqnarray}

where $E_z$ is at the quarter-wave field strength in a two-crystal RTP cell system, approximately given by
\begin{equation}
 |E_{\lambda/4}| = \frac{\lambda}{4 L_0 (n_{z0}^3 r_{33}-n_{y0}^3 r_{23})} 
\end{equation}
and where the corresponding quarter wave voltage is given by $V=-E_z d$
\begin{equation}
 |V_{\lambda/4}| = \frac{d \lambda}{4 L_0 (n_{z0}^3 r_{33}-n_{y0}^3 r_{23})} 
\end{equation}
where d is the width of the crystal. For 780 nm, L=10 mm,  d=12 mm, we have $V_{\lambda/4} = 1491$ V and $E_{\lambda/4} = 124.3$ V/mm. %\footnote{Empirically observed $V_{\lambda/4} \sim 1600V$ implies  $E_{\lambda/4} = 133.3$V/mm }
%-------

%---------
Regarding the intensity asymmetry in an RTP cell (Equ. \ref{equ:PITAvoltageequtation}), it can be derived from the $\Delta$-phase and by considering the total phase shifts from the two crystals
\begin{eqnarray}
 \Delta= && \frac{-1}{2} (\delta^R + \delta^L)  = \frac{-1}{2}( \delta^{R}_{tot}+  \delta^{L}_{tot}) =  \nonumber\\
  && \pi /\lambda( (n^3_{0,y} r_{23} - n^3_{0,z} r_{33})(E_{\Delta1}L_1 + E_{\Delta2} L_2) \nonumber\\
 && + (n_{0,y}-n_{0,z})(L_1-L_2))
 \end{eqnarray}
  where for each crystal 
  \begin{eqnarray}
 E_{z1}^{R(L)} = &&\mp (|E_{\lambda/4}| + E_{\alpha1}) - E_{\Delta1} =   \mp E_{z1,0} - E_{\Delta1}\nonumber\\
 E_{z2}^{R(L)} =&& \pm (|E_{\lambda/4}| + E_{\alpha2}) + E_{\Delta2} =   \pm E_{z2,0} + E_{\Delta2}\nonumber\\
  V_{z1}^{R(L)} = &&\pm (|V_{\lambda/4}| + V_{\alpha1}) + V_{\Delta1} =   \pm V_{z1,0} +V_{\Delta1}\nonumber\\
 V_{z2}^{R(L)} = &&\mp (|V_{\lambda/4}| + V_{\alpha2}) - V_{\Delta2} =   \mp V_{z2,0} - V_{\Delta2}
  \end{eqnarray} 
and where $E_{z1} \approx -E_{z2} \approx E_z$ and these are defined in relation to the overall cell quarter-wave fields as 
  \begin{eqnarray}
\Delta  \approx  && \frac{2 \pi L_0 }{\lambda}  (n^3_{0,y} r_{23} - n^3_{0,z} r_{33})E_{\Delta} \approx - \frac{\pi}{2 |V_{\lambda/4}|} V_\Delta 
  \end{eqnarray} 
    \begin{eqnarray}
E_z^{R(L)} = \frac{E_{z1}^{R(L)} - E_{z2}^{R(L)}}{2} = \mp E_{0} - E_{\Delta}\\
 E_{0} = \frac{E_{z1,0} - E_{z2,0} }{2} ,\;\; E_{\Delta} =  \frac{E_{\Delta1} + E_{\Delta2} }{2}\\
 V_{0} = \frac{V_{z1,0} - V_{z2,0} }{2} ,\;\; V_{\Delta} = \frac{V_{\Delta1} + V_{\Delta2} }{2}
  \end{eqnarray}

\subsubsection{\label{sec:level3AqangleS1}Angle dependence Intensity asymmetry-S1}

When aligning the Pockels cell in $\lambda/4$ -configuration, angular adjustments are critical for minimizing HCBA (and for maximizing the  extinction ratio in $\lambda/2$ systems). The angle-dependence of the Pockels Cell polarization asymmetry can be derived from the extra birefringence induced when the beam path through the crystal in changed. In general, any passive birefringent element inserted after the Pockels cell, when operated in $\lambda/4$ -configuration, adds a phase shift $\Delta$ to the circular polarization states, producing polarization asymmetry $\Delta$ between right and left states along the direction of birefringence. When analyzed along the asymmetry direction, this gives rise to an intensity asymmetry of magnitude $A_I= - \Delta$. 

In the case of an X-cut RTP Pockels cell, the beam propagates mainly along the crystal x-axis and the fundamental refractive indices $n_y$ and $n_z$ which the transverse polarization are exposed to are quite different.  At 780nm, the refractive indices of RTP are shown in Table \ref{tab:indexRTP}.
%caryn edit 
%$n_x = 1.7739$ ,$n_y = 1.7832$, $n_z = 1.8673$ (from Sellmeier' s equation \cite{CristalLaser}). 
\begin{table}%[H]
%\centering
\captionsetup{justification=raggedright,singlelinecheck=false}
\begin{tabular}{|r|r|}
 \hline
$n_x$ & 1.7739\\\hline
$n_y$ & 1.7832\\\hline 
$n_z$ & 1.8673 \\\hline
\end{tabular}
\caption[RTP refractive indices at 780nm]{\label{tab:indexRTP} RTP refractive indices at 780nm calculated from Sellmeier' s equation \cite{CristalLaser} }
\end{table} 

When laser beam propagation is at a slight angle, this will mix $n_x$ into $n_z$ or $n_y$ slightly, as well as lengthen the beam propagation distance through the crystal. The angle-dependence of the Pockels Cell intensity asymmetry can be derived from the extra birefringence induced phase shift in the beam path through the crystal by (1) the extra crystal length when it's tilted at an angle and (2) the effective refractive index mixing with the longitudinal x-axis, which alters the effective birefringence.  Both of these mechanisms work together and can be modeled as effectively adding an angle-dependent extra birefringent passive element, which produces an intensity asymmetry

\begin{eqnarray} \label{eq:Aqangledep}
A_I(\xi_{x0},\xi_{y0}) = && \frac{\epsilon}{T} \sin(2(\eta-\psi)) \frac{\pi L_0}{\lambda} (n_z+n_y) \nonumber\\
&&\times (\frac{1}{n_x^2}- \frac{1}{n_z n_y})(2 \xi_{x0} \xi_{y0}) 
\end{eqnarray}
where $\xi_{x0}$ is the yaw tilt of the Pockels cell relative to the crystal's primary x-axis and  $\xi_{y0}$ is the pitch tilt relative to the crystal's primary x-axis.  In S1 ($\eta-\psi=45^o$), the sensitivity to angle is maximal ($\sin(2(\eta-\psi))=1$). For a 100\% analyzer and 1cm long crystals, the angle sensitivity is predicted to be \cite{palatchithesis}
\begin{eqnarray}
  \frac{d^2 A_I}{d \xi_{x0} d \xi_{y0}} &&= - 2  \frac{\pi L_0}{\lambda} (n_z+n_y) (\frac{1}{n_z n_y} -\frac{1}{n_x^2}) \nonumber\\
  &&=5137ppm/mrad^2 
 \end{eqnarray}

\subsection{Intensity Asymmetry - S2}

\begin{figure}%[H]
%\centering
\captionsetup{justification=raggedright,singlelinecheck=false}
\includegraphics[width=0.5\textwidth]{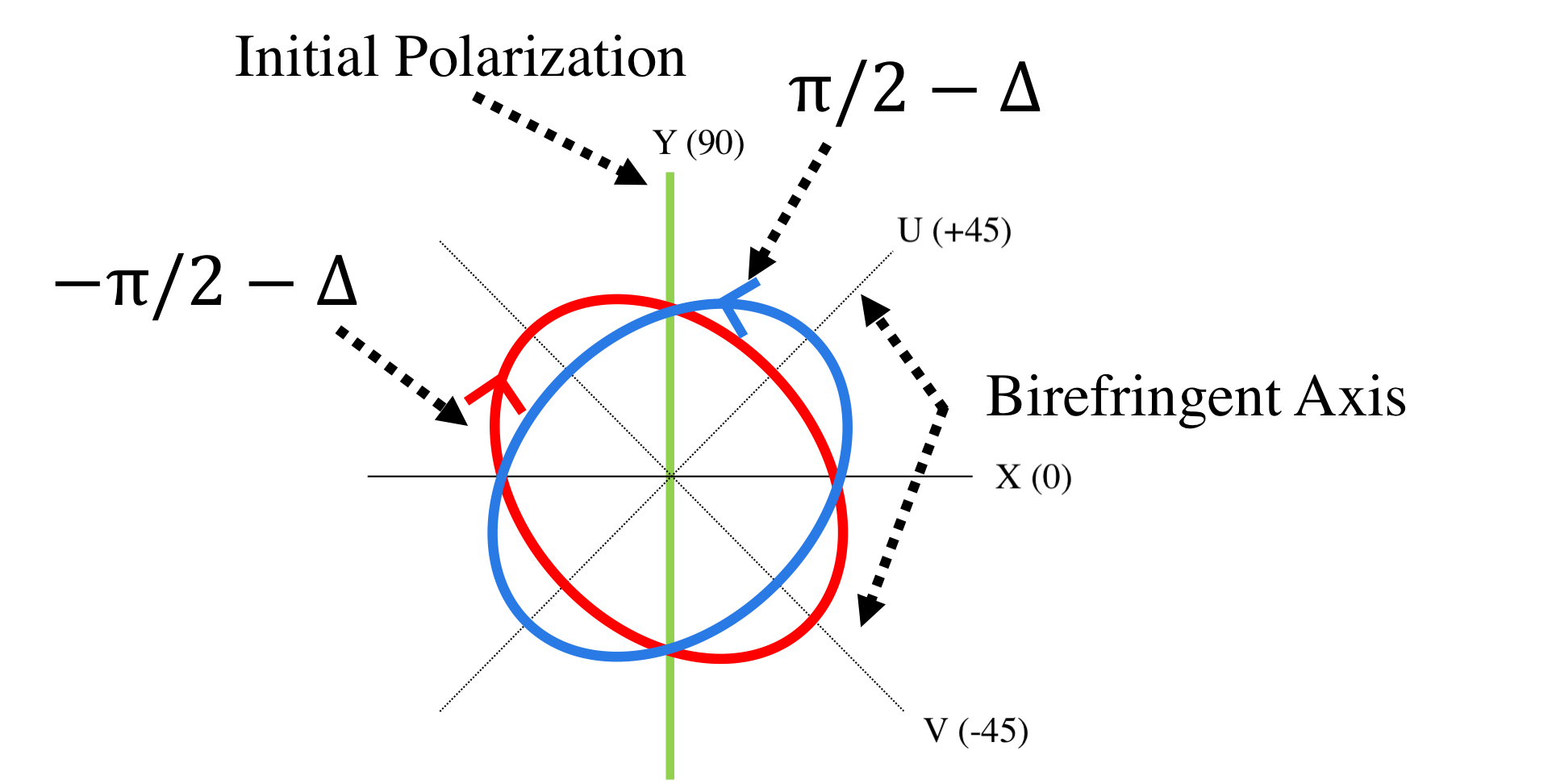}
\caption{\label{fig:AqinS2}Polarization ellipses for asymmetric S2 components on mostly right and left circular polarization states after initially vertically polarized light passes through the Pockels cell. The birefringence axis of the Pockels cell is along the diagonal U($+45^o$) and V($-45^o$) directions. A $\Delta$-phase along S2 is defined as an anti-symmetric polarization component, resulting in residual linear polarization along diagonal U and V complimentary axes between the two helicity states of light.\cite{SilwalThesis}}
\end{figure}

In addition to there being an asymmetric degree of linear polarization along the vertical/horizontal axes for the two left and right helicity states, there can also be an asymmetric degree of linear polarization along the diagonal axes. In addition to a polarization asymmetry oriented along the Stokes parameter S1, there can also be an asymmetry along the Stokes parameter S2, signifying the degree of linear polarization along diagonal axes $+45^o$ and $-45^o$, as shown in Fig. \ref{fig:AqinS2}. 

This S2 polarization asymmetry can arise in KD*P cells from an angular misalignment of the cell, but in RTP cells, due to the large intrinsic birefringence of the crystal, this angular dependence is suppressed in S2 (as can be seen by Eq. \ref{eq:Aqangledep}).  However another kind of misalignment in RTP cells can give rise to this S2 polarization asymmetry: the relative roll angle between the two crystals. In commercial RTP cells, the two crystals are mounted such that they are fixed permanently relative to one another. To obtain the high degree of symmetry required by a parity experiment application, it was necessary to design a cell system with the two crystals mounted independently, allowing for precise control over the relative roll angle between the crystals.

 An advantage to this adjustable cell design is that this Pockels cell has control over both S1 polarization and S2 polarization states, which make it a device capable of completely controlling the polarization state of the outgoing light, 
whereas ordinary Pockels cell designs only have control over S1 polarization. In other accelerator systems, complete control over both S1 and S2 was obtained by the use of two Pockels cells in succession \cite{SLAC2cellsystem}. This system uses the two crystals in an RTP cell to obtain the same degree of control with a single system, using relative roll degree of freedom. 

\subsubsection{Angle Dependence of Intensity asymmetry - S2}

The S2 polarization asymmetry in RTP does not arise from pitch or yaw angular misalignment. Examining the equation for angle dependence Eq. \ref{eq:Aqangledep}, we note that that $A_I \approx 0$ in S2 when $\eta=\psi = 45^o$; meaning the dependence on angular adjustments is negligible when analyzing along $\pm45^o$. The reason for this lack of S2 angle dependence in RTP, as opposed to in KD*P which does have S2 angle dependence is that RTP is a biaxial crystal and the cell is transverse while KD*P is uniaxial and the cell is longitudinal. 

Unlike the uniaxial KD*P, which has nearly equal transverse refractive indices, in RTP the fundamental refractive indices $n_y$ and $n_z$ which the transverse polarization are exposed to are quite different. The refractive indices of RTP at 780nm are $n_x = 1.77$ (the beam propagation axis), $n_y = 1.78$, $n_z = 1.87$ (the transverse axes) \cite{CristalLaser}. By comparison, in KD*P the refractive indices are $n_z\approx1.5$ (the beam propagation axis), $n_x=n_y=1.47$ (the transverse axes). In KD*P, propagation at a slight tilt angle mixes the longitudinal index $n_z$ into the nearly equal transverse indices $n_x \approx n_y$, altering the direction of the effective transverse fast and slow axes. By contrast, in RTP when the laser beam propagation is at an slight tilt angle, while this will mix $n_x$ into $n_z$ or $n_y$ slightly, it is insufficient to alter the direction of the effective fast and slow axes. The effective fast and slow axis very nearly remain along the original $y,z$ directions of $\pm45^o$ regardless of a small tilt angle (because $n_z$ and $n_y$ are very different to start with).  For an RTP crystal, the slow axis is $\eta = y = 45^o$ regardless of small ($\xi_{x0}$, $\xi_{y0}$) tilt angles. 

In the RTP cell two crystal design, since the primary indices are along $\pm45^o$, the effective indices along $0,90^o$ are identical and hence there can be virtually no birefringence along S2 orientation, and no angle dependent asymmetry.  Furthermore, since there is virtually no asymmetry in S2, there are no higher order asymmetry gradients in S2, and hence RTP suffers less less from position differences and spot-size asymmetries than KD*P when the analyzer is oriented along S2. 

\subsubsection{Roll Dependence of Intensity asymmetry - S2}

The S2 polarization asymmetry in RTP, while insensitive to angular misalignment, is sensitive to another kind of misalignment specific to transverse, two crystal cell systems: the relative roll angle between the two crystals. The effect of a relative roll misalignment on S2 asymmetry can be understood as follows. 
An extra birefringent element gives rise to a polarization asymmetry, the direction which depends on the orientation of the effective fast and slow axes of the birefringent element. An extraneous birefringence with fast/slow axes along $+45^o$/$-45^o$ and along the RTP crystals y/z axes gives rise to an asymmetry along S1 which can be corrected with Pockels cell PITA voltage. But if there is an extraneous birefringence with a fast/slow axes along x/y, this gives rise to an asymmetry along S2 which cannot be corrected with Pockels cell voltage. Since there are two crystals, if one crystal has its y/z axes slightly rotated relative to the other crystal, it is effectively acting as an extraneous birefringent element which (a) no longer perfectly cancels out the birefringence of the other crystal and (b) has a component of birefringence along x/y which can't be corrected with Pockels cell voltage. 

\begin{figure}%[H]
%\centering
\captionsetup{justification=raggedright,singlelinecheck=false}
\includegraphics[width=0.3\textwidth]{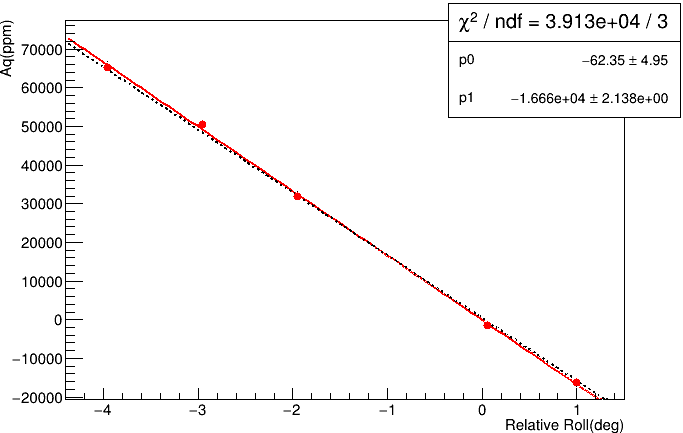}
\caption[Relative Roll and Aq in S2]{\label{fig:RelativeRoll} Relative Roll and Aq in S2. Charge asymmetry Aq as measured by photodiode after passing through Pockels cell and polarizer oriented along $45^o$ on the y-axis. Relative roll angle between the two crystals in the RTP cell on the x-axis.  These measurements were performed by setting rotating the two RTPs relative to each other, and examining the helicity correlated beam asymmetry when analyzing along S2 (along the crystal's primary axis direction). Red dots are data points, red line is a linear fit to the data, black line is the predicted sensitivity based on wavelength, RTP refractive indices,  and the assumption of quarter wave voltage}
\end{figure}

The cell system presented here is designed with the two crystals mounted independently, allowing for precise control over the relative roll angle between the crystals.  In aligning our RTP cell, we use relative roll between the crystals to minimize the polarization asymmetry along the S2. We predicted by Jones calculus, with two birefringent elements with approximately quarter wave operation, that the S2 polarization dependence on relative roll would result in an intensity asymmetry of $16,200 \pm 580 ppm/deg$ when analyzed with a 100\% polarizer along S2 (the error bar is for $\pm1^o$ overall cell roll angle). This prediction was confirmed by measurement where the roll angle of one crystal was varied while the roll angle of the other crystal was fixed and obtained $16665 \pm2 ppm/deg $ within the predicted range in Fig. \ref{fig:RelativeRoll}. 

\subsection{\label{sec:PosDiffsS1}Position differences: Analyzing-like}

A gaussian beam with power distribution $P(x)=P_0 e^{-2x^2/w^2}$ which encounters a gradient in transmission $T=(T_0+\frac{dT}{dx}x)$ undergoes a shift in the beam central position characterized by $\langle x \rangle =\frac{\int x P dx}{\int  P dx}=\frac{\frac{dT}{dx} w^2}{4}$ where the change in first moment is proportional to the transmission gradient. This effect is a shift in central laser beam position.  Analogously, as illustrated in Fig. \ref{fig:asymtypes} , a polarization asymmetry gradient in the Pockels cell, when analyzed, gives rise to an intensity asymmetry gradient which in turn produces position differences between right and left polarization states,

A gaussian beam exposed to a helicity dependent gradient $T^{R(L)}= T_0 (1\pm\frac{dA}{dx}x)$ will have helicity dependent power distribution 
\begin{equation}
P'^{R(L)}=P_0 e^{-2x^2/w^2}(1\pm\frac{dA}{dx}x)
\end{equation}
 which gives rise to a position difference 
 \begin{equation}
 D_x = x^R - x^L =  \frac{\frac{dA}{dx} w^2}{2} = - \frac{\epsilon}{T}\frac{\frac{d\Delta}{dx} w^2}{2}\cos(2\psi) 
 \end{equation}
% \[D_x = w_{cathode} w_{PC}}{2} (\frac{\frac{dA}{dx} + \frac{\frac{d^2A}{dx^2} x+\frac{\frac{d^2A}{dy dx} y) \]
 where w is the beam waist (2$\sigma$) at the Pockels cell and where $\pm$ correspond to right- and left-handed helicity states, $\epsilon/T$ is the analyzing power, and $d\Delta/dx$ is the $\Delta$-phase polarization gradient. %\footnote{This gradient also gives rise to a helicity correlated asymmetric beam shape as well as a shift in the central position(position difference). Theoretically this beam shape difference could couple in to other HCBAs in beam transport.}. 
 We refer to this type of position difference as `analyzing-like' since the position differences only appear in proportion to the analyzing power. For the electron beam, the spot size of the laser once it reaches the photocathode is important. The position differences scale with spot-size, so at the cathode we take:
 \begin{eqnarray}
  D_x^{cathode} = &&\frac{w_{cathode}}{w} D_x \nonumber\\
  &&=- \frac{\epsilon}{T}\frac{\frac{d\Delta}{dx} w_{cathode} w}{2}\cos(2\psi) 
  \end{eqnarray}

Polarization gradients and the consequential `analyzing-like' position differences are seen to arise from birefringence gradients in the Pockels cell (other optical elements can also contribute). These birefringence gradients in the Pockels cell have 3 potential sources (1) electric field non-uniformity (2) crystal length variation due to imperfect crystal face cuts (3) intrinsic birefringence gradients in the crystals due crystal imperfections in the growth process or stress in the crystals. These sources of birefringence gradients (that ultimately contribute to asymmetry gradients) can be derived by taking derivatives of the equation describing the birefringence $\Delta$. The asymmetric birefringence component $\Delta$ can be written as:

\begin{eqnarray}
 \Delta =    &&\pi /\lambda( (n^3_{0,y} r_{23} - n^3_{0,z} r_{33})(E_{\Delta1}L_1 + E_{\Delta2} L_2) \nonumber\\
 &&+ (n_{0,y}-n_{0,z})(L_1-L_2))
 \end{eqnarray}
where each crystal contributes
\begin{eqnarray}
\Delta_1 =  \frac{\pi L_1 }{\lambda}( (n^3_{0,y} r_{23} - n^3_{0,z} r_{33})E_{\Delta1}+ n_{0,y}-n_{0,z})&&\\
\Delta_2 =  \frac{\pi L_2 }{\lambda}( (n^3_{0,y} r_{23} - n^3_{0,z} r_{33})E_{\Delta2} - n_{0,y}+n_{0,z})&&
\end{eqnarray}

The position difference is a vector which can be described by components $D_x$,$D_y$ (or by $45 ^o$ rotated coordinate system components $D_u$,$D_v$ which is more natural, being along the crystals primary axes, where $u=\frac{x-y}{\sqrt{2}}$ and $v=\frac{x+y}{\sqrt{2}}$):
\begin{equation}
\vec{D_r} = D_x \hat{x} + D_y \hat{y} =D_u \hat{u} + D_v \hat{v}
\end{equation}
where $y_1,z_1$ are the y,z-axes for crystal 1 and $y_2,z_2$ are the y,z-axes for crystal 2 as shown in Fig.  \ref{fig:CristalLaserXcut}. Each crystal contributes to the position difference through its own gradients along primary axes $D_{z1}\propto \frac{d\Delta_1}{d z_1}$, $D_{y1}\propto \frac{d\Delta_1}{d y_1}$, where the position differences proportional to the phase-gradient that gives rise to the asymmetry gradient. Each of crystal contributions combine to form the net position difference induced by the Pockels cell:
\begin{eqnarray}
-D_u = D_{z1}+D_{y2}\\
D_v=D_{y1}+D_{z2}
\end{eqnarray}
Here, for simplicity, we consider just the position difference caused by crystal 1, and isolate the component only along the crystal's primary axes $z_1$(oriented at $-45^o$). There exist corresponding expressions for crystal 2 and for the other transverse direction along $-45^o$. The position difference has three main contributions: field gradients, length gradients, and refractive index gradients which can be independently described as $D_{\partial E}$ 
%\footnote{we note that $D_{\partial E}$ is difficult to measure in S1 \cite{elog860} due to quad-photodiode detector Aq-Dx-Dy coupling}, $D_{\partial L} $, and $D_{\partial n}$ \footnote{we note there are also additional terms but they are negligible (<1nm) for PITA voltages $<$QWV ~1600V, specifically gradients due to applied PITA voltage $V_\Delta$ coupled to the length or birefringence gradient of the Pockels cell. 
%\[ D_{\Delta,\partial L,z1} =  -\frac{w^2 \pi}{2 \lambda} (n^3_{0,y} r_{23} - n^3_{0,z} r_{33})  E_{\Delta1} \theta_{fc,z1}  \;\;\; \frac{d D_{\Delta,\partial L}}{d V_{\Delta1}} = 0.00026-0.0026nm/V  \]
%\[  D_{\Delta,\partial n,z1} =  -\frac{w^2 \pi}{2 \lambda} (3n^2_{0,y} r_{23} \frac{dn_{0,y}}{dz_1} - 3n^2_{0,z} r_{33}\frac{dn_{0,z}}{dz_1})(E_{\Delta1}L_1) \] }
:
\begin{eqnarray}
D_{z1} \approx &&D_{\partial E,z1} + D_{\partial L,z1} + D_{\partial n,z1} \\
 D_{\partial E,z1} = && -\frac{w^2 \pi}{2 \lambda} (n^3_{0,y} r_{23} - n^3_{0,z} r_{33})(\frac{d E_{\Delta1}}{dz_1} L_1) \\
&& \approx 6-13nm V^{-1} V_{\delta pos} \nonumber\\
D_{\partial L,z1} = && -\frac{w^2 \pi}{2 \lambda} (n_{0,y}-n_{0,z})\theta_{fc,z1}  \\
&&\approx 1.7-17um \nonumber\\
D_{\partial n,z1}= && -\frac{w^2 \pi}{2 \lambda} \frac{d(n_{0,y}-n_{0,z})}{dz_1} L_1 \\
&&\approx 20 - 40um \nonumber 
\end{eqnarray}
where for our RTP cell and laser system, $L_1=10$mm , $d =12$mm,  $w\approx1$mm, $\lambda =780$nm;  we have measured face cut parallelism of $\theta_{fc,z1} = \frac{d L_1}{d z_1} =0.01-0.1$mrad;  we have estimated by simulation the helicity correlated electric field gradient $\frac{d E_{\Delta1}}{dz_1}$ induced by a voltage shift $V_{\delta pos}$ in our  cell design to be $\frac{d E_{\Delta1}}{dz_1}  \approx 3100\pm1100 m^{-2} V_{\delta pos}$; and where the intrinsic refractive index gradient in RTP has been measured to be on the order of $\frac{dn_z}{dz}=1-2\times10^{-5}/cm>>\frac{dn_y}{dz}$ \cite{RothDraft} \cite{Roth2003}.

\subsubsection{Reducing Analyzing-like Position differences }

We note that the largest birefringence gradients come from the intrinsic refractive index non-uniformity in RTP. While no crystal is perfectly uniform, RTP crystals suffer from greater non-uniformity than KD*P crystals which can be more easily grown to extremely large sizes. Unlike in KD*P cells, where position differences can be reduced by finding the electric-center in the crystal, in RTP cells, the crystal imperfection is inherent to the system, and tend to compromise extinction ratios in half-wave voltage (HWV) systems and create position differences in QWV systems. Innovation in the Pockels cell design was necessary to overcome these position differences due to non-uniformity. There are several potential solutions to counteract this non-uniformity: (1) reduce the laser spot-size (2) increase the laser divergence and make cell angular alignment adjustments  (3) cut the crystal faces with small wedges (4) 2 crystal system orientation (5) use cell design with ability to control E-field gradients.

\subsubsection*{ Laser spot-size}

We note that having a modest laser beam spot-size in the crystal can help to minimize the gradient experienced by the beam distribution and reduce the resulting position differences. However, when using RTP crystals, we cannot reduce the beam size significantly ($2\sigma<1$mm for $\sim$1 Watt) or else thermal gradients induced by the laser absorption ($0.75\%/cm-4\%/cm$) \cite{RaicolSpecSheet} over a small space with high intensity could create additional birefringence non-uniformity leading to position difference drift ($\sim0.1\mu m-0.6\mu$m) and interfere with Pockels cell performance. 

Having a modest laser beam spot-size on the photcathode (when used to make electron-beam), can help minimize the analyzing-like position differences in the generated electron beam. 
%\footnote{During Qweak, the laser spot size on cathode was 0.5mm for run1 and 1mm for run2 (which implies $4\sigma=1.7$mm for run2). Run1 suffered from cathode degradation and polarimetry problems. Run2 was fine in this regard. The $4\sigma$ spot size for Qweak run2 is a good spot size to aim for $4\sigma=1.7$mm} . 
Generally, the position differences are linearly proportional to the spot-size on the photocathode. 

\subsubsection*{\label{sec:level3AngDepPosDiff} Angular Dependence of Position Differences} 

When a laser beam with slight divergence ($\sim 1$mrad) at the Pockels Cell is used, the asymmetry gradients can be canceled out with Pockels cell angular alignment. This is due to the fact that in addition to Aq gradients with respect to position on the crystal $\frac{dA_q}{dX}$, there are also Aq gradients with respect to angle $\frac{dA_q}{d \theta}$. An angle dependent gradient in Aq, when combined with a beam divergence, can produce position differences which in principle can cancel the position differences caused by $\frac{dA_q}{dX}$. 

The angle dependence of Aq with respect to Pockels cell pitch and yaw is a saddle function. If the angle of the Pockels cell is not aligned so it is centered on the saddle-point, there is a 1st order gradient $\frac{dA_q}{d \theta}$ which couples via beam divergence $ \theta_{div} = \frac{d w}{d z}$, through radius and angle coupling, into an effective position difference: 
\begin{equation}
D_x = \frac{\frac{dA}{d\theta_x} w \theta_{div}}{2} = - \frac{\epsilon}{T}\frac{\frac{d\Delta}{d\theta_x} \theta_{div} w}{2} cos(2\psi) 
\end{equation}
These position differences due to Aq angle dependence are only minimized at the extremum saddle-point, so we refer to centering on the pitch/yaw saddle-point as ``angular centering". It is important to align the Pockels cell's angle such that its primary longitudinal axis is parallel to the beam propagation direction, this is where the $dA/d\theta_x=dA/d\theta_y=0$ and where the position differences are minimized. It is also important that the divergence of the laser beam passing through the crystals be small, otherwise the position differences will be very sensitive to slight angular misalignment and could be quite large. 

For a $2\sigma$ spot size of 1mm and 1 mrad divergence we obtain
\begin{eqnarray}
 D_x \approx (2.57\mu m/mrad) \theta_y \nonumber\\ 
 D_y \approx (2.57\mu m/mrad)\theta_x  
\end{eqnarray}
for $\epsilon/T=100\%$, with $Aq \approx (5137$ppm/mrad$^2) \theta_x \theta_y$.
In order to minimize analyzing-like position differences, we can set the Pockels cell angle such that the position differences caused by the angle dependent gradient in Aq $\frac{dA_q}{d\theta}$   cancel the position differences caused by $\frac{dA_q}{dX}$. We refer this optimization, canceling one type of gradient in Aq with another, as ``angular alignment". This slight angular adjustment, away from the central saddle-point, on the mrad-level, can cancel position differences in a $\lambda/4$-wave system, and obviously improve extinction ratios in $\lambda/2$-wave  systems.

\subsubsection*{\label{sec:level3} Crystal wedge cuts}

In principle, one could improve the extinction ratio in $\lambda/2$-wave configuration and the analyzing like position differences in $\lambda/4$-wave configuration if the RTP crystals were cut with a 0.1-0.2 mrad wedge deliberately to cancel out the refractive index gradient along the crystal z-axis. However, such small angle cuts are outside the realm of typical precision.

\subsubsection*{ 2 crystal system \& crystal orientations}

The intrinsic refractive index non-uniformity in RTP, predominantly along z, could be partially counteracted in a 2 crystal system by testing various orientations,  flipping each over so the z and y axis are in opposite directions so that the partly gradients cancel. This arrangement can reduce the net gradient under the assumption that the gradients in each crystal are similar, so they can cancel well when combined. %\footnote{This seems to be the case for the crystal pair we purchased that were grown from the same batch and cut from the same block. Of course there are other gradients present which may not cancel perfectly, but the bulk could be reduced in this scheme.}. %Such a system design is shown in Fig. \ref{fig:4crstalRTPcell}.  

%The intrinsic refractive index non-uniformity in RTP, predominantly along z, could be counteracted in a 4 crystal system, where each crystal (1 and 2) is cut in half to form 4 crystals (1a,1b,2a,2b) and (a,b) haves are flipped over so the z and y axis are in opposite directions (i.e. $z_{1a}=-z_{1b}$, $y_{1a}=-y_{1b}$ ), so that the gradients cancel ($\frac{dn_z}{dz_{1a}} + \frac{dn_z}{dz_{1b}} \approx 0$), while the opposite voltages are applied to achieve additive active induced birefringence. This arrangement can reduce the net gradient under the assumption that the gradients in each crystal are similar, so they can cancel well when combined \footnote{This seems to be the case for the crystal pair we purchased that were grown from the same batch and cut from the same block. Of course there are other gradients present which may not cancel perfectly, but the bulk could be reduced in this scheme.}. Such a system design is shown in Fig. \ref{fig:4crstalRTPcell}.  

%\begin{figure}%[H]
%\centering
%\includegraphics[width=0.6\textwidth]{4crstalRTPcell.png}
%\caption[4 crystal RTP cell]{\label{fig:4crstalRTPcell} 4 crystal RTP cell:  $z_{1a}=-z_{1b}$, $y_{1a}=-y_{1b}$) so that the gradients cancel $\frac{dn_z}{dz_{1a}} + \frac{dn_z}{dz_{1b}} \approx 0$ , while retaining the thermal compensation design}
%\end{figure}

To reduce the combined gradients in our 2 crystal system, we flipped the orientations of the two crystals relative to one another, in several orientations, to try to cancel out the gradients on one crystal with the gradients of the other crystal as much a possible. While this helped to some extent, empirically we observed an asymmetry gradient along $v \equiv \frac{x+y}{\sqrt{2}}$ due to both crystals combined of $ \frac{dA_q}{dv} \sim 20,000-50,000$ppm/mm which would imply, for $w=1$mm, gives position differences of magnitude: 
\begin{eqnarray}
 D_v =  D_{y1}+ D_{z2} = \frac{\frac{dA_q}{d v} w^2}{2} = 10 -25 \mu m\\
 D_x \sim D_y \sim D_v/\sqrt{2} = 7.1-17.7\mu m
\end{eqnarray}

\subsubsection*{ E-field gradient control}

\begin{figure}%[H]
    %\centering
      \begin{subfigure}{0.2\textwidth}
        \includegraphics[width=\textwidth]{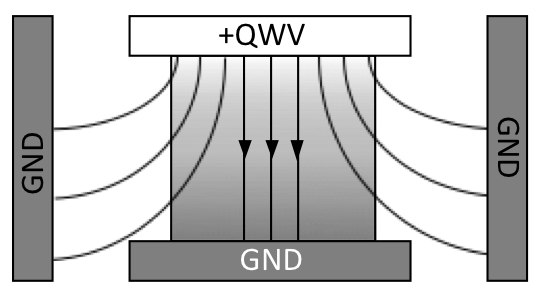}    
         \caption{+ shift}
          \label{fig:cartoonplus}
      \end{subfigure}
        \begin{subfigure}{0.1\textwidth}
        \includegraphics[width=\textwidth]{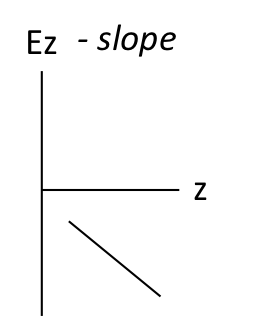}
         %\caption{}
          \label{fig:cartoonplusplot}
      \end{subfigure}                  
        \begin{subfigure}{0.2\textwidth}
        \includegraphics[width=\textwidth]{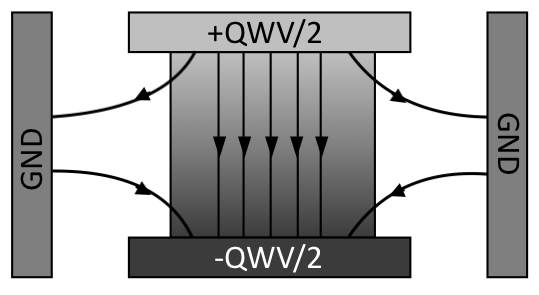}   %{EgradCartoon0.png}   
         \caption{no shift}
          \label{fig:cartoon0}
      \end{subfigure}
                 \begin{subfigure}{0.1\textwidth}
        \includegraphics[width=\textwidth]{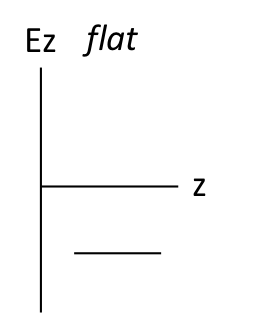}
         %\caption{}
          \label{fig:cartoon0plot}
      \end{subfigure}                    
     \begin{subfigure}{0.2\textwidth}
        \includegraphics[width=\textwidth]{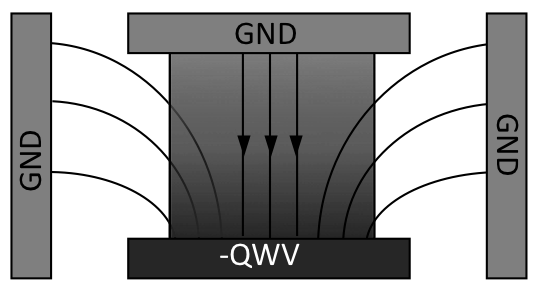}    
         \caption{- shift}
          \label{fig:cartoonminus}
      \end{subfigure}
           \begin{subfigure}{0.1\textwidth}
        \includegraphics[width=\textwidth]{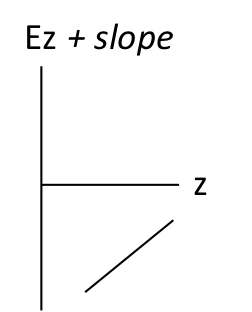}
         %\caption{}
          \label{fig:cartoonminusplot}
      \end{subfigure}                
     % \hfill
     \captionsetup{justification=raggedright,singlelinecheck=false}
\caption[Conceptual Diagram of E-field gradient control in Grounded Side-Panel Design]{
\label{fig:CartoonEfield} Conceptual Diagram of E-field gradient control in Grounded Side-Panel Design: The electric potential across crystal face is illustrated in grey-scale and electric field lines are shown conceptually. Electric field magnitude of Ez with respect to z is also illustrated. +z is up in this diagram. (a) Positive shift in electrodes voltage relative to ground. Entire quarter wave voltage is applied to top electrode. Bottom electrode is grounded. Electric field gradient is negative with respect to z. (b) No shift in electrodes voltage relative to ground. Quarter wave voltage is divided equally between top and bottom electrodes. Electric field gradient is fairly constant with respect to z. (c) Negative shift in electrodes voltage relative to ground. Entire quarter wave voltage is applied to bottom electrode. Top electrode is grounded. Electric field gradient is positive with respect to z.  }
\end{figure}

Ultimately we chose to use a cell design with the ability to control the electric-field gradients in the crystals. In order to counteract the crystal intrinsic non-uniformity and make the two helicity states of light passing through the crystals symmetric, we used grounded side panels to induce fringe-electric fields. A cartoon of the basic cell design is shown in Fig. \ref{fig:CartoonEfield}. 3D electric field modeling informed the finalized design presented in Sec. \ref{sec:8HVdesign}. Shifting the voltage of the top and bottom plates, while keeping the voltage difference between them at QWV, allows us to induce an electric field gradient $\frac{dE_z}{dz}$, while approximately maintaining the appropriate value of the electric field $E_{QWV}$ near the center of the crystal.

By controlling the electric field gradients for each helicity state, in both of the crystals, the asymmetric position motion of the light can be suppressed. For each helicity state, we can choose to have equal and opposite voltage shifts or have the same voltage shift. Correspondingly, we have the freedom to induce the same electric field gradient for both helicity states $\frac{dE^R}{dz}=\frac{dE^L}{dz}$, producing a gradient in $\Delta E$, $\frac{d E_{\Delta}}{dz}$, or we can also induce equal and opposite gradients in each helicity state $\frac{dE^R}{dz}=-\frac{dE^L}{dz}$, producing a gradient in $E_{0}$ ($\frac{d E_{0}}{dz}$). The gradient in $\Delta E$ ($\frac{d E_{\Delta}}{dz}$) has a small effect on the analyzing-like position differences, expressed in Sec \ref{sec:PosDiffsS1}. The gradient in $E_{0}$ ($\frac{d E_{0}}{dz}$) controls another type of position difference, what we refer to as ``steering" position differences, through a much stronger, dominant position difference effect, expounded on in Sec. \ref{sec:Steering}. In operation, rather than adjusting the gradient in $\Delta E$ ($\frac{d E_{\Delta}}{dz}$) we tend to control the gradient in $E_{0}$ ($\frac{d E_{0}}{dz}$) using these ``steering" position differences to cancel out the analyzing-like position differences.

\subsection{\label{sec:Steering}Position differences: Steering} 

\subsubsection{Describing Steering} 

Position differences also arise simply through an angular deviation via GRIN (gradient-index) effects and Snell's Law. This type of position difference is referred to as helicity correlated beam `steering'  an is entirely independent of analyzing power. Steering is a helicity correlated change in angle of the outgoing laser beam after having passed through the Pockels Cell. It produces a position difference between right and left helicity states which increases with throw distance, hence steering is referred to as an `angle-like' position difference which does not depend on analyzing power.

\begin{figure}%[H]
    %\centering
      \begin{subfigure}{0.3\textwidth}
        \includegraphics[width=\textwidth]{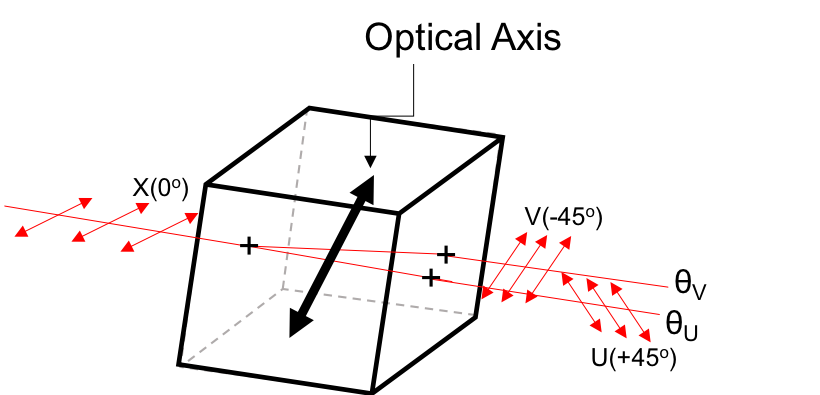}    
         \caption{}
          \label{fig:2raysrtp}
      \end{subfigure}
%      \hfill
      \begin{subfigure}{0.15\textwidth}
        \includegraphics[width=\textwidth]{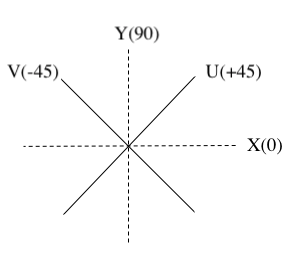}
         \caption{}
          \label{fig:coordsrtp}
      \end{subfigure}             
      \captionsetup{justification=raggedright,singlelinecheck=false}     
\caption[Birefringence Ray Separation]{
\label{fig:2rayscoords} Birefringence Ray Separation and Coordinate Axes (a) Ray Separation due to double refraction. Horizontally polarized beam splits into two components in the birefringent crystal along the primary axes(angles not to scale) (b) Crystal centric coordinate system. Primary crystal axes in the RTP Pockels cell are along the diagonal with respect to horizontal and vertical polarizations. }
\end{figure}

Since the crystal in the Pockels cell is birefringent, the beam should be viewed as 2 separate rays: one with a polarization along the diagonal $U(45^o)$ and the other along $V(-45^o)$, the primary axes of the RTP as shown in Fig. \ref{fig:2rayscoords}. These rays can separate via a difference in the refractive index combined with an angled face cut or a difference in the refractive index gradient $\frac{dn_{z}}{dx_i}$ and $\frac{dn_{y}}{dx_i}$. In RTP, the optoelectric effect $n_{z} = n_{z0}-\frac{1}{2} r_{33} E_z n_{z0}^3$ and $n_{y} = n_{y0}+\frac{1}{2} r_{23} E_z n_{y0}^3$ causes the 2 rays (one for each helicity state) to separate into 4 rays for different primary axes 
%as shown in Fig. \ref{fig: 4 Ray Separation rtp} . 
The difference $\theta^R-\theta^L$ is the helicity correlated angular steering. 

%\begin{figure}%[H]
%\centering
%\includegraphics[width=0.4\textwidth]{4raysrtpe.png}
%\caption[4 Ray Separation: Steering]{\label{fig: 4 Ray Separation rtp} 4 Ray Separation:  the steering behavior of RTP.  Horizontally polarized beam splits in the crystal along the primary axes, and voltage application split the beam further. Angles not to scale, but the relative ordering with grater deviation from $z$ than from $y$ for a given crystal is correct. }
%\end{figure}

Steering can only arise through either a gradient in the average refractive index or a length gradient. If the RTP crystal is cut with a slight wedge $\theta_{fc}=\theta_{cut1}-\theta_{cut2}$, with component $\theta_{fc,x_i}$ along $\hat{x_i}$,  the beam will experience an angular deviation upon exiting the crystal $\theta \sim (n_{eff}-1) \theta_{fc,x_i} = n_{eff} \frac{dL}{dx_i}$ proportional to the face cut wedge angle or length gradient through Snell's Law. Equivalently, a gradient in the effective refractive index will cause the beam will experience an angular deviation $\theta_{x_i} = \frac{dn_{eff}}{dx_i} L$ upon exiting the crystal through GRIN effects. We can generally describe the angle induced from any phase gradient with $\theta_{x_i} = \frac{d}{dx_i} (n_{eff} L) =\frac{\lambda}{2\pi} \frac{d\phi}{dx}$ . In the case of steering, the laser beam is bent by the crystal in a helicity correlated manner, induced by the voltage applied:
\begin{equation}
\Delta \theta _{x_i}= \theta^R-\theta^L = \theta_{fc,x_i} \Delta n + L \frac{d \Delta n}{dx_i} 
\end{equation}
where $\Delta n = n^R-n^L \sim \Delta E_z^R - E_z^L$

Steering can be modeled as arising from electric field gradients, length gradients/face cuts angles/curvature, and intrinsic refractive index gradients from crystal growth. RTP crystals actually have electro-optic prism applications where they are intentionally cut with a large wedge and used to control beam position by application of voltage \cite{ligonotes} \cite{eoprism}. We have effectively taken this application and incorporated it into our Pockels cell system. As discussed below, we use the electric field gradient to induce steering, providing precise control of the helicity correlated position differences.

Steering is a strongly input polarization dependent effect. Formerly, steering in KD*P Pockels cells was thought to be a polarization-independent effect since both horizontal and vertical input polarizations appeared to exhibit the same steering behavior (the ``Skew" effect  \cite{SkewPaschkeeffect}). However, we have since shown that steering is, in fact, quite dependent on input polarization in both KD*P and in RTP Pockels cells \cite{palatchithesis}. When the incident beam polarization is along the RTP crystal's primary axes at $\pm 45^o$ (Fig. \ref{fig:CristalLaserXcut}), the steering induced for z-axis $+45^o$ input polarization differs significantly from the steering induced for y-axis $-45^o$ input polarization.  Placing an analyzer along S2, along the diagonal $U(45^o)$ or along $V(-45^o)$, probes the steering for the primary axis polarization states. For a single RTP crystal, analyzing along the diagonal $U(45^o)$ isolates  $n_{z}$ (or $n_{y}$) rays so that $\frac{dn_{z}}{dx_i}$ will change for +$E_z$ and -$E_z$ producing position differences which are angle-like and grow as throw distance increases. S2 steering depends on $\theta_z$.% as illustrated in Fig. \ref{fig: 4 Ray Separation rtp}.

When the  input polarization is horizontal or vertical, upon entering the X-cut RTP crystal, the light is split into 2 rays, one with polarization along the z axis, and the other with polarization along the y axis. For RTP, the incident H/V polarized light propagates along the crystal x-axis direction with nearly equal components along the $n_z$, $n_y$ directions which are the crystals primary axes. In general, for a birefringent material, the refractive indexes mix to form an effective index and n is more generally described as n($\theta$,$\phi$), where $(\theta,\phi)$ describe the polarization direction, and can be computed through the index ellipsoid of the crystal, which is defined by the surface $\frac{x_1^2}{n_1^2}+\frac{x_2^2}{n_2^2}+\frac{x_3^2}{n_3^2}=1$. We calculate n($\theta$,$\phi$) by noting
\begin{eqnarray}
 x_1=n(\theta,\phi)\sin(\theta)\cos(\phi)\nonumber\\
 x_2=n(\theta,\phi)\sin(\theta)\sin(\phi)\nonumber\\
 x_3=n(\theta,\phi)\cos(\theta) 
 \end{eqnarray}
 Hence,
 \begin{eqnarray}
 && \frac{1}{n^2(\theta,\phi)}=\frac{\sin^2(\theta)\cos^2(\phi)}{n_1^2}+\frac{\sin^2(\theta)\sin^2(\phi)}{n_2^2}+\frac{\cos^2(\theta)}{n_3^2}\nonumber
 \end{eqnarray}  
    This effective refractive index in the RTP crystal is approximately given by the average of $n_z$ and $n_y$ when the incident polarization state is horizontal or vertical. Consequently, the steering for H and V input polarizations is simply the average of the steering along the $\pm45^o$, y and z primary crystal axes. With no analyzer present (or with the analyzer along the horizontal), there is steering given by the average gradients  
  \begin{eqnarray}   
 &&   \frac{d}{dx_i}\bigg (\frac{n_{z}+n_{y}}{2}\bigg) \\
&&    = \frac{d}{dx_i} \bigg (\frac{n_{z0}-\frac{1}{2} r_{33} E_z n_{z0}^3+n_{y0}+\frac{1}{2} r_{23} E_z n_{y0}^3 }{2}\bigg)\nonumber
     \end{eqnarray}  
    which does not cancel and has a dependence on the sign of the E-field. There is helicity-correlated beam steering steering for H and V polarization in S1 as well as for no analyzer. %Fig. \ref{fig: 4 Ray Separation rtp} reflects the conclusion that 
No-analyzer steering in RTP does not cancel and depends on the difference between $\frac{\theta_{z,0}+\theta_{y,0}}{2}-\frac{\theta_{z,1}+\theta_{y,1}}{2}$. 

\subsubsection{Derivation} 

We examine the helicity correlated phase gradients for each input polarization state component. These polarization components undergo phase shifts 
\begin{equation}
\phi^{R(L)}_i = 2 \pi n^{R(L)}_i L^{R(L)} /\lambda 
\end{equation}
where R(L) indicates right and left circular polarization states determined by the sign of the voltage applied to the Pockels Cell as controlled by the helicity signal. 

The phase shift in each of the two crystals, for each $\pm45^o$ polarization component, is given by (see Fig. \ref{fig:CristalLaserXcut})
\begin{eqnarray}
&& \phi^{R(L)}_{45^o,1} =  2 \pi  n^{R(L)}_{z1} L_1 /\lambda \;,\; \phi^{R(L)}_{-45^o,1} =  2 \pi  n^{R(L)}_{y1} L_1 /\lambda \\
&& \phi^{R(L)}_{45^o,2} =  2 \pi  n^{R(L)}_{y2} L_2 /\lambda  \;,\; \phi^{R(L)}_{-45^o,2} =  2 \pi  n^{R(L)}_{z2} L_2 /\lambda 
 \end{eqnarray}
The total phase shift for the two crystals combined, for each $\pm45^o$ polarization component, is given by:
\begin{eqnarray}
 \phi^{R(L)}_{45^o,tot} =&&  2 \pi  (n^{R(L)}_{z1}L_1+n^{R(L)}_{y2}L_2)/\lambda \\
 =&&2 \pi/\lambda ( n_{0,z}L_1 + n_{0,y}L_2  \nonumber\\
 &&- \frac{1}{2} n^3_{0,z} r_{33} E_{z1}^{R(L)}L_1  - \frac{1}{2} n^3_{0,y} r_{23} E_{z2}^{R(L)}L_2) \nonumber\\
 \phi^{R(L)}_{-45^o,tot} = && 2 \pi  (n^{R(L)}_{y1}L_1+n^{R(L)}_{z2}L_2) /\lambda \\
 =&& 2 \pi/\lambda ( n_{0,y}L_1 + n_{0,z}L_2 \nonumber\\
 &&- \frac{1}{2} n^3_{0,y} r_{23} E_{z1}^{R(L)}L_1  - \frac{1}{2} n^3_{0,z} r_{33} E_{z2}^{R(L)}L_2) \nonumber
 \end{eqnarray}
Averaging the phase shifts for polarization components along $\pm45^o$, we obtain a general equation for the overall phase shift:
\begin{eqnarray}
 \phi^{R(L)}_{avg} = &&(\phi^{R(L)}_{45^o,tot}+\phi^{R(L)}_{-45^o,tot})/2 \\
 = &&\pi/\lambda (( n_{0,z}+ n_{0,y})(L_1 + L_2) \nonumber\\
&&- \frac{1}{2} (n^3_{0,z} r_{33} + n^3_{0,y} r_{23})  (E_{z1}^{R(L)}L_1+ E_{z2}^{R(L)}L_2) 
  )  \nonumber
  \end{eqnarray}

As stated above, since in one crystal the electric field $E_z$ is \textit{along} the primary crystal axis $+\hat{z}$ , and in the other crystal the electrics field $E_z$ is \textit{opposite} the primary crystal axis $-\hat{z}$, the effect of the electric field will increase the refractive indices $n_y,n_z$ in one crystal and decrease the refractive indices $n_y,n_z$ in the other crystal. So the total average phase shift is, largely, the same for both helicity states. However the \textit{gradient} of the average phase shift is not the same for both helicity states as we will show. 

The opto-electric effect in RTP results in a change in refractive index to the two primary axes y and z of 
\begin{eqnarray}
\Delta n_z = - n^3_{z0} r_{33} (E_z^R-E_z^L) \\
 \Delta n_y = - n^3_{y0} r_{23} (E_z^R-E_z^L) 
\end{eqnarray}
with $(E_z^R-E_z^L) = 2 E_{z0}$. The helicity correlated refractive index gradient is then given by
\begin{eqnarray}
 \frac{d \Delta n_j}{dx_i} = - 3 n^2_{j0} r_{j3} \frac{d n_{j0}}{dx_i}E_{z0} -n^3_{j0} r_{j3} \frac{d E_{z0}}{dx_i}   
\end{eqnarray}
 where $j = 2,3$ (y,z) refers to the input polarization state of the beam along $\pm45^o$, the primary fast and slow axes of the RTP crystal.  The parameter $r_{23}$ is the electro-optic coupling coefficient between  $n_y$ and $E_z$ and $r_{33}$ is the electro-optic coupling coefficient between  $n_z$ and $E_z$. A gradient in the electric field or intrinsic refractive indices in each crystal gives rise to a voltage dependent phase gradient which steers the beam as a wedge does, producing an angle-difference (and hence position difference for throw distance D)
 \begin{equation}
 \theta_{air,defl}=\theta_{fc,x_i} (n-1) + L \frac{dn}{dx_i} 
\end{equation}
The steering due to these gradients in a single RTP crystal is given by
\begin{eqnarray}
\Delta \theta_{x_i} \big\rvert_{j} =&& 
-n^3_{j0} r_{j3} (\frac{d E_{z0}}{dx_i} L\nonumber\\
&&+(\theta_{fc,x_i} + 3 \frac{L}{n_{j0}}\frac{d n_{j0}}{dx} r_{j3})E_{z0} ) 
\end{eqnarray}
\begin{eqnarray}
D_x \big\rvert_{i}= &&\Delta_x \theta \big\rvert_{i} D= -n^3_{j0} r_{j3}D( \frac{d E_{z0}}{dx_i} L\nonumber\\
&&+(\theta_{fc,x_i}   + 3 \frac{L}{n_{j0}}\frac{d n_{j0}}{dx} r_{j3})E_{z0} )
\end{eqnarray}

We simply take the average of the steering along $\pm45^o$, the y and z primary crystal axes, to obtain the steering for H and V input polarizations:
\begin{eqnarray}
D_{x_i} \big\rvert_{H,V}=&& \frac{1}{2}(D_{x_i} \big\rvert_{y}+D_{x_i} \big\rvert_{z})= D_{x_i,fixed} \nonumber\\
&&-\frac{1}{2}(n^3_{y0} r_{23}+n^3_{z0} r_{33})D \frac{d E_{z0}}{dx_i} L 
\end{eqnarray}
where $D_{x_i,fixed}$ is the unchangeable intrinsic steering position difference from 1 crystal given by
\begin{eqnarray}
D_{x,fixed} =&& -n^3_{y0} r_{23}D((\theta_{fc}   + 3 \frac{L}{n_{y0}}\frac{d n_{y0}}{dx} r_{23})E_{z0} \nonumber\\
&& -n^3_{z0} r_{33}D((\theta_{fc}   + 3 \frac{L}{n_{z0}}\frac{d n_{z0}}{dx} r_{33})E_{z0}
\end{eqnarray}
The steering due to each of the two RTP crystals combines additively, resulting in a total angle difference averaged over y and z polarizations and summed over each crystal.
%\FloatBarrier

\subsubsection{Simulation} 

\begin{figure*}%[H]
    %\centering
      \begin{subfigure}{0.31\textwidth}%{0.28\textwidth}
                \caption{Grounded}
        \includegraphics[width=\textwidth]{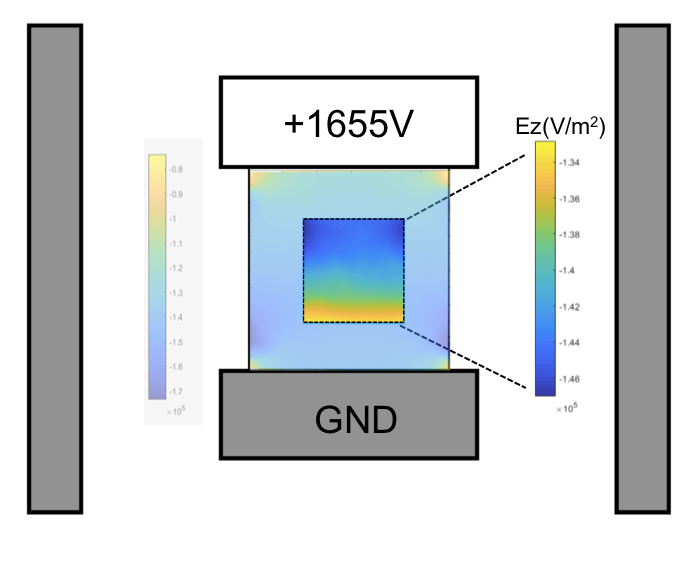}    
          \label{fig:heatmapGND}
      \end{subfigure}
        \begin{subfigure}{0.31\textwidth}
                \caption{Balanced}  
        \includegraphics[width=\textwidth]{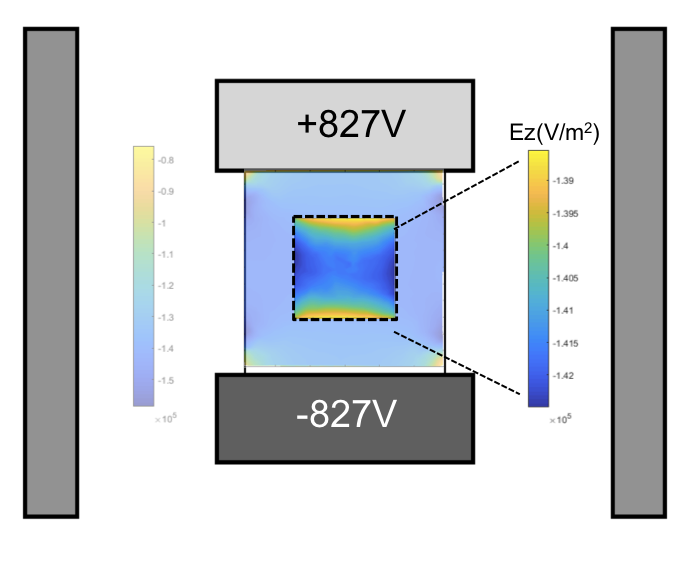}  
                    \label{fig:heatmapBalanced}
      \end{subfigure}
                 \begin{subfigure}{0.31\textwidth}
                 \caption{Reverse Grounded}
        \includegraphics[width=\textwidth]{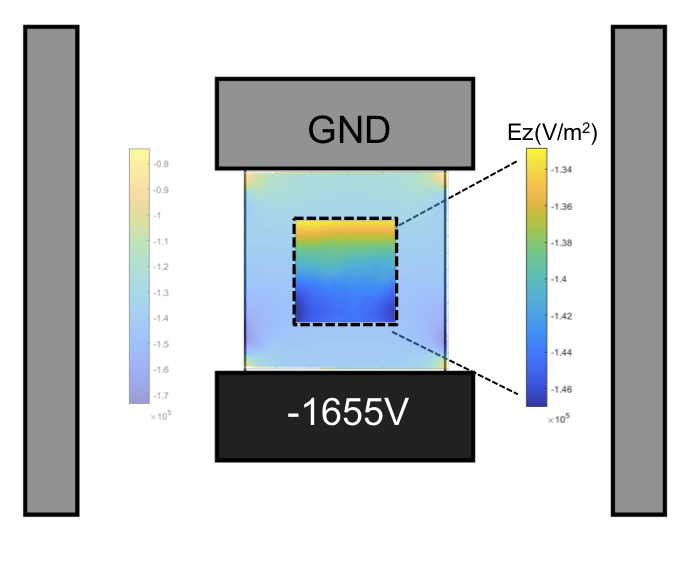}    
          \label{fig:heatmapRGND}
      \end{subfigure}
      
%      \hfill
      \begin{subfigure}{0.31\textwidth}%{0.19\textwidth}
                      \caption{Grounded}
        \includegraphics[width=\textwidth]{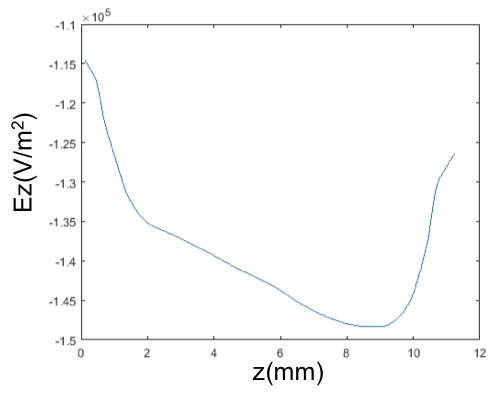}
         % \caption{Grounded}
          \label{fig:EzGND}
      \end{subfigure}        
            \begin{subfigure}{0.31\textwidth}
                            \caption{Balanced}  
        \includegraphics[width=\textwidth]{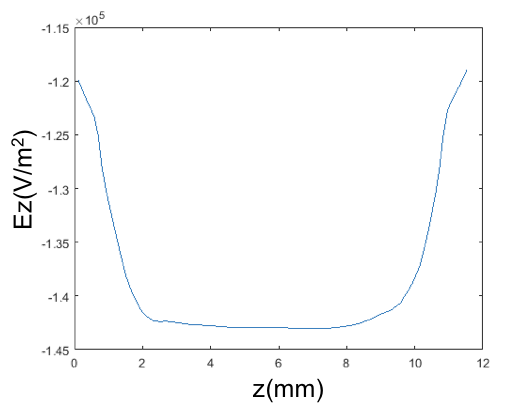}
         % \caption{Grounded}
          \label{fig:EzBalanced}
      \end{subfigure}   
            \begin{subfigure}{0.31\textwidth}
                             \caption{Reverse Grounded}
        \includegraphics[width=\textwidth]{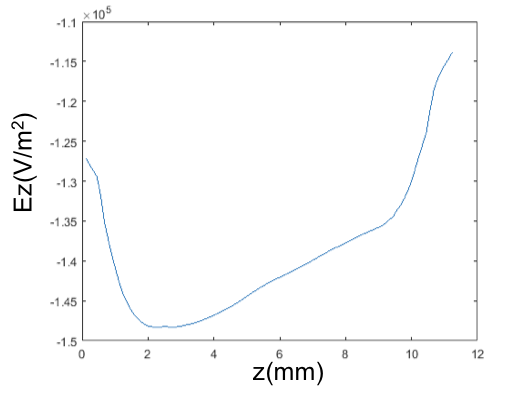}
         % \caption{Grounded}
          \label{fig:EzRGND}
      \end{subfigure}   
      
            \begin{subfigure}{0.31\textwidth}%{0.2\textwidth}
                            \caption{Grounded}
        \includegraphics[width=\textwidth]{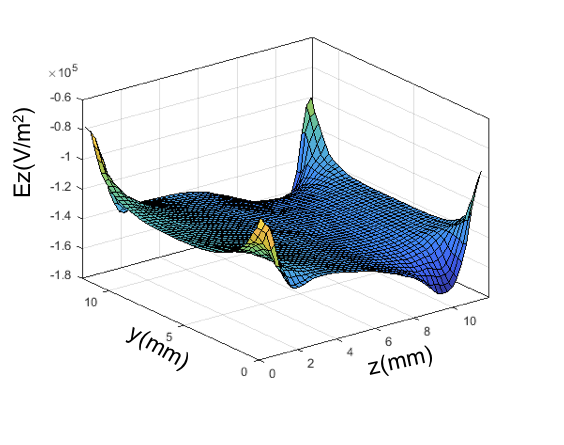}
        % \caption{Grounded}
          \label{fig:3DGND}
      \end{subfigure}
                      \begin{subfigure}{0.30\textwidth}
                                      \caption{Balanced}  
        \includegraphics[width=\textwidth]{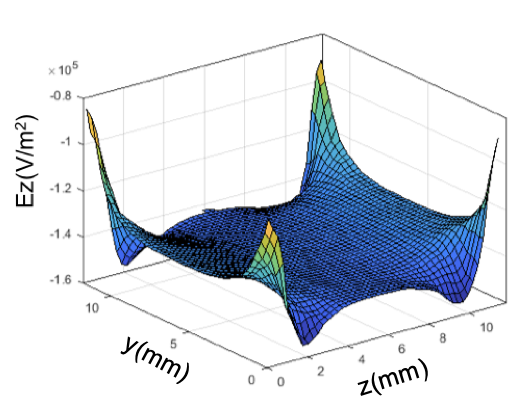}
        % \caption{Balanced}
          \label{fig:3DBalanced}
      \end{subfigure}
                       \begin{subfigure}{0.31\textwidth}
                                        \caption{Reverse Grounded}
        \includegraphics[width=\textwidth]{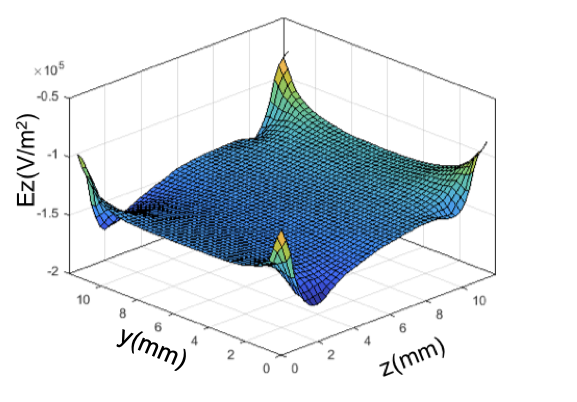}
       %  \caption{Reverse Grounded}
          \label{fig:3DRGND}
      \end{subfigure}
%      \hfill

      \begin{subfigure}{0.31\textwidth}%{0.2\textwidth}
                      \caption{Grounded}
        \includegraphics[width=\textwidth]{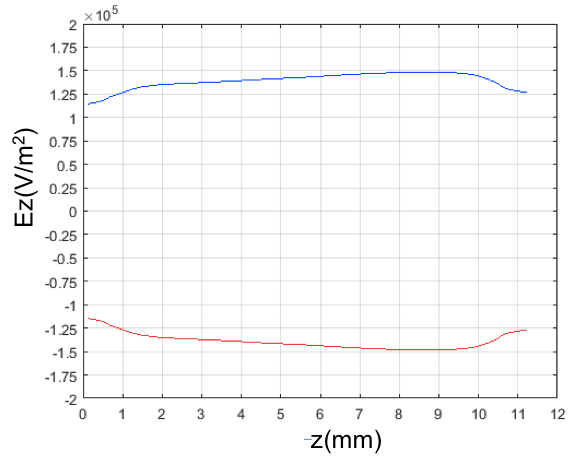}
         % \caption{Grounded}
          \label{fig:EzhelicityGND}
      \end{subfigure}            
      \begin{subfigure}{0.31\textwidth}
                      \caption{Balanced}  
        \includegraphics[width=\textwidth]{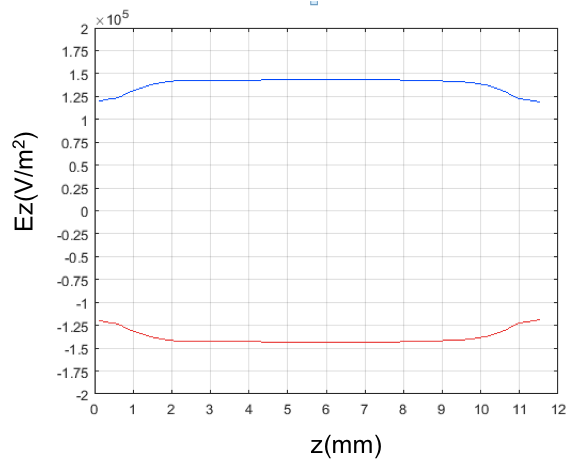}
         % \caption{Grounded}
          \label{fig:EzhelicityBalanced}
      \end{subfigure}        
      \begin{subfigure}{0.31\textwidth}
                       \caption{Reverse Grounded}
        \includegraphics[width=\textwidth]{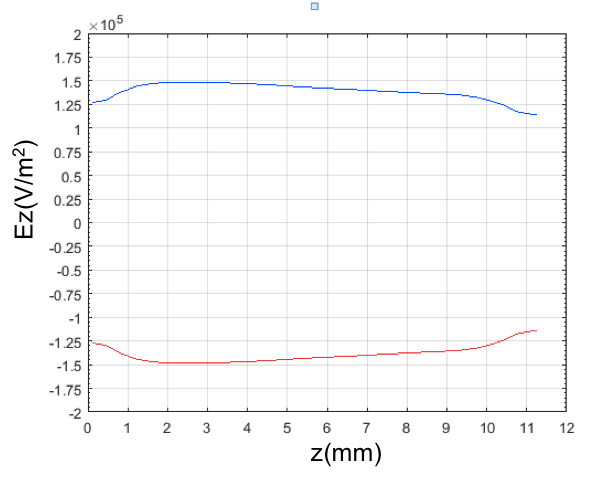}
       %   \caption{Grounded}
          \label{fig:EzhelicityRGND}
        \end{subfigure}
        \captionsetup{justification=raggedright,singlelinecheck=false}
\caption[RTP E-field Simulations]{
\label{fig:EfieldSim} Electric field calculation for RTP cell design. 
%Calculations for three configurations are shown. 
(a) Grounded configuration: in which one plate (bottom) is grounded for both helicity states. (b) Balanced configuration: in which voltage is applied symmetrically to the top and bottom plates (c) Reverse grounded configuration: in which the opposite plate (top) is grounded. (a),(b),(c) Electric field component Ez across crystal is illustrated as a color map. (d),(e),(f) Ez with respect to z (g),(h),(i) 3D plot of Ez with respect to both z and y (j),(k),(l) For each configuration, electric field with respect to z (vertical dimension) are shown for each helicity state as red and blue curves. %The slope of Ez with respect to z flips sign with helicity state in the red and blue curves.
}
\end{figure*}

We simulated the induced electric field gradients along the z-axis of each crystal in our RTP cell design. Fig. \ref{fig:EfieldSim} shows the electric field $E_z$ for three different voltage configurations with the same voltage difference between the top and bottom plate but differing in the voltage offset, $V_{\alpha pos}$ of the plates relative to the grounded side panels. We refer to these states as `grounded' configuration with a positive voltage shift, `balanced' configuration with no voltage shift, and `reverse grounded' configuration with a negative voltage shift. Fig. \ref{fig:heatmapGND} shows the electric field of the `grounded' configuration with a positive voltage shift $V_{\alpha pos}=+V_{\lambda/4}/2$, when the bottom plate is grounded at the same voltage as the side panels, inducing a maximal gradient in the +z direction.  Fig. \ref{fig:heatmapBalanced} shows the electric field of the `balanced' configuration, with no voltage shift $V_{\alpha pos}=0$, when the top and bottom plates have equal and opposite voltages relative to the grounded the side panels, inducing a symmetric field with zero first order term at the crystal center. Fig. \ref{fig:heatmapRGND} shows the electric field of the `reverse grounded' configuration, with a negative voltage shift $V_{\alpha pos}=-V_{\lambda/4}/2$, when the top plate is set the same voltage as the side panels, inducing a maximal gradient in the -z direction. By keeping the voltage difference between the plates constant, and changing the voltage shift relative to the grounded side panels, we control the electric field gradient without changing the electric field $E_z$ magnitude, and hence control the position differences without changing the $\delta$-phase much. The voltages are set independently for each helicity state such that when the polarity of the electric field is switched, the sign of the induced electric field gradient reverses as well. 
%Caryn you have to add captions to all the subfigures

%Caryn you need to adjust your pngs to have at least 600dpi

\begin{figure}%[H]
    %\centering
      \begin{subfigure}{0.23\textwidth}
        \includegraphics[width=\textwidth]{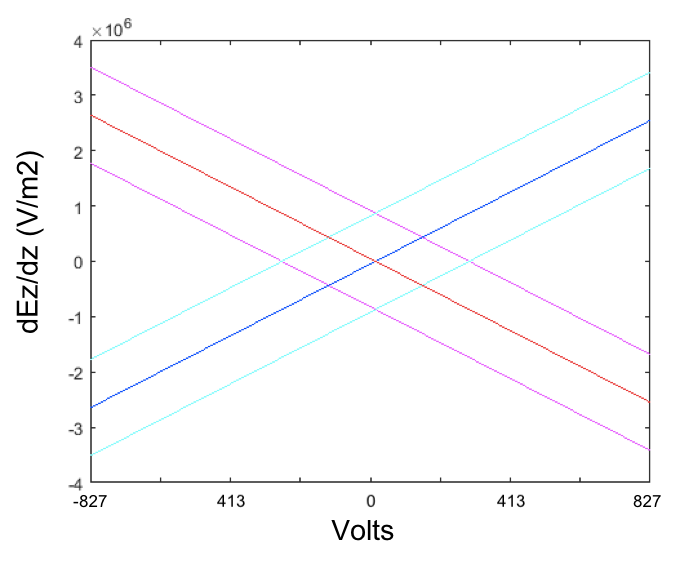}    
         \caption{}
          \label{fig:EgradientRL}
      \end{subfigure}
%      \hfill
      \begin{subfigure}{0.23\textwidth}
        \includegraphics[width=\textwidth]{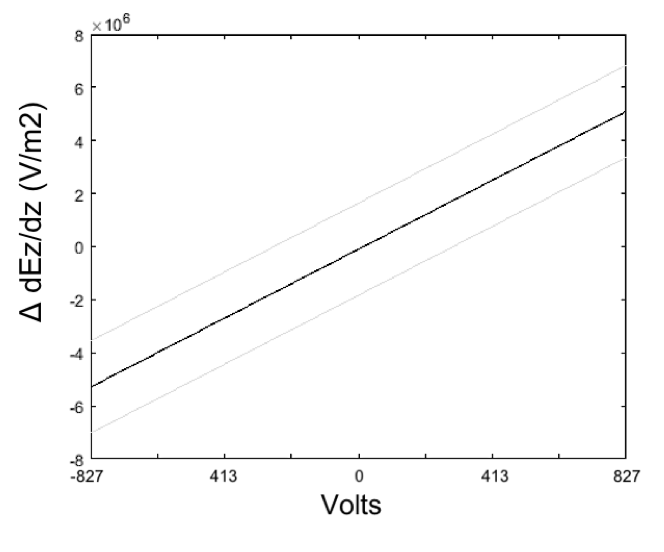}
         \caption{}
          \label{fig:DeltaEgradients}
      \end{subfigure}
      \captionsetup{justification=raggedright,singlelinecheck=false}                  
\caption[Simulation - Steering Gradient]{
\label{fig:EGradientSim} Electric Field Simulations. Simulations of steering gradient in RTP cell design with respect to steering voltage $V_{\alpha pos}$ spanning Grounded(827V), Balanced(0V) and Reverse Grounded(-827V) configurations. (a) Induced electric field gradient $d(dE_z/dz)$ on the y-axis vs. $V_{\alpha pos}$ on the x-axis. Red curve represents right helicity state, while blue curve represents left helicity state. Pink and aqua error bands represent uncertainty of 1mm in beam position across crystal face. (b) Helicity correlated difference in electric field gradient between right and left states as a function of steering voltage $V_{\alpha pos}$. Grey error bands represent uncertainty of 1mm in beam position across crystal face.}
\end{figure}

Fig. \ref{fig:EgradientRL} shows the electric field gradient $\frac{d E_z}{d z}$ for each helicity state as a function of voltage shift $V_{\alpha pos}$, corresponding to scanning from the grounded configuration, across the balanced configuration ($V_{\alpha pos}=0$), to the reverse grounded configuration.  The red line with pink uncertainties corresponds to positive helicity state, and the blue line with green uncertainties corresponds to the negative helicity state. The uncertainty in $\frac{d E_z}{d z}$ is determined by the resolution of the mesh in the 3D simulation; also included in the $\frac{d E_z}{d z}$ error bar is the uncertainty from centering the laser position on the crystal within $\pm1mm$ of center. Fig. \ref{fig:DeltaEgradients} corresponds to the helicity correlated change in the electric field gradient  $\Delta \frac{d E_z}{d z} =\frac{d E_z^R}{d z} -\frac{d E_z^L}{d z} = 2 E_{z,0}$. We find $\frac{d E_{z,0}}{dz} =(3100 \pm1100 V/m2/V_{\alpha pos}) V_{\alpha pos}$. 

%Caryn this figure has yellow in it, you should probably change that color. and indicate what the heck the colored bands mean 

Examining the steering due to the first RTP crystal along the primary axis $z_1$, for $\lambda=780$nm,$L=10$mm, $r_{33}=35$pm/V, $r_{23}=12.5$pm/V, $n_y = 1.7832$, $n_z = 1.8673$, for horizontal input polarization
\begin{equation}
\Delta \theta_{z1} = D_{z1} /D =  -\frac{1}{2}(n^3_{y0} r_{23}+n^3_{z0} r_{33}) L \frac{d E_{z1,0}}{dz_1} 
\end{equation}
for $\frac{d E_{z1,0}}{dz_1} =(3100 \pm1100 V/m2/V_{\alpha pos}) V_{\alpha pos}$, we obtain the steering sensitivity to voltage
\begin{eqnarray}
\frac{d \Delta \theta_{z1}}{d V_{\alpha pos}} &&=  -\frac{1}{2}(n^3_{y0} r_{23}+n^3_{z0} r_{33})L \frac{d E_{z1,0}}{dz_1}\nonumber\\
&& = -4.6 \pm1.7 nrad/V %4.633 3.0 to -6.3nrad/V
\end{eqnarray}
where each polarization component along $z_1$ and $y_1$ was steered by:
\begin{eqnarray}
\frac{d \Delta \theta_{z1}}{d V_{\alpha pos}} \big\rvert_{z1}= && -n^3_{z0} r_{33}L \frac{d E_{z1,0}}{dz_1} \nonumber\\
= &&-7.1 \pm2.5 nrad/V\\
\frac{d \Delta \theta_{z1}}{d V_{\alpha pos}} \big\rvert_{y1}=&&  -n^3_{z0} r_{33}L \frac{d E_{z1,0}}{dz_1} \nonumber\\
= &&-2.2 \pm0.8 nrad/V 
\end{eqnarray}

\section{Design}

\subsection{Eight-High-Voltage Design} \label{sec:8HVdesign}

All crystals suffer from some degree of non-uniformity. In order to counteract this non-uniformity and minimize helicity correlated beam asymmetries, an innovation in the design of the RTP Pockels Cell was required. We used grounded plates to induce fringe-electric fields. By controlling the electric field gradients, the helicity correlated position differences could be suppressed. 

\begin{figure}%[H]
%\centering
\begin{subfigure}{0.2\textwidth}
\includegraphics[width=\textwidth]{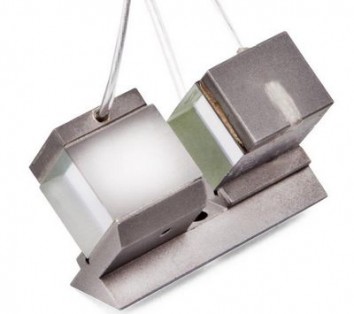}
\caption{\label{fig:raicolpic} }
\end{subfigure}
\begin{subfigure}{0.25\textwidth}
\includegraphics[width=\textwidth]{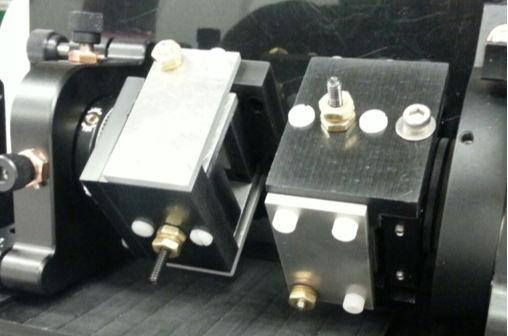}
\caption{\label{fig:mypic}  }
\end{subfigure}
\captionsetup{justification=raggedright,singlelinecheck=false}
\caption{RTP Pockels Cell Designs (a) Commercial RTP cell (Raicol \cite{RaicolSpecSheet}) with common voltage applied to both crystals. (b) Side Panel Design. Crystals and electrodes are contained within Delrin\textsuperscript{\textregistered} housing on independent mounts with independent voltages for each crystal. Grounded Aluminum side panels are mounted along the sides of each crystal. }
\end{figure}

\begin{figure*}%[H]
%\centering
\includegraphics[width=0.8\textwidth]{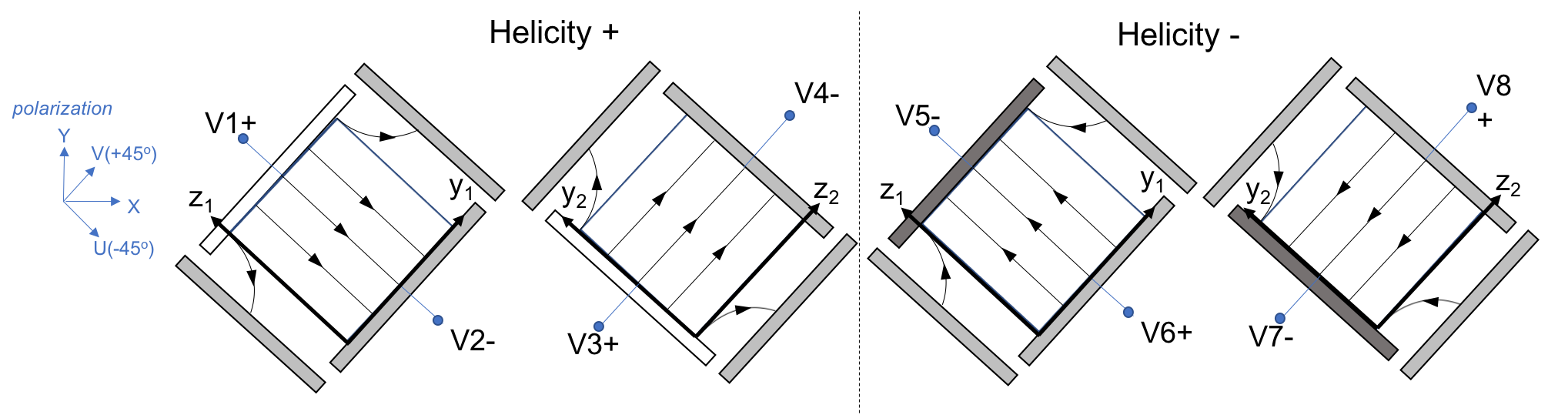} %8HVsystem.png
      \captionsetup{justification=raggedright,singlelinecheck=false}  
\caption[Electric field gradients]{\label{fig:8HVsystem}  Defining axes of RTP cell and configuration of 8 HVs. Polarization defining axes X,Y,U,V are shown on the far left in blue. For + helicity state (left), votatges V1,V2 are applied to crystal 1 and voltages V3,V4 are applied to crystal 2. For - helicity state (right), voltages V5,V6 are applied to crystal 1 and voltages V7,V8 are applied to crystal 2. Figure illustrates conceptual electric lines case of $V_{\alpha pos, U}= V_{\alpha pos, V}=V_{\lambda/4}>0$ with grounded side panels. The primary axes of each crystal y,z are also shown.}
\end{figure*}

 In commercial designs, the 2 crystals have a common grounded plate and two high voltage (HV) plates on the top of each crystal as shown in Fig. \ref{fig:raicolpic} and a common HV voltage is applied to the top plates of both crystals. We gain the ability to control the electric field gradients by a simple design change. We separated the two crystals' ground plates, so each crystal electric field is independently controlled, added grounded side-panels near the sides of each crystal, and added more several HV power supplies so that the voltage on each of the 4 HV plates can be independently set to different values as illustrated in Fig. \ref{fig:mypic} and Fig. \ref{fig:8HVsystem}. There are 8 independent voltages in total, for each of the 4 HV plates on the two crystals for both helicity states. We also designed the cell mount with additional degrees of freedom to have control over the relative pitch, yaw, roll, horizontal and vertical translation between the two crystals as well as the overall pitch, yaw, roll, horizontal and vertical translation degrees of freedom. The Pockels cell is designed with two RTP crystals (Raicol 12x12x10mm, AR coated), with full angular and translational control of each crystal, and grounded side-panels in addition to 4 HV plates. 8 high voltages are applied to the cell, driven with a opto-diode switch. The laser polarization states are switched within 12$\mu$s, while controlling the direction and intensity of the laser beam to keep it extremely symmetric (at nm-level and ppm-level). 

This design allows for a straightforward control over the electric field gradient along the z-axis of each crystal without changing the magnitude of the E-field very much. Since the voltage setting controls the electric field gradient along the z-axis of each crystal, it controls the steering along the z-axis for each crystal. The first RTP crystal's electric field gradient controls and the steering along $-45^o$ and the second crystal's electric field gradient controls the steering along $+45^o$ as shown by the orientations in Fig. \ref{fig:8HVsystem}.

We have the ability, with our 8 independent voltages, to control the delta-phase, the alpha-phase, and to create a delta-phase gradient or alpha-phase gradient. While position differences in S1 are caused by a delta-phase gradient, steering is controlled by an alpha-phase gradient. The voltage shift in each crystal which induces steering we refer to as $\alpha$-position voltage, because it controls the alpha-phase gradient. 
%\footnote{referred to as `PITA'-position voltage in elogs.  The name ``PITA-position-voltage" is a misnomer and we should actually call it ``alpha-position-voltage" because it induces an alpha-phase gradient and it is that which causes steering.} 
The first crystal, with its z-axis along U, controls the steering along the U-direction with ``alpha-position-U" voltage $V_{\alpha pos,U}$.  The second crystal, with z-axis along V controls the steering along the V-direction, with ``alpha-position-V" voltage $V_{\alpha pos,V}$.  
 
The equations describing how the voltage settings are used to induce steering-like position differences via $V_{\alpha pos, U/V}$ are shown in Table \ref{tab:8HVlist} .

\begin{table*}%[H]
%\centering
\begin{tabular}{|r|r|r|r|r|r|}
 \hline
$_{HV}$ &$_{hel.}$&  $_{crystal}$ &$_{pol.}$ & $_{orien.,dir.}$ &$_{equation}$  \\\hline
1&+&1&+&$_{+z1,+U}$&$V_1 = V_{\lambda/4} +V_{\alpha,1} +V_{\Delta,1} +V_{\alpha pos, U} + V_{\delta pos, U}$\\\hline
2&+&1&-&$_{-z1,-U}$&$V_2 = -V_{\lambda/4} - V_{\alpha,1} - V_{\Delta,1} +V_{\alpha pos, U} +V_{\delta pos, U}$\\\hline
3&+&2&+&$_{-z2,-V}$&$V_3 = V_{\lambda/4} +V_{\alpha,2} + V_{\Delta,2} + V_{\alpha pos, V} + V_{\delta pos, V}$\\\hline
 4&+&2&-&$_{+z2,+V}$&$V_4 = -V_{\lambda/4} -V_{\alpha,2}- V_{\Delta,2} + V_{\alpha pos, V} +  V_{\delta pos, V}$\\\hline
5&-&1&-&$_{+z1,+U}$&$V_5 = -V_{\lambda/4} - V_{\alpha,1}+ V_{\Delta,1} - V_{\alpha pos, U}+ V_{\delta pos, U}$\\\hline
6&-&1&+&$_{-z1,-U}$&$V_6 = V_{\lambda/4} +V_{\alpha,1}- V_{\Delta,1}  - V_{\alpha pos, U} + V_{\delta pos, U}$\\\hline
7&-&2&-&$_{-z2,-V}$&$V_7 = -V_{\lambda/4} -V_{\alpha,2}+V_{\Delta,2} - V_{\alpha pos, V} +  V_{\delta pos, V}$\\\hline
8&-&2&+&$_{+z2,+V}$&$V_8 = V_{\lambda/4} +V_{\alpha,2}- V_{\Delta,2} - V_{\alpha pos, V} +  V_{\delta pos, V}$\\\hline
\end{tabular}
      \captionsetup{justification=raggedright,singlelinecheck=false}  
\caption{\label{tab:8HVlist} Correspondence between each of the 8HVs and helicity state, which crystal the HV is applied to, the polarity of the HV, the orientation of the electrode where the HV is applied, and the equation that determined the HV value as a function of relevant variables. }
\end{table*} 

The voltages relate to the quarter-wave voltage via:
\begin{eqnarray}
V_{\lambda/4}  = &&(V_1-V_2 +V_3-V_4-V_5+V_6-V_7+V_8)/8\nonumber\\ 
&&- (V_{\alpha,1} + V_{\alpha,2})/2
\end{eqnarray}
Our 8 degrees of freedom translate into:
\begin{description}
\item[1,2]
  An alpha-phase for crystal 1 $V_{\alpha,1}$ and 
%\item[2]
  An alpha-phase for crystal 2 $V_{\alpha,2}$ (where we tend to keep these equal, and combine them into to one parameter $V_{\alpha} =V_{\alpha,1}=V_{\alpha,2} $)
\begin{eqnarray}
&&V_{\alpha,1} = (V_1-V_2-V_5+V_6)/4 - V_{\lambda/4}   \\
&& V_{\alpha,2} = (V_3-V_4-V_7+V_8)/4 - V_{\lambda/4} 
\end{eqnarray}
\item[3,4]
 A delta-phase for crystal 1 $V_{\Delta,1}$ and 
% \item[4]
  a delta-phase for crystal 2  $V_{\Delta,2}$ (where we tend to keep these equal, and combine them into to one parameter $V_{\Delta} =V_{\Delta,1}=V_{\Delta,2} $)
  \begin{eqnarray}
  &&V_{\Delta,1} = (V_1-V_2+V_5-V_6)/4 \\
 &&V_{\Delta,2} = (V_3-V_4+V_7-V_8)/4
\end{eqnarray}
\item[5,6]
An alpha-phase-gradient for crystal 1, along its z axis, used for steering along $+45^o$ and 
%\item[6]
an alpha-phase-gradient for crystal 2 along its z axis, used for steering along $-45^o$
\begin{eqnarray}
&&V_{\alpha pos, U} = (V_1+V_2-V_5-V_6)/4 \\
&& V_{\alpha pos, V} = (V_3+V_4-V_7-V_8)/4
\end{eqnarray}
\item[7,8]
 A delta-phase-gradient for crystal 1, along its z axis $+45^o$, and finally a delta-phase-gradient for crystal 2, along its z axis $-45^o$, though we tend to not use the delta phase gradient and  keep $V_{\delta pos, V}=V_{\delta pos, U}=0$. %\footnote{In principle we could also induce a delta-phase gradient, but we do not currently do this. It would alter the Aq gradient in S1, and for no analyzer ,it would have the effect of shifting the average position of the beam, but not induce position differences. }
 \begin{eqnarray}
&&V_{\delta pos, U} = (V_1+V_2+V_5+V_6)/4 \\
&&V_{\delta pos, V} = (V_3+V_4+V_7+V_8)/4
\end{eqnarray}
 \end{description}
 
In table \ref{tab:8HVlist}, helicity ``+" corresponds to the Pockels cell acting as a QWP with fast axis along $+45^o$ relative to the horizontal %\footnote{X in a right-handed XYZ coordinate system where Z is the beam propagation axis}
, such that horizontal input polarization $ \vec{H} =  \begin{tiny} \begin{bmatrix}
    1     \\
    0      
\end{bmatrix} \end{tiny}$
 (IHWP out) becomes right-circularly polarized $ \vec{R} =\frac{1}{\sqrt{2}} \begin{tiny} \begin{bmatrix} 
    1     \\
    -i      
\end{bmatrix} \end{tiny} $, carrying spin-angular momentum $-\hbar$.  Helicity ``-" corresponds to the Pockels cell acting as a QWP with fast axis along $-45^o$ relative to the horizontal, such that horizontal input polarization $ \vec{H} = \begin{tiny} \begin{bmatrix}
    1     \\
    0      
\end{bmatrix} \end{tiny}$
 (IHWP out) becomes left-circularly polarized $ \vec{R} =\frac{1}{\sqrt{2}} \begin{tiny} \begin{bmatrix} 
    1     \\
    i      
\end{bmatrix} \end{tiny} $, carrying spin-angular momentum $+\hbar$. In the photocathode, upon core-level excitation by circularly polarized light, the angular momentum of the light, or helicity, is transferred to the emitted photoelectron. The angular momentum of the emitted photoelectron is the sum of the helicity and the orbital magnetic quantum number of the initial state \cite{Fumihiko2018}. %\footnote{A Mott measurement (on the HallB laser) determined the sign of the beam polarization to be -86.6\% for IHWPout and for 87.2\% for IHWPin \cite{RiadMott}}
 
\begin{figure}%[H]
%\centering
\includegraphics[width=0.49\textwidth]{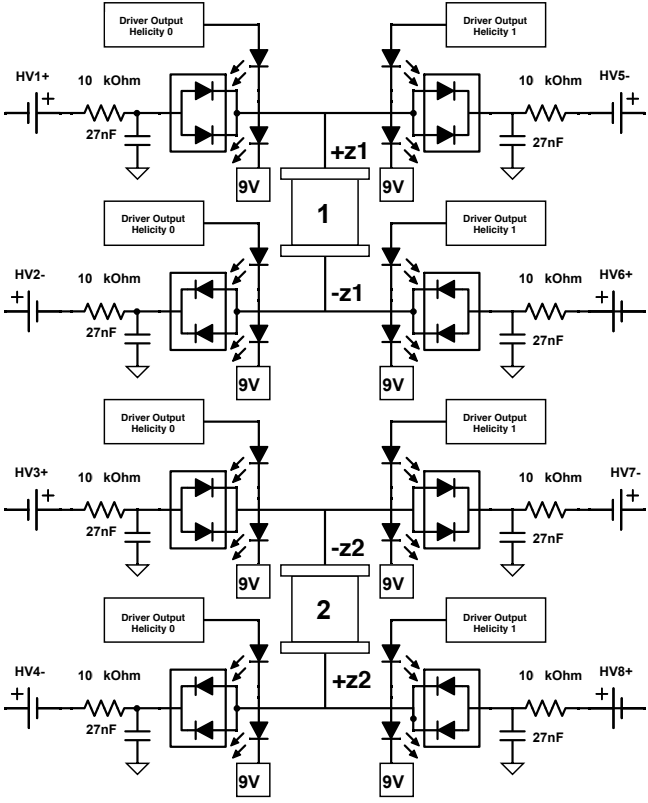}%{8HVdriverRTP.png}
      \captionsetup{justification=raggedright,singlelinecheck=false}  
\caption[Diagram of the 8HV driver configuration with optodriver]{\label{fig:8HVdriverRTP} Diagram of the 8-HV driver configuration with optodriver. Each of the 4 electrodes attached to crystals 1 and 2 has its own driver that switches between positive and negative polarity HV triggered by helicity signal. Each HV supply is buffered by an RC circuit before passing through 2 optodiodes in parallel, triggered by 2 LEDs in series powered by 9V. }
\end{figure}

The HV driver is composed of an optocoupler system. Each of the 8 HV's are driven by two high-voltage optodiodes in parallel. %\footnote{An optodiode is a high-voltage diode which switches states upon exposure to IR light and can be controlled using LEDs. The optodiodes are made by a Voltage Multipliers Inc. (VMI) part number OZ100SG. Our design has two optodiodes in parallel to allow for more current to flow to the Pockels cell for a faster transition}. 
Each of the two optodiodes switches states upon exposure to IR light and can be controlled using two 1W LED's in series, pulsed at the beginning of each helicity cycle for $\sim20\mu s-40 \mu s$, powered by $8V-9V$ of DC voltage ($4-4.5V$ on each LED). The circuit diagram is shown in Fig. \ref{fig:8HVdriverRTP}. 

\begin{figure}%[H]
  %\centering
 %   \begin{subfigure}{0.45\textwidth}
   %     \includegraphics[width=\textwidth]{RTPmount.png}
   %       \caption{}
   %       \label{fig:RTPmount}
  %    \end{subfigure}
%    \begin{subfigure}{0.3\textwidth}
        \includegraphics[width=0.4\textwidth]{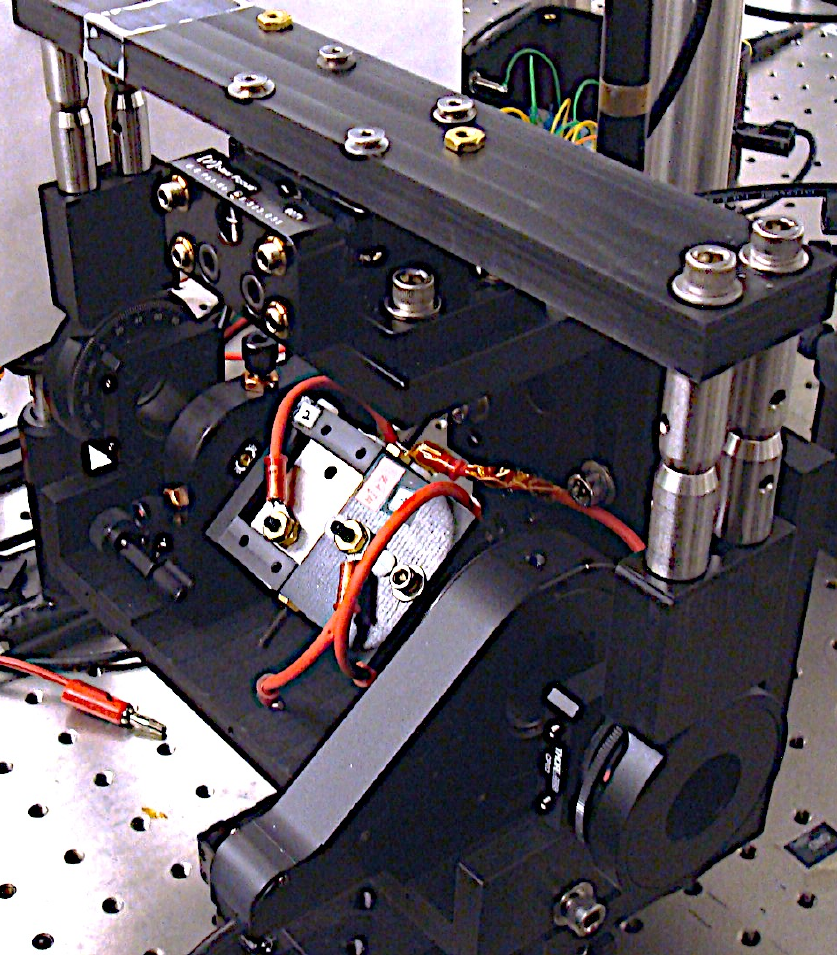}
   %       \caption{}
   %       \label{fig:RTPcellpic}
   %   \end{subfigure}
         \captionsetup{justification=raggedright,singlelinecheck=false}  
      \caption{
 \label{fig:RTPmountpic} Assembled RTP mount. Each crystal and its electrodes are contained within Delrin\textsuperscript{\textregistered} housing. Al grounded side panels are located on the sides of each crystal. Crystal 1 is mounted on a rotation stage for independent roll control. Crystal 2 is mounted on a 6-axis mount for independent angle and position control. The base of the dual crystal assembly is mounted onto rotation stages for overall roll control. The entire assembly is attached to a 4axis angle mount for overall angle control as well as motorized translation stages for position control. }
  \end{figure}

 Fig. \ref{fig:RTPmountpic} shows the components of the cell system and the physically assembled cell. The two crystals have individual Al plates, Al side panels, and are held together by a DuPont Delrin\textsuperscript{\textregistered} sandwich design, each mounted individually to a rotation stage and a 6axis optics mount. The whole assembly is mounted on a 4axis angle mount which adjusts pitch and yaw, a rotation stage which adjusts roll, and two translation stages for X and Y control. To potentially increase Pockels cell lifetime, our cell design has no silver cement on the crystals or electrodes.
%A full list of components required is found in \cite{RTPcomponentlistelog},  the CAD file can be found in \cite{RTPCADelog} and 
%Caryn maybe include this footnote in the main text
%Raicol crystals specs are located in
The crystals are Raicol \cite{RaicolSpecSheet} RTP Matched Pair with 12x12mm aperture, 10mm length crystals, length mismatch between crystal pair $<$2 $\mu$m, extinction ratio  $>$23 dB in a clear aperture of 10 mm, face cut parallelism $<$10arcsec, AR coated at 780 nm with R$<$0.2\%, and capacitance 4-6 pF. The HV power supplies are IGES 4W 2kV BPS series. For the optodiode driver switched by LEDs, we used OSRAM-Opto-Semiconductors Platinum Dragon LEDs.  The optocouplers are VMI OZ100HG. Out mount includes Thorlabs PT1-Z8 translation stages, Newport 9071 4-axis mount, Thorlabs PRM1Z7 motorized rotation mount, and Thorlabs K6XS 6axis mirror mount.

%\footnote{ The crystals are Raicol RTP Matched Pair 2x(12x12x10 mm) at 780 nm, 12x12mm aperture, 10mm length crystals, with length mismatch between crystal pair $<$2 $\mu$m, with extinction ratio  @>23 dB in clear aperture 10 mm , face cut parallelism $<$10arcsec, AR coated at 780 nm, R<0.2\%, and capacitance 4-6 pF. The power supplies are IGES 4W 2kV BPS series. The LED's are OSRAM-Opto-Semiconductors Platinum Dragon.  The optocouplers are VMI OZ100HG. The mount includes thorlabs PT1-Z8 translation stages, Newport 9071 4-axis mount, thorlabs PRM1Z7 motorized rotation mount, thorlabs K6XS 6axis mirror mount, and an Edmund optics 3x3 right angle bracket.} . 
%Several spares may need to be made when the system is installed at JLab, especially since Moller will runs for such a long time. 
%Caryn maybe don't include the pretty pics because they are too hard to explain

%\begin{figure*}%[H]
%\centering
%\includegraphics[width=0.9\textwidth]{RTPprettypics.png}
%\caption[RTP cell design]{\label{fig:Prettypics} RTP cell design}
%\end{figure*}

%Caryn you cut out the optodriver section 
%\subsection{Opto-driver}
%Caryn you took out the transition times section
%\subsection{Considerations for Fast Switching}
%Caryn you removed to 4peak effect section
%\subsection{Pockels Cell History Effect} \label{Sec:4peak}

%JLab observed crystal degradation in a commercial RTP cell (with metal cement) which had the symptom of decreased responsiveness to voltage, ultimately failure to hold voltage, and crystal graying and transmission reduction as shown in Fig. \ref{fig:RTPDCdamageDarkening}. This occurred when a DC voltage was (on average, periodically) applied for a long period of time. A possible explanation is ion migration from silver oxide across the crystal face \cite{ionmigration} from the silver cement. To potentially increase Pockels cell lifetime, our cell design has no silver cement on the crystals or electrodes.

\section{Characterization}

The RTP cell was fully characterized as regards the intensity asymmetry, position differences, and spot-size asymmetry dependence on position, angle, roll and voltage. 

\begin{figure}%[H]
%\centering
\includegraphics[width=0.15\textwidth]{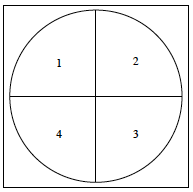}% Here is how to import EPS art
      \captionsetup{justification=raggedright,singlelinecheck=false}  
\caption[Quad-photodiode layout]{\label{fig:QPD} Quad-photodiode layout \cite{SilwalThesis}. 4 quadrants of photodiodes detect beam power as well as beam position.  }
\end{figure}

Polarized light was incident on the RTP Pockels cell, which was set to flip between $\pm$QWV, alternating helicity states, and the transmitted light was detected downstream of the Pockels cell (as in Fig. \ref{fig:LayoutLaserTable}). A quad-photodiode (Fig. \ref{fig:QPD}) detected the transmission and beam position for each helicity state, our data acquisition system (DAQ) integrated over each helicity window, and our data analysis code formed an asymmetry (or pair difference) between the signals for right and left helicity states. Measured parameters were formed by taking difference between successive states of opposite helicity. The intensity asymmetry $A_I$, position difference in x $D_x$, and position difference in y $D_y$, were defined as:
\begin{eqnarray}
&&A_I = \frac{I^R-I^L}{I^R+I^L} \;,\; D_x = x^R - x^L \;,\; D_y = y^R-y^L 
\end{eqnarray}

The intensity of the transmitted beam is proportional to the sum of the photodiode pad signals and the horizontal and vertical positions of the beam are computed through taking differences between the pad signals as follows:
\begin{eqnarray}
I_{sum}= I_1+I_2+I_3+I_4\\
x = (x-const) \frac{(I_2+I_3)-(I_1+I_4)}{I_1+I_2+I_3+I_4} \\
 y = (y-const) \frac{(I_1+I_2)-(I_3+I_4)}{I_1+I_2+I_3+I_4}  
\end{eqnarray}

where $I_1$, $I_2$, $I_3$ and $I_4$ are the responses of the pads 1,2,3 and 4 respectively to the incident light and x-const, y-const are the calibrated proportionality factors. The calibration procedure \cite{SilwalThesis} determined the relative gain and pedestal(accounting for non-linearity in response) for each pad: $I_i = gain_i\times(S_i-ped_i)$. 

Multiple lasers were used in the Pockels cell characterization as described in subsequent sections. These laser table studies were done at several helicity flip rates including 30Hz, with $\sim33ms$ integration windows, and at 240Hz, with $\sim4ms$ integration windows. To minimize 60Hz noise, the helicity flip rate was line synced to 60Hz and quartet patterns $+--+$ (or octet patterns $+--+-++-$) were used to cancel out 60Hz noise \cite{HelicityControlBoard} with randomly sequenced helicity patterns.

\subsection{\label{sec:translationscan}Translation scan}

The Pockels cell was mounted on motorized stages which allowed for horizontal and vertical position control. The cell sensitivity to position was measured by translating the cell with the Thorlabs stages. A LD785-SED30 diode CW laser was used %(18.10$^o$C,180mA,75mW) 
and sent through a single mode fiber and Glan-Taylor polarizer, producing 4.3 mW of horizontally polarized 785nm light. The 785nm horizontally polarized light was incident on the RTP Pockels cell, which was set to flip between $\pm$QWV, and the transmitted light analyzed with a vertical polarizer downstream of the Pockels cell (as in Fig. \ref{fig:LayoutLaserTable}), and detected on a quad-photodiode. The cell transverse position was scanned over an X/Y grid pattern and the intensity asymmetry dependence on cell position was measured. 

\begin{figure}%[H]%make this 3D so looks like TranslationScanRun2652.png
%  \centering
   % \begin{subfigure}{0.45\textwidth}
      %  \includegraphics[width=\textwidth]{Run4340_AqvsXplusY.png}
      %    \caption{Run 4340 \cite{elog798}}
     %     \label{fig:translation}
   %   \end{subfigure}
   % \begin{subfigure}{0.45\textwidth}
        \includegraphics[width=0.45\textwidth]{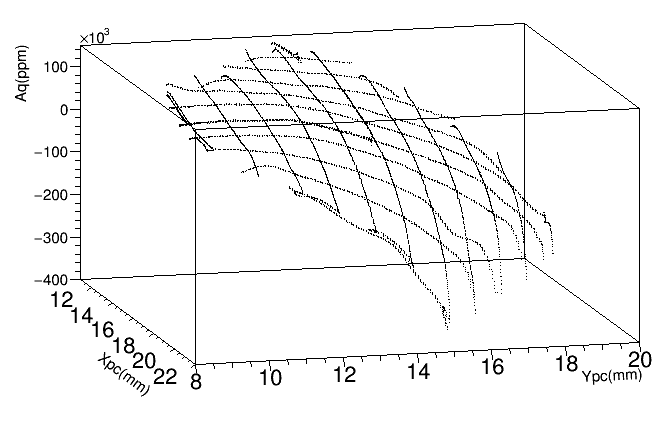}
        %  \caption{Run 4340 (approximate reproduction in 3D)}
      %    \label{fig:translationredo}
   %   \end{subfigure}
         \captionsetup{justification=raggedright,singlelinecheck=false}  
      \caption{
 \label{fig:translationscan}Translation scan. The Pockels cell is scanned by translation stages horizontally along X and vertically along Y and the intensity asymmetry Aq in S1 (after polarized beam passes through vertical analyzer) is detected and shown along the z-axis. }
  \end{figure}

The translation scan, Fig. \ref{fig:translationscan}, shows the intensity asymmetry $A_I$ (in S1) as a function of Pockels cell transverse position in the X/Y place perpendicular to the beam propagation axis.  This measurement of $A_I$ with respect to position shows the birefringence gradient inherent in the system. Empirically we observed an asymmetry gradient along $v \equiv \frac{x+y}{\sqrt{2}}$ due to both crystals combined of $ \frac{dA_q}{dv} \sim 20,000-50,000ppm/$mm, which for $w=1$mm gives position differences of magnitude: 
\begin{eqnarray}
 D_v =  D_{y1}+ D_{z2} = \frac{\frac{dA_q}{d v} w^2}{2} = 10 -25 \mu m\\
 D_x \sim D_y \sim D_v/\sqrt{2} = 7.1-17.7\mu m
\end{eqnarray}
The parallelism of the crystals' face cuts; the electric field non-uniformity; the variation in intrinsic refractive index along the crystals growth axis; and stress gradients in the crystal all affect the birefringence gradient. This translation scan motivated the new cell design: because these gradients are intrinsic to the RTP system, and give rise to analyzing-like position differences which cannot be minimized by translational alignment, it became necessary redesign the cell to zero out position differences using voltage induced steering.

We note that in an RTP cell system, it is very challenging to obtain a flat asymmetry across the cell face, especially since the crystal is so highly birefringent ($\delta n \sim 0.1$), and there is an intrinsic refractive index gradient along the z-axis growth direction. Even a slight non-parallelism in face cuts (a slight length gradient) can lead to a significant gradient in the asymmetry. We have also measured face cut parallelism of each crystal by examining the back reflections of the front and back crystal faces, and obtained for the 1st crystal
\begin{eqnarray}
&&\theta_{fc,z1} = \frac{d L}{d z_1} =-0.079\pm 0.01mrad\\
&& \theta_{fc,y1} = \frac{d L}{d y_1} =0.024\pm 0.01mrad 
\end{eqnarray}
and for the 2nd crystal
\begin{eqnarray}
&& \theta_{fc,z2} = \frac{d L}{d z_2} =-0.055\pm 0.01mrad\\
&& \theta_{fc,y2} = \frac{d L}{d y_2} =0.015 \pm 0.01mrad 
 \end{eqnarray} 
The bound on the parallelism 
\begin{equation}
\theta_{fc} = \frac{d L}{d x_i} \approx 0.01-0.1mrad 
\end{equation}
in these RTP crystals indicates for a 1mm beam, the induces position differences is
\begin{equation}
D_{\partial L,x_i} =  -\frac{w^2 \pi}{2 \lambda} (n_{0,y}-n_{0,z})\theta_{fc} \approx 1.7-17um 
\end{equation}
 The results indicate the major contribution to the birefringence gradient comes more from the intrinsic refractive index gradient than from face cut non-parallelism.

\subsection{\label{sec:Anglescans}Angle scans}

\begin{figure}%[H]
\includegraphics[width=0.49\textwidth]{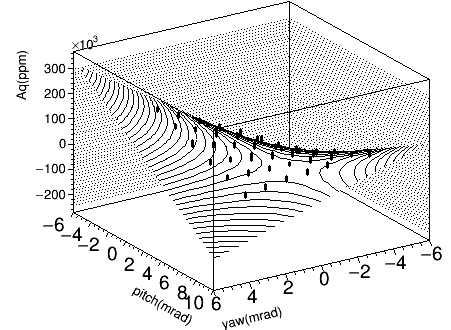}% Here is how to import EPS art
      \captionsetup{justification=raggedright,singlelinecheck=false}  
\caption[RTP pitch-yaw scan]{\label{fig:angledep2}  Angle scan. The Pockels cell angle scanned in pitch and yaw (x and y axes in figure) and the intensity asymmetry Aq in S1 (after polarized beam passes through vertical analyzer) is detected and shown along the z-axis. Data points are fit with a saddle function. }
\end{figure}

The cell sensitivity to angle was measured by using a Newport 9071 Four-Axis Tilt Aligner, to control the pitch and yaw of the Pockels cell at the mrad-level. % \footnote{In the past, in our group's work, it was assumed this mount has the same pitch and yaw sensitivity in terms of mrads/turn. This is not the case. Additionally there was a set screw left in which was supposed to be removed, but which remained in previously. For these reasons, previous year's measurements of angle sensitivity should be called into question. }. 
For this measurement, a picosecond pulsed laser (Edinburgh Instruments EPL-785) was used, with central wavelength 783.0nm, 66.5ps pulse width, 3.4nm bandwidth, and 20MHz pulse repetition rate. The laser was coupled through a single mode fiber the output of which was $\sim 10.3\mu W$, refocused to have $\sim1$mrad divergence, and cleaned up with a Glan-Tayler polarizer which transmitted $\sim 6 \mu W$ of horizontally polarized light to the cell. The helicity flip rate for this measurement was 240Hz, with randomized helicity pair pattern. The light transmitted through the Pockels cell was analyzed with a vertical polarizer (S1), and detected on a  fast photodiode. The cell angle scanned over several pitch/yaw positions in a $\sim8x8$mrad$^2$ grid. The yaw control on this mount is 8 mrad/turn and pitch control is 4.6 mrad/turn. The intensity asymmetry $A_I$ dependence on cell angle was measured and is shown is Fig. \ref{fig:angledep2}.  A saddle function was fit to the data of functional form 
\begin{equation}
A_I = k(\theta_{pitch}-p_0)(\theta_{yaw}-y_0)+A_0
\end{equation}
and the measured angle sensitivity was obtained: $k^{meas}=4527\pm942$ppm/mrad$^2$  with $20.8\%$ error. The dominant errors are from angle measurement calibration and Aq measurement calibration.  The predicted angle sensitivity is $k=5137$ppm/mrad$^2$  (Sec. \ref{sec:level3AqangleS1}) which is consistent with the measured sensitivity. 
%AqvspyFit pd anal17cm 8mrad4p6mradwithAqrescaling

%\begin{figure}[H]
%\centering
%\includegraphics[width=0.2\textwidth]{Run5202_RTP_analS1_pitchyawscan_plain.png}% Here is how to import EPS art
%\caption{\label{fig:angledep2}  \cite{elog858}}
%\end{figure}

\begin{figure}%[H]
        \begin{subfigure}{0.23\textwidth}
        \includegraphics[width=\textwidth] {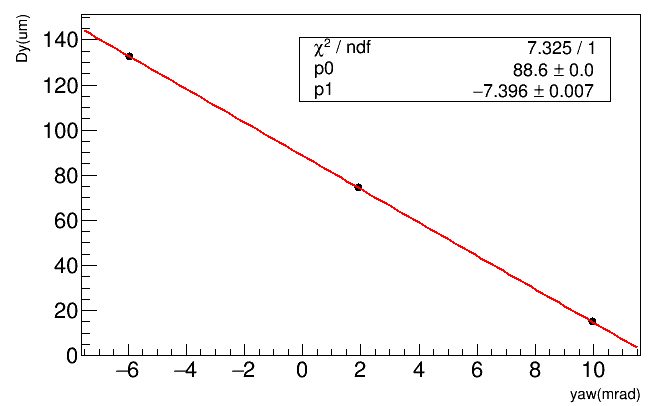}
          \caption{ }
          \label{fig:Dyyaw}
      \end{subfigure}
             \begin{subfigure}{0.23\textwidth}
        \includegraphics[width=\textwidth] {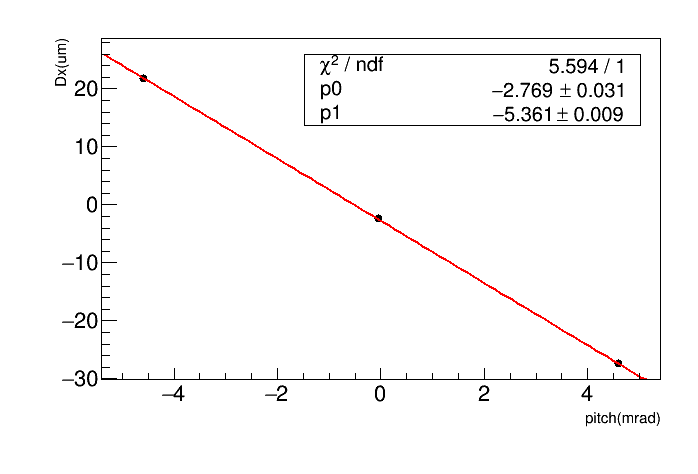}
          \caption{}
          \label{fig:Dxpitch}
      \end{subfigure} 
            \captionsetup{justification=raggedright,singlelinecheck=false}  
\caption[$D_x$,$D_y$ angle dependence in S1]{
\label{fig:angledeppos}$D_x$,$D_y$ angle dependence in S1. The Pockels cell angle scanned in pitch and yaw (x axes in figures) and the intensity asymmetry Aq in S1 (after polarized beam passes through vertical analyzer) is detected and shown along the y-axes. Data points are fit with linear functions in red. (a)Vertical position difference Dy with respect to yaw Pockels cell angle (b) Horizontal position difference Dx with respect to pitch Pockels cell angle. Data points are shown in black. Linear fit is shown in red with tfit parameters p0(offset) and p1(slope) shown in figure text. }
\end{figure}

An additional study on the angle dependence of analyzing-like position differences was performed with the qpd detector. This measurement is, in theory (Sec. \ref{sec:level3AngDepPosDiff}), very dependent on the laser beam size and divergence at the cell. The beam conditions were measured to be 
\begin{eqnarray}
&&\theta_{wx} = 1.40\pm0.05mrad, \theta_{wy}=1.30\pm0.05mrad\\
&&w_{PCx}=1.29\pm0.03mm, w_{PCy}=1.45\pm0.11mm 
\end{eqnarray}
where $w=2\sigma$ and $\theta_w = \frac{dw}{dz}$.
 These measured beam parameters imply a prediction (see Sec. \ref{sec:level3AngDepPosDiff}) on the angle dependence of
 \begin{eqnarray}
  |D_x|= (4.6+-0.2um/mrad)\theta_{pitch}\\
   |D_y|=(4.8+-0.4um/mrad)\theta_{yaw}
  \end{eqnarray}
   The results of scanning pitch and yaw are show in Fig. \ref{fig:angledeppos}. As predicted based on the saddle function in $A_I$, yaw couples predominantly to the position difference in x and pitch couples predominantly to the position difference in y. The angle dependence was measured to be 
    \begin{eqnarray}
   D_x=(-5.36 \pm 0.91 \mu m/mrad)\theta_{pitch}\\
   D_y=(-7.37 \pm 1.25 \mu m/mrad)\theta_{yaw}
     \end{eqnarray}
    are from angle measurement calibration ($8\%$ on the scale factor of mrad/turn) and qpd measurement calibration ($15\%$ 
    %\footnote{40mV signal on each pad with 20mV pedestal, so an error of  TR $20mV/(40mV+20mV)\sim 30\%$ is hypothetically possible on each pad which translated into a calibration error on x-const of $\sim15\%$}
    ). The measurement is consistent with the calculated value within $1\sigma$ for pitch sensitivity and within $2\sigma$ for yaw sensitivity.  This measurement demonstrates that indeed angle adjustments to the Pockels cell can counteract the analyzing-like position differences caused by birefringence gradients with a modest beam divergence of $\sim1$mrad for a $1$mm beam spot size.

\subsection{\label{sec:level2Steeringcontrol}Steering control}
 Steering is an angle difference between right and left helicity states, a helicity correlated change in angle of the outgoing laser beam after having passed through the Pockels Cell. The RTP cell design was intended to allow for straightforward control over helicity correlated beam steering via the electric field gradient along the z-axis of each crystal. The first crystal has its z-axis along U ($-45^o$ direction) and it controls the steering along the U-direction with ``alpha-position-U" voltage $V_{\alpha pos,U}$. The second crystal has its z-axis along V($+45^o$ direction) and it controls the steering along the V-direction with ``alpha-position-V" voltage $V_{\alpha pos,V}$.
 
For the characterization of the RTP cell's steering control, we used JLab's HallA pulsed diode laser which is a frequency doubled with a PPLN crystal (focused before the crystal with an lens f=40mm, and refocused after the PPLN with an f=35mm lens), has a 30-50ps pulse duration, 500MHz repetition rate, central wavelength of 776.5nm, and 0.2nm bandwidth. The study was performed with horizontal input polarization before the cell, no analyzer downstream of the cell, and qpd detector to measure the beam position differences. Because steering produces a position difference which increases with throw distance, and steering is referred to as an `angle-like' position difference which does not depend on analyzing power, no analyzer was used in the study and the throw distance from the Pockels Cell center to the qpd was measured to be 140cm. 

As predicted in Sec. \ref{sec:Steering}, for horizontal input polarization, there is a linear dependence between $V_{\alpha pos, U}$ applied to the first crystal, inducing a field gradient $\frac{d E_{z1,0}}{dz_1}$, and the steering angle $\Delta \theta_{U}$ along $U=-45^o= \frac{x-y}{\sqrt{2}}$. The steering control of position differences with voltage was simulated to be approximately:
\begin{eqnarray}
 |\Delta \theta_{U}| =&&  |\frac{1}{2}(n^3_{y0} r_{23}+n^3_{z0} r_{33})L \frac{d E_{z1,0}}{dz_1} V_{\alpha pos U} |\nonumber\\
 = &&4.6 \pm1.7 nrad/V V_{\alpha pos U} %4.633 3.0 to -6.3nrad/V
\end{eqnarray}

\begin{figure}%[H]
    \centering
        \includegraphics[width=0.3\textwidth]{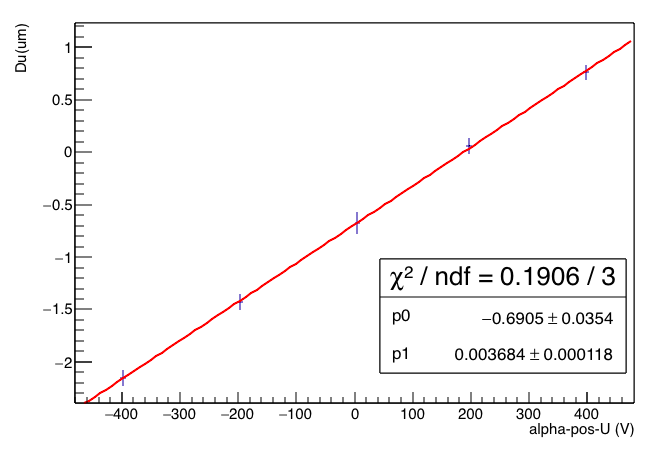}
              \captionsetup{justification=raggedright,singlelinecheck=false}  
         \caption[Alpha-phase-U Voltage Scan]{\label{fig:alphaposU} $V_{\alpha pos,U}$ Voltage Scan:  The steering voltage $V_{\alpha pos,U}$ is scanned and the position difference along U($-45^o)$ is measure by a quad-photodiode 140cm downstream of the Pockels cell. Scan was performed with  analyzer (see Fig. \ref{fig:LayoutLaserTable}) out. Data points are shown in black. Linear fit is shown in red with tfit parameters p0(offset) and p1(slope) shown in figure text. }
   \end{figure}

Fig. \ref{fig:alphaposU}  shows the results of a scan of voltage $V_{\alpha pos U}$ at JLab. The results show a position difference sensitivity to voltage of $\frac{D_u}{d V_{\alpha pos U} }= 3.68\pm0.118nm/V$ at a throw distance of 140cm, corresponding to a measured steering control of 
\begin{equation}
\frac{d \Delta \theta_{U}}{d V_{\alpha pos U}} =  2.63\pm0.084nrad/V
\end{equation}
 The laser steering dependence on applied voltage was found to be linear as predicted, and while on the low end of simulated $4.6\pm 1.7nrad/V$, more than sufficient to control and zero out any position differences intrinsic to the crystal system. %\footnote{The range of control is $-800V<V_{\alpha pos U}<800V$, $-800V<V_{\alpha pos V}<800V$ which corresponds to $D_u = \pm 2.1 \mu$m, $D_v = \pm 2.1 \mu$m, $D_r = \sqrt{D_u^2 + D_v^2}=3.0\mu$m at a 1m throw distance. When used in conjunction with a $<7\%$ analyzing element, such as a photocathode, the S1 analyzing-like position differences for a 1mm beam are at most $7\%$  on $10-25 \mu$m which is $0.7-1.75 \mu$m, well within the range of the steering control.}

\subsection{RHWP scan characterization}\label{goodrhwpscans}

%\subsection{RHWP scans}
%Caryn on your equations, you need to follow the {[( formatting thing
To assess the Pockels Cell alignment, we examine the asymmetry between polarization states using a rotating-HWP (RHWP). 
%By rotating the HWP, we rotate the polarization state of light. 
 Analyzing power dependent helicity correlated beam asymmetries (HCBAs) are caused by asymmetric linear polarizations. A RHWP changes the orientation of this asymmetric linear polarization. The transmission through an analyzer, measured with respect to RHWP angle determines the degree and orientation of asymmetric linear polarization. The RHWP is inserted downstream of the Pockels cell, before subsequent elements in the beamline. For laser beam studies, subsequent elements include a 100\% analyzer (Glan-Taylor polarizer), followed by mirrors and a photodiode detector. For the electron beam, subsequent elements include the vacuum windows (with some birefringence of their own) and photocathode with slight analyzing power. The electron beam is then characterized by beam current and position monitors.  RHWP scans assess the level of alignment of the Pockels cell. If the RHWP and an additional retardation plate downstream of the RHWP (such as the vacuum windows) is inserted between the Pockels cell and the analyzer (in this case the photocathode), the HCBA becomes \cite{SilwalThesis}
 \begin{eqnarray} \label{RHWPequ}
A_q = &&- \frac{\epsilon}{T}  \bigg[ \beta \sin(2\rho - 2\psi) + \gamma \sin(2\theta-2\psi) \nonumber\\
&&+\Delta_{S1}\cos(4\theta-2\psi) +\Delta_{S2}\sin(4\theta-2\psi) \bigg]  
\end{eqnarray}
where $\theta$ is the RHWP angle, $\epsilon/T$ is the analyzing power, $\psi$ is the analyzing direction, $\beta$ and $\rho$ are respectively the phase shift and orientation of an additional retardation plate (i.e. vacuum windows), $\gamma$ is due to the RHWP's deviation from being a perfect $\lambda/2$-plate, and $\Delta$ is an anti-symmetric phase shift described previously which can arise either in S1 or S2. In aligning the Pockels Cell, minimizing the $4\theta$ terms is desirable. We can also minimize the impact of the vacuum windows on the offset term by rotating the photocathode direction such that its analyzing power matches the vacuum window birefringence $\psi=\rho$. The $2\theta$ term corresponds to an imperfect RHWP and can be reduced by using a RHWP very well matched to the laser wavelength. 
%Caryn you might want to just get rid of the footnotes totally
RHWP scans are also used to assess the position differences and steering. The position differences $D_{x}, D_y$ with respect to RHWP angle can be described by
% \;\;\;\;\;\;\;\;\;\;\;\;\;\;\;\;\;\;\;\;\;\;\;\;\;\;\;\;\;\;\;\;\;\;\;\;\;\;\;\;\;\;\;\;\;\;\;\;\;\;\;\;\;\;\;\;\;\;\;\;
\begin{eqnarray}
D_{xi} = && D^{steer}_{xi} - \frac{\epsilon}{T} \bigg[ \partial_{\beta i} \sin(2\rho - 2\psi) + \partial_{\gamma_i} \sin(2\theta-2\psi)\nonumber\\
&&+D^{S1}_{xi}\cos(4\theta-2\psi) +D^{S2}_{xi}\sin(4\theta-2\psi) \bigg] 
\end{eqnarray}
where $\partial_{\beta i}$ corresponds to the birefringence gradient in the vacuum windows, $\partial_{\beta i}$ is the birefringence gradient in the RHWP which is typically negligible, $D^{S1}_{xi}$ is the 4$\theta$ term from S1 analyzing-like position differences %\footnote{We note that in RTP RHWP we have observed the $4\theta$ position difference term to be in phase with the Aq $4\theta$ term, indicating $D^{S2}_{xi})^2$ is small compared with $D^{S1}_{xi}$ as in Fig. \ref{fig:Run3250RHWPscanIHWPout_shiftDvsteering} }
. A comprehensive $4\theta$ term amplitude is given by $D^{4\theta}_{xi}=\sqrt{(D^{S1}_{xi})^2 + (D^{S2}_{xi})^2}$. The 4$\theta$ terms can be inferred from the individual measured position differences $D^{\theta}_{xi}$ where the analyzer is probing the polarization along $\theta$ and corresponding position difference:
\begin{eqnarray}
D^{S1}_{xi}=\frac{D^{0^o}_{xi} - D^{90^o}_{xi}}{2} \;\;,&&\;\; D^{S2}_{xi}= \frac{D^{45^o}_{xi} - D^{-45^o}_{xi}}{2}\\
D^{steer}_{xi} =&& \frac{D^{45^o}_{xi} + D^{-45^o}_{xi}}{2}
\end{eqnarray}
Each crystal 1,2 contributes individually to position differences $D^{pol,1}_{xi}$ and $D^{pol,2}_{xi}$. We note that in RTP, due to the values of the opto-electric coefficients, typically $D^{45^o,1}_{xi} \approx 3 D^{-45^o,1}_{xi}$ and $D^{-45^o,2}_{xi}\approx 3 D^{45^o,2}_{xi}$. %An illustration of an RHWP scan and how to interpret it is shown in Fig. \ref{fig: RHWPscanRTP}.

\begin{figure*}%[H]
    \centering
        \begin{subfigure}{0.45\textwidth}
        \includegraphics[width=\textwidth]{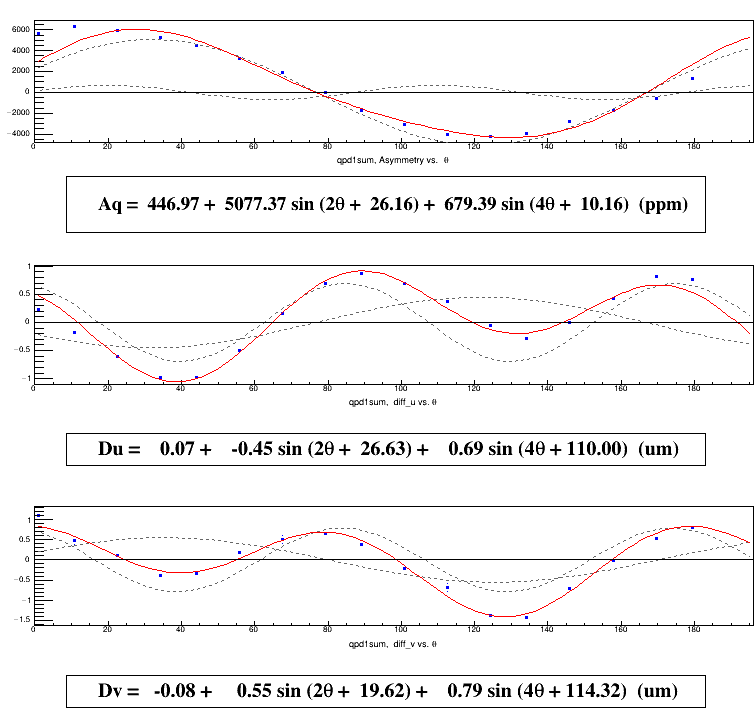}
          \caption{IHWP out}
          \label{fig:RHWPscanIHWPout}
      \end{subfigure}
 %             \begin{subfigure}{0.45\textwidth}
  %      \includegraphics[width=\textwidth]{Run5080_findingHallAS2_IHWPout.png}
   %       \caption{IHWP out, PITA offset}
   %       \label{fig:RHWPscanIHWPoutoffset}
  %    \end{subfigure}
              \begin{subfigure}{0.45\textwidth}
        \includegraphics[width=\textwidth]{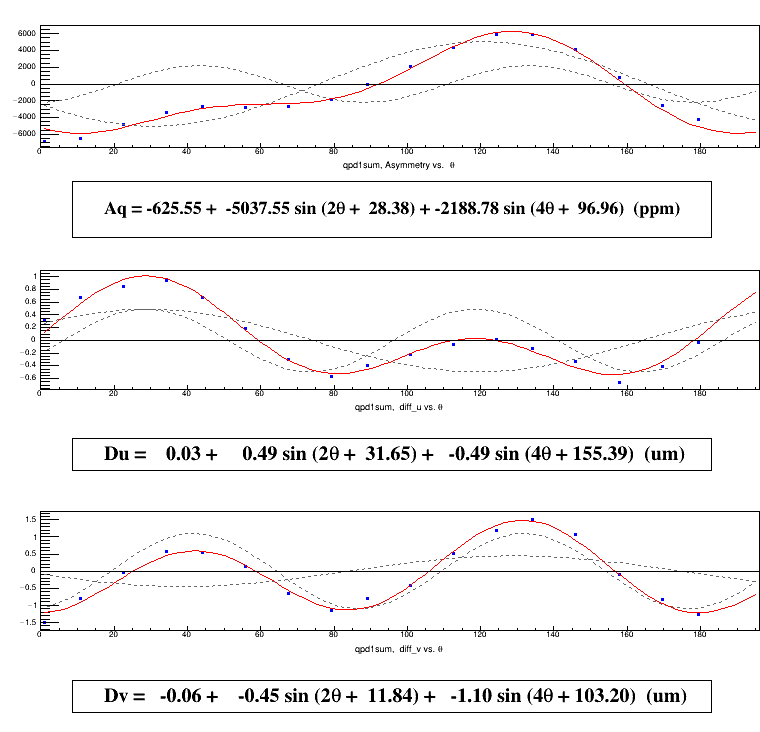}
          \caption{IHWP in}
          \label{fig:RHWPscanIHWPin}
      \end{subfigure}   
   %                 \begin{subfigure}{0.45\textwidth}
  %      \includegraphics[width=\textwidth]{Run5081_findingHallAS2_IHWPin.png}
  %        \caption{IHWP in, PITA offset}
  %        \label{fig:RHWPscanIHWPinoffset}
  %    \end{subfigure}
        \captionsetup{justification=raggedright,singlelinecheck=false}  
\caption[RHWP scans with QPD]{\label{fig:RHWPscans} RHWP scans: RHWP (see Fig. \ref{fig:LayoutLaserTable}) is scanned while beam is analyzed with a vertical polarizer and detected by a quad-photodiode which measures both intensity asymmetry and position differences. Top plot shows intensity asymmetry Aq in ppm with respect to RHWP angle in degrees. Center and lower plots show position differences along U and V diagonal directions, Du and Dv, in $\mu$m with respect to RHWP angle. Data points taken with quad photodiode are shown in blue. Data is fit with Equ. \ref{RHWPequ} and fit results are shown in red and in the figure text. 4$\theta$ terms and 2$\theta$ term contributions to the data fits are shown as dotted lines. (a) IWHP out, PITA voltage set to minimize Aq in S1 , PITA=-25V, $V_{\alpha pos U}$=45V, $V_{\alpha pos V}$=77V.
%(b) IHWP out, PITA voltage offset from optimal, Run5080 (c)
(b) IHWP in, PITA voltage set to minimize Aq in S1 , PITA=-12V, $V_{\alpha pos U}$=118V, $V_{\alpha pos V}$=48V.
%(d) IHWPin, PITA voltage offset from optimal, Run5081
  }
\end{figure*}

The results of RHWP scans performed with the quad-photodiode are shown in Fig. \ref{fig:RHWPscans}. To minimize the steering offset terms in these RHWP scans, the steering voltages $V_{\alpha pos,U}$ and $V_{\alpha pos,V}$ were optimized with independent values for IHWPout and IHWPin states. Each crystal has had its electric field gradient altered in order to shift the steering-like position difference along the ``U"($-45^o$) direction or ``V"($-45^o$) direction to minimize the offset terms in these RHWP scans. The steering term can changed at the $\mu$m-level for a $\sim1$m throw distance by using voltages. To minimize the position difference 4$\theta$ terms, position differences in S1 were reduced with Pockels cell pitch and yaw. The RHWP scans show steering terms of $<$70nm at a 1.5m throw distance and 4$\theta$ terms of $<1\mu$m for a $4\sigma=2.6$mm spot size.
% \footnote{Conditions: Captain Expando lens ($f \sim 1m$, named by John Hansknecht) inserted 46inches upstream of cell. Spot size at PC center $2\sigma=w=0.60mm(Horizontal), 0.65mm(Vertical)$, divergences at PC center $dw/dz= -0.39$mrad. Effective throw from PC to qpd 1.5m. At qpd, spot size $4\sigma X/Y= 1.32/1.29$mm with calibration constants $x-const=0.442$mm, $y-const=0.445$mm. Path to cathode: 2m steering lens, cathode spot size $2\sigma=1.2mm(Horiz), 1.3mm(Vert)$, distance to cathode $\sim3.1m$, distance to steering lens $\sim1.067m$, effective throw from PC to cathode $\sim2.015m$, cathode analyzing power $\sim 5\%$  \cite{polog3641295}}.

%Caryn you removed the linear array stuff
%\subsection{Spot size asymmetries - linear array measurements}
\subsection{\label{sec:Temperature}Temperature Sensitivity}

In the thermal compensation design, any change in the refractive indices due to temperature affects both crystals equally. Because the crystals are aligned with opposing y,z axis, the net birefringence remains near zero, even when the overall temperature of both crystals changes.  While the thermal compensation design does a great deal to mitigate temperature effects on the Pockels Cell performance, some small thermal fluctuations are still empirically observed. %as shown in Fig. \ref{fig:AqvstimeTemp}. 

 %   \begin{figure}%[H]
  %   \centering
  %    \includegraphics[width=0.49\textwidth]{Run4883_stabilty.png} 
   %   %AqTfluctuation3hours.png
   %     \caption[Aq stability]{\label{fig:AqvstimeTemp} Aq fluctuation (Run4883 3hours)}
   %  \end{figure}

Fluctuations in the intensity asymmetry $A_I$ through a 100\% analyzer are approximately $\pm20,000$ppm over the course of several hours. The wavelength stability of the laser precluded the possibility that this fluctuation was due to wavelength changes, leaving temperature fluctuations as the main potential contributor to instability. Theoretically, we expect the overall temperature to have very little effect because of the thermal compensation design 
%\footnote{If the crystals are mismatched in length, by dL, we have 6.65e4ppm/$^o$C/mm. So a mismatch of dL=10mm, returns us back 6.65$\times 10^{5}$ppm/$^o$C for one crystal above. A mismatch of 2um gives us 133ppm/$^o$C where the temperature is the absolute temperature of both crystals. 20um mismatch gives 1330ppm/$^o$C.}
, especially since the two crystals are very nearly equal in length to within $2\mu$m according to specifications. However, the effect of a temperature difference, in particular,  between the two RTP crystals is not cancelled out by the thermal compensation design. The dominant effect is the temperature difference between the crystals. When crystal 1 has a higher temperature than crystal 2, the temperature difference $T_1-T_2$ gives rise to a net temperature induced birefringence $\Delta \phi_{T12} \sim T1-T2$, which in QWV operation gives (and 100\% analyzing power) rise to an intensity asymmetry $A_{I,T12} \approx \Delta \phi_{T12}$

We calculated the sensitivity to a temperature differences based on the refractive index temperature dependence \cite{Yutsis}. At $25^o$C, 
\begin{eqnarray}
dn_z/dT\sim1.22\times10^{-5}/^oC \\
 dn_y/dT\sim 3.88\times10^{-6}/^oC
\end{eqnarray}
 implies a birefringence temperature sensitivity 
 \begin{equation}
d\Delta n/dT\sim0.831\times10^{-5}/^oC
\end{equation}
which gives rise to an asymmetry with temperature dependence
\begin{eqnarray}
Aq= \Delta \phi =2 \pi \Delta n L/\lambda  \\
dA_q/dT \sim0.67/^oC\sim6.65\times10^{5}ppm/^oC. 
\end{eqnarray}

%Obtain dAq/dT ~ 6.7e5ppm/$^o$C where just increasing the temperature of one crystal for 10mm thick crystal, around 70-80degF, at 780nm (21-27degC). This implies 1.5mC/1000ppm. This temperature corresponds to a difference between the two crystal temperatures.

\begin{figure}%[H]
%    \centering
%      \hfill
   % \begin{subfigure}{0.3\textwidth}
    %  \includegraphics[width=\textwidth]{AqvsTempRTP.png}%AqvsT1minusT2.png
     %   \caption{Calculated \cite{elog847}}
     %   \label{fig:T1minusT2}
    % \end{subfigure}
 %    \begin{subfigure}{0.4\textwidth}
      \includegraphics[width=0.49\textwidth]{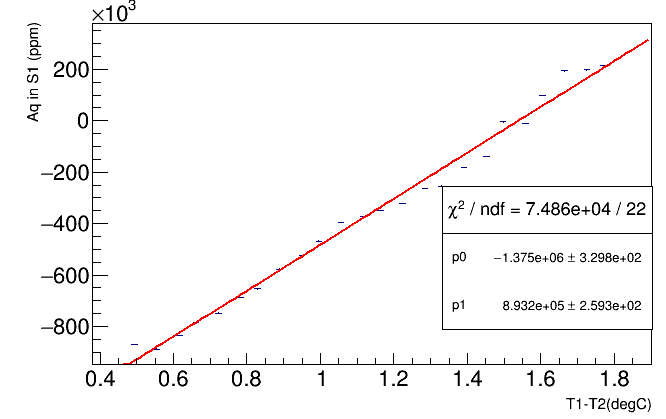}%AqvsT1minusT2.png
   %     \caption{ Measured \cite{elog846}}
  %      \label{fig:T1minusT2}
  %   \end{subfigure}
%          \begin{subfigure}{0.4\textwidth}
 %     \includegraphics[width=\textwidth]{Run5082_AqvsT1minusT2degC.png}%AqvsT1minusT2.png
  %      \caption{ Measured \cite{elog846}}
  %      \label{fig:T1minusT2}
   %  \end{subfigure}
         \captionsetup{justification=raggedright,singlelinecheck=false}  
  \caption[Temperature difference between crystals]{\label{fig:Aqtemp} RTP cell temperature sensitivity. One of the two RTP crystals is heated inducing a temperature difference between the crystals. The intensity asymmetry Aq in S1 is measured with a photodiode and shown on the y-axis. Temperature difference T1-T2  between the crystals is shown on the x-axis. Data points are shown in blue and linear fit is shown in red with fit parameters p0(offset) and p1(slope) shown in figure text.  }
   \end{figure}

The temperature dependence of the RTP cell was measured between 21-27$^oC$. Resistive heaters ($\sim 2.7$Watts) were attached to the Delrin\textsuperscript{\textregistered} housing on the outside of each of the two crystal mounts along with two thermocouples to monitor the temperature of each mount. The heater on the first crystal's mount was switched on, allowing one crystal to heat up more than the other, and then switched off, allowing the crystals to cool down together at different rates.  The thermocouple readings for the two mounts were periodically recorded while the intensity asymmetry was continuously measured. A few measurements of temperature dependence were taken as shown in Fig. \ref{fig:Aqtemp}. Taken together, the measurement results give 
\begin{equation}
\frac{dA_q}{dT}^M = 6.75 \pm 2.18\times10^5ppm/^oC
\end{equation}
 for the RTP cell sensitivity to temperature difference between the crystals, consistent with the calculated value of $\sim6.65\times10^5$ppm$/^o$C.

This confirmed temperature sensitivity indicates the observed $\pm20,000$ppm fluctuations correspond to a fluctuation of $\sim \pm 30$mK temperature difference between the crystal pair. It is hard to control the temperature difference between the crystals at the milli-Kelvin level, but the intensity asymmetry can easily be corrected with voltage, since $20$kppm can be zeroed out with a small  $\sim20V$ PITA-voltage adjustment. The temperature induced birefringence is well within PITA-voltage induced birefringence adjustment range.So, during operation, we simply correct temperature fluctuation with a PITA-voltage feedback loop, rather than trying to force two crystals to maintain milli-Kelvin temperature differences.

This temperature difference sensitivity is one reason why it is important to not focus the laser down tightly for high powers in RTP cells. If one crystal absorbs slightly more power than the other, this leads to a temperature difference in the shape of the beam. The smaller the beam, the greater the gradients induced. This can lead to degradation of performance. %At JLab, with 4 laser beams on one crystal, it could lead to thermal effects from one Hall's laser interfering with the performance of the Pockels cell on another Hall's beam, as well as a sensitivity of HCBA to whether or not some Hall's beam have tripped, are in use, and changes to other Hall's currents. 
When using RTP crystals for parity experiments, we cannot reduce the beam size significantly ($2\sigma<1$mm for $\sim$1Watt) or else thermal gradients induced by the laser absorption ($0.75\%/cm-4\%/cm$) \cite{RaicolSpecSheet} over a small space with high intensity could create additional position differences ($\sim0.1\mu m-0.6\mu$m) and interfere with Pockels cell performance.

%For further discussion on temperature sensitivity fluctuations and feedback see Appendix \ref{app:tempfeedback}.

\section{\label{sec:ebeammeasurements}Results: Electron beam measurements}
%\subsection{Charge Asymmetries}

\begin{figure}%[H]
    \centering
        \includegraphics[width=0.3\textwidth]{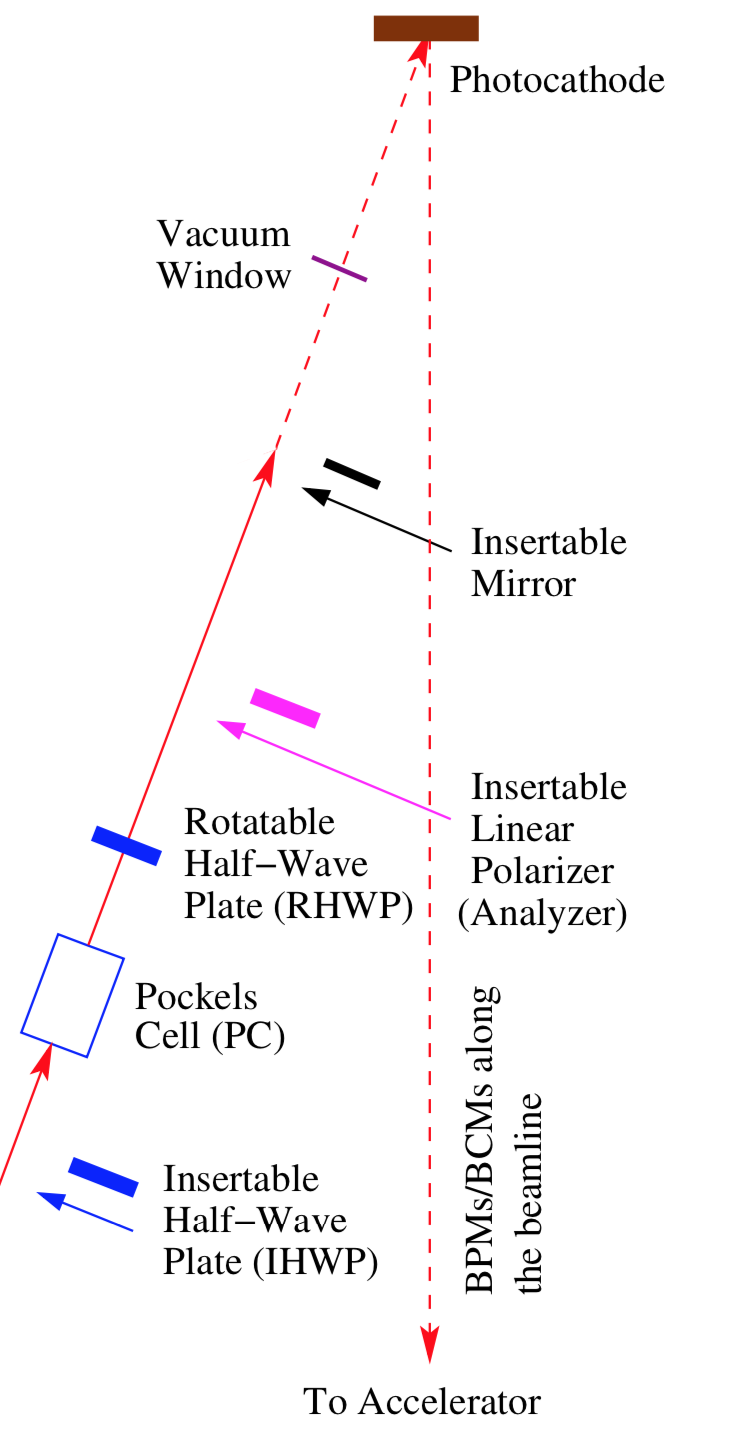}
              \captionsetup{justification=raggedright,singlelinecheck=false}  
         \caption{\label{fig:JLabtable} Jefferson lab injector laser table layout \cite{SilwalThesis}}
   \end{figure}
%JLabTableLayout.png

The RTP cell was installed in the polarized injector source at Jefferson Laboratory.  Polarized electron beam was produced with the JLab HallA laser described in Sec. \ref{sec:level2Steeringcontrol}. %\footnote{paragraph 2}.
 The reduce the laser spot size, a 50cm lens was inserted 1m upstream of the cell. The divergence at the Pockels cell center was measured to be 
\begin{equation}
\theta_{w,x}=dw/dz=0.51mrad \;,\; \theta_{w,y}=0.66mrad 
\end{equation}
and the laser spot size ($2\sigma$) at the cell was measured to be
\begin{equation}
 w_{PCx}=0.825mm\;,\; w_{PCy}=0.94mm
 \end{equation}
  Downstream of the cell, in-between the cell and the photocathode, there is a rotatable half-wave-plate (RHWP) for rotating the polarization state relative to the analyzing direction of the photocathode, a `steering' lens (1.067m downstream of the cell) to refocus the laser beam onto the cathode (3.1m downstream of the cell), and vacuum windows which have slight birefringence gradients. The cathode steering lens used typically at JLab is an f=2m lens, and the effective throw distance from the cell to the cathode is  
  \begin{equation}
  D_{eff}=D_{tot}-(D_{tot}-D_{lens})\times D_{lens}/f \sim2.015m
  \end{equation}
   However, under these conditions the spot-size on the cathode was quite large, 
     \begin{equation}
   w_x=1.45mm\;,\;w_y=1.505mm, 
     \end{equation}
   compared with spot-size at the cathode during previous parity experiments at JLab. 
   %For example, during Qweak, the spot-size at the cathode was $w\sim0.85$mm (and $w\sim0.425$mm during qweak Run1
 %\footnote{this spot size was too small at the high $>100\mu A$ currents and lead to issues with the QE degradation on the cathode which correlated to degradation in the degree of polarization and the need for multiple polarimetry measurements}).
  So, for this study, to mimic experimental conditions, the steering lens was changed to a f=75cm lens which reduced the spot-size at the cathode and changed the effective throw distance from the cell to the cathode to:
       \begin{equation}
  w_x = 1.135mm\;,\;   w_y=1.055 \;,\;  D_{eff}\sim20cm 
    \end{equation}
 
\begin{figure}%[H]
    \centering
        \includegraphics[width=0.46\textwidth]{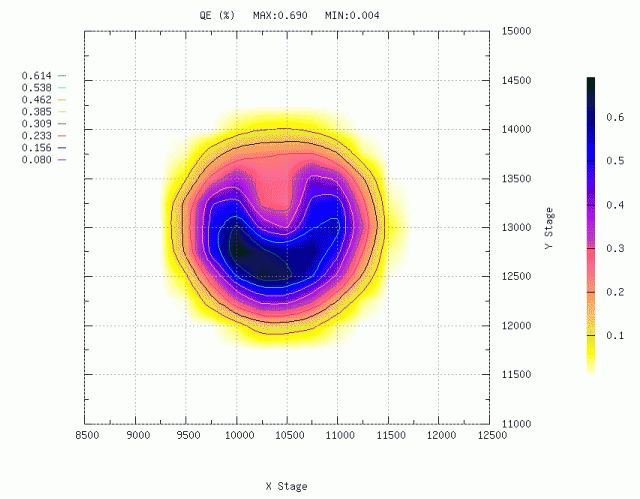}
              \captionsetup{justification=raggedright,singlelinecheck=false}  
         \caption{\label{fig:QEscan} QE scan. Laser beam position on photocathode is scanned in a grid pattern and electron yield is measured. Quantum efficiency \% (QE) is calculated for each (X,Y) position (arb. units) and shown as a color map.  }
   \end{figure}

 The cathode used during this study was measured to have quantum efficiency $QE\sim0.69\%$, as shown in Fig. \ref{fig:QEscan}, and an analyzing power of $7\%$, producing $\sim 90\%$ polarized electron beam. 

%steering lens now 75cm got 4sigma 2.27mmV, 2.11mmH. Leff=Ltot-L1*L2/f = 3.1-1.067*2.033/0.75=0.2m=20cm 
%2sigma: w=0.825mm Horizontal, w=0.94mm Vertical at  PC center, 
%Measured divergences: dw/dz 0.51mrad horizontal, 0.66mrad vertical at PC center, 
%steering lens was previously 2m, cathode 4sigma 2.9mm Horiz, 3.1mm Vert, distance to cathode ~3.1m 1/19/17, distance to steering lens ~1.067m, effective throw from PC to cathode ~2.015m, Leff=Ltot-L1*L2/f 
% qweak 4sigma 1.7mm  (from quoted '1mm' size during run2), and 4sigma 0.85mm (from quoted '0.5mm' spot size during run2)
%Pulse Duration 30-50ps , 0.2nm linewidth, central value  776.5nm, 

Just after the cathode, the beam position of the 130keV electron beam was detected by multiple beam position monitors (BPMs) in its transport along the beamline. Each BPM consists of four RF antenna, detecting 500MHz repetition rate e-beam pulses, and the beam position is computed by comparing the magnitude of the four wire-channel responses (denoted $X^+,X^-,Y^+,Y^-$). Analogous to the quad-photodiode measurement, the BPM measurement of beam position in x and y consists of wire-channel signal differences and sums:
\begin{eqnarray}
&&bpmx = \kappa (X^+-X^-)/ (X^++X^-) \\
&& bpmy = \kappa (Y^+-Y^-)/ (Y^++Y^-)
\end{eqnarray}
where $\kappa$ is a proportionality factor (13.7mm in this region of the accelerator) based on the geometry of the BPM. 

 Measurements were taken at a helicity flip rate of 240Hz ($\sim4ms$ helicity windows) to more quickly converge to nm-level statistical precision on electron beam position monitors. To suppress 60Hz noise, the helicity flip rate was line synced and pseudo-random octet helicity patterns: 
 
 ($+--+-++-$ OR $-++-+--+$) 
 
 were used to cancel 60Hz noise. The helicity signal reporting to the DAQ was delayed by 16 helicity windows to prevent electronics pickup \cite{HelicityControlBoard}. 
 
 \subsection{Charge Asymmetry}
 
 We performed feedback on charge asymmetry $A_q$ with Pockels cell PITA voltage. As previously stated in Sec. \ref{sec:Temperature}, the RTP suffers from some slow drifts due to fluctuation in the temperature difference between the two RTP crystals of $\sim 30mK$. This is correctable with PITA voltage feedback, we can adjust PITA voltage to keep intensity asymmetry minimized
%Caryn make these figures not grey
\begin{figure}%[H]
    \centering
     \begin{subfigure}{0.23\textwidth}
      \includegraphics[width=\textwidth]{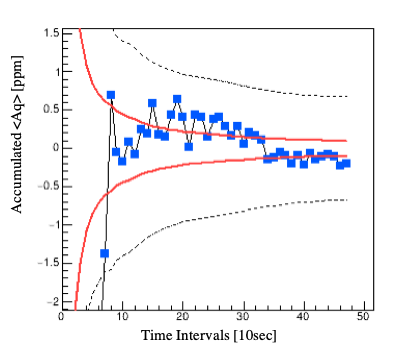}
        \caption{Aq feedback S1}
        \label{fig:AqfeedbackS1}
     \end{subfigure}
     \begin{subfigure}{0.23\textwidth}
      \includegraphics[width=\textwidth]{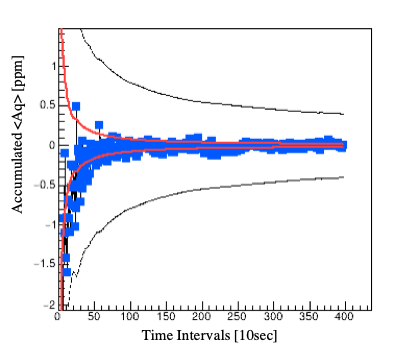}
        \caption{Aq feedback $\sim$S2}
        \label{fig:AqfeedbackS2}
     \end{subfigure}    
           \captionsetup{justification=raggedright,singlelinecheck=false}  
      \caption[Aq feedback]{\label{fig:Aqfeedback}Charge asymmetry feedback: electron beam Aq is suppressed through PITA voltage feedback. Accumulated Aq is shown on the y-axis with respect to time in units of 10sec feedback intervals on the x-axis. Convergence rates curves RMS/$\sqrt{N}$ is shown in black and RMS/N is shown in red. (a) Aq feedback for RHWP angle set to $67.5^o$ such that photocathode analyzing power is along S1(b) Aq feedback for RHWP angle set to $45^o$ such that photocathode analyzing power is near S2}
   \end{figure}

$A_q$ feedback was performed, using the RTP PITA voltages to minimize the measured charge asymmetry on a BPM in the 130keV region of the injector. Feedback was performed in approximately 10 second intervals at two different RHWP settings. One RHWP angle ($67.5^o$) was set to align the Pockels cell S1 polarization direction along the analyzing axis of the photocathode, maximizing the sensitivity of $A_q$ to PITA voltage and temperature fluctuation. The other RHWP angle ($45^o$) reduced the sensitivity to PITA voltage and temperature fluctuations by a factor of 10X. The charge asymmetries, accumulated over time in both case are shown for each feedback interval in Fig. \ref{fig:Aqfeedback}. After approximately 10 minutes (Fig. \ref{fig:AqfeedbackS2}), the accumulated $A_q$ converged to $<1$ppm (for RHWP angle set to S1 at $67.5^o$). After approximately 2 hours (Fig. \ref{fig:AqfeedbackS2}), the accumulated $A_q$ converged to $<0.1$ppm (for RHWP angle set to near S2 at $45^o$). Statistically, the theoretical limit for the rate of convergence $A_q$ is between $RMS/\sqrt{N}$ to $RMS/N$ depending on the type of noise being measured, where N is the number of feedback intervals and RMS is the root-mean-square charge asymmetry noise. In both measurements, the charge asymmetry converged faster than $RMS/\sqrt{N}$ (shown as the black dotted line), and nearly as fast as $RMS/N$. Despite slow drifts from temperature fluctuation, the RTP Pockels cell successfully controlled charge asymmetry in the electron beam as well as the previously used KD*P cell.

 \subsection{Position Differences}
 
\begin{figure}%[H]
    \centering
        \includegraphics[width=0.49\textwidth]{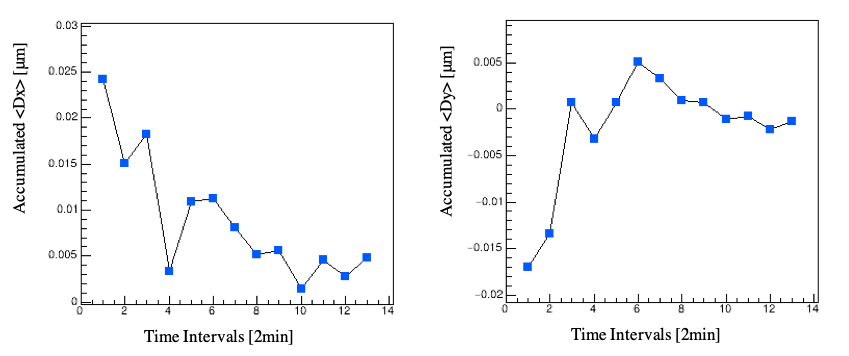}
              \captionsetup{justification=raggedright,singlelinecheck=false}  
         \caption{\label{fig:DxDyconverge} Position Difference Feedback: electron beam position differences (as measured by an electron beam position monitor a few meters from the photocathode) are suppressed through steering voltage feedback. Accumulated position differences Dx and Dy are shown on the y-axis with respect to time in units of 2 minute feedback intervals on the x-axis. Using RTP Pockels cell voltages, position differences converged to 1-5nm within 30minutes. }
   \end{figure}

  In addition to demonstrating control over charge asymmetry, we demonstrated control over position differences. Position difference feedback was performed using the RTP steering control voltages $V_{\alpha pos, U}$ and $V_{\alpha pos, V}$, to minimize the measured position differences on a BPM (``1I04") in the 130keV region of the injector, shortly after the cathode, before acceleration. Feedback was performed every 2 minutes, after sufficient precision was obtained on the position difference measurement to make the correction meaningful at the level of 20nm during each interval. The feedback coefficients used were on the order of $\sim 2nm/V$. The  x and y position differences, $D_x$ and $D_y$, accumulated over time are shown for each feedback interval in Fig. \ref{fig:DxDyconverge}. After approximately 30 minutes, the accumulated average horizontal position difference $D_x$ converged to $<5$nm and the accumulated average horizontal position difference $D_y$ converged to $<1$nm. This RTP Pockels cell successfully controlled position differences in the electron beam with nm-level precision. The steering control voltages $V_{\alpha pos, U}$ and $V_{\alpha pos, V}$ ultimately used for position difference corrections were small $<$50-170V out of the $\pm800V$ range.

\begin{figure}%[H]
%    \centering
 %     \begin{subfigure}{0.45\textwidth}
  %      \includegraphics[width=\textwidth]{Run4017_1I04DxDy_firs10bpms.png}
  %       \caption{\label{fig:DxDy70nma}  $D_x$,$D_y$<70nm}
%  \end{subfigure}
%    \begin{subfigure}{0.45\textwidth}
   \includegraphics[width=0.49\textwidth]{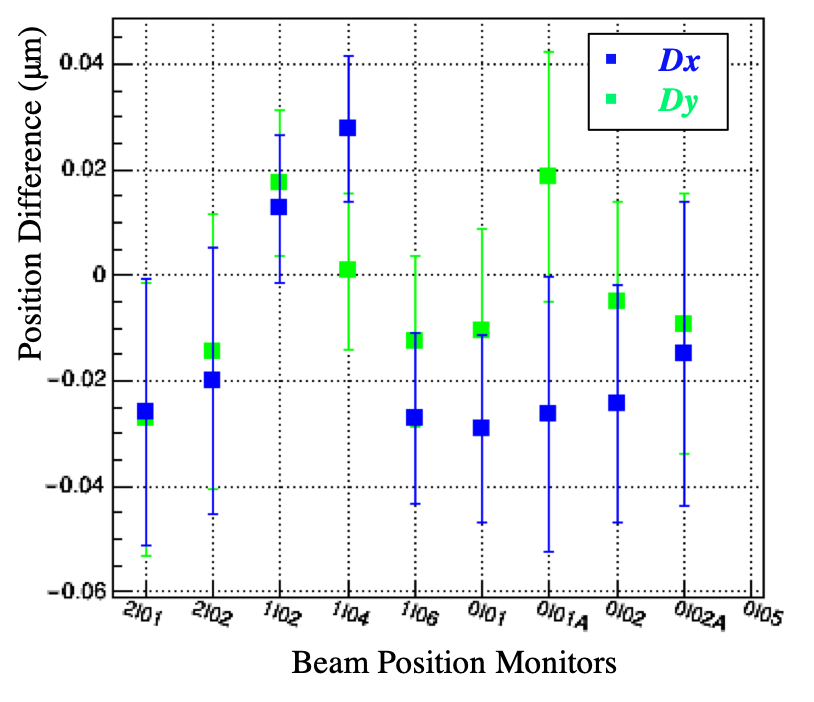}
        % \caption{\label{fig:DxDy30nm} $D_x$,$D_y$$<$30nm }
   %\end{subfigure}
         \captionsetup{justification=raggedright,singlelinecheck=false}  
    \caption[<30nm Position Differences in 130keV region]{\label{fig:DxDy70nm}   %(a) Position differences $<$70nm in 130keV region, Run4017  \cite{polog3584281} 
      Position differences $<$30nm: electron beam position differences $D_x$(blue) and $D_y$(green) as measured bu $1^{st}$ 10 beam position monitors in the 130keV region of the injector at JLab. Beam position monitor names (arb. convention) are shown on the x-axis. The y-axis corresponds to measured position differences in units of $\mu$m  }
   \end{figure}

After feedback was performed on one BPM in the injector (BPM ``1I04"), the position differences at other BPMs %\footnote{These BPMs include 2 newly installed bpms just after the cathode: 2I01, 2I02. These are modified M-20 cans which have calibration coefficients $\kappa=25.67$mm. The other BPMs in this plot are M15-mini cans with $\kappa=13.7$mm, or according to the Goubau line scanner $\kappa=15.59$mm. Further down in the injector beamline, the bpms are M15 cans with $\kappa=18.76-18.81$mm or according to the Goubau line scanner $\kappa=18.4$mm}
 throughout the beamline were measured as shown in Fig. \ref{fig:DxDy70nm}. We use multiple BPMs in the beamline because the e-beam varies in size and has nodes as well as undergoing rotation, and the BPMs have different sensitivities to beam position/angle. The RTP Pockels Cell successfully produced e- beam which achieved $<$30nm position differences in the $1^{st}$ 10 beam position monitors in the 130keV region of the injector at JLab. This measurement shows the smallest position differences that have historically been observed in this region of the JLab accelerator. The RTP Pockels cell demonstrated the best ever control of position differences in this region. 

%\begin{figure}[H]
 %   \centering
 %       \includegraphics[width=0.6\textwidth]{Run4017_1I04DxDy_firs10bpms.png}
 %        \caption{\label{fig:DxDy70nm} }
 %  \end{figure}
   
%We compare this measurement with that of the position differences measured during Qweak \cite{ManolisThesis}. Despite the spot-size on the cathode being a factor of 2X larger in our recent measurement, the position differences during Qweak were $<$200nm in the 130keV region, whereas the RTP achieved $<$70nm in the same region. 

%\begin{figure}%[H]
  %  \centering
   %     \includegraphics[width=0.95\textwidth]{QweakInjector.png}
   %      \caption[Qweak horizontal position differences]{\label{fig:QweakInj}  Qweak horizontal position differences. These are defined as $\frac{1}{2}(x^R-x^L)$ and so to compare to our measurements, they must be multiplied by 2X. \cite{ManolisThesis} }
  % \end{figure}

%-Emphasis precision - nm level
%-Maybe show injector 8 bpm plot, but must explain why we use multiple BPMs (position, angle, e-beam rotation, nodes, sensitivity). But possibly show this plot a a ratio of Dx/(laser-beam-spot-size-on-cathode). 
%Compare with older plot (like Qweak?) to show we have set a record for smallest pos diffs observed in this system ever.
%-Maybe show PITA pos scan demonstrating control at nm-level for an individual bpm.

%Caryn you cut out these sections 
%\subsection{Spot-size Asymmetries}
%\subsection{Longitudinal Asymmetries}
\section{Conclusion}

We have presented an innovative ultra-fast switch RTP cell design which uses electric field gradients to counteract crystal non-uniformities, leading to improved extinction ratios (in $\lambda/2$-wave configuration) and minimization of voltage dependent beam steering (in $\lambda/4$-wave configuration). This RTP Pockels cell design has been demonstrated to be capable of producing precisely controlled polarized electron beam at Jefferson Laboratory and control beam steering down to the nm-level with voltage feedback. Additionally, the cell has an increased number of degrees of freedom and can control both S1 and S2 polarizations. The precision reached with the RTP cell offered sufficient control over and minimization of intensity asymmetry, position differences, and spot-size asymmetry to perform PREX II \cite{PREX2paper}, a recent parity violation electron scattering experiment at JLab. The position differences achieved with electron beam were $<$30nm. The RTP Pockels cell system will provide fast flipping and suitable control of position differences and  parity quality beam for the future MOLLER experiment \cite{MOLLERcd1}, providing an unprecedented precision on the electron weak charge and electroweak mixing angle. Future studies will explore the importance of accelerator beam transport in maintaining small helicity correlated beam asymmetries achieved by the Pockels Cell.

\bibliography{mytemplate}

\providecommand{\noopsort}[1]{}\providecommand{\singleletter}[1]{#1}%
\begin{thebibliography}{17}%
\makeatletter
\providecommand \@ifxundefined [1]{%
 \@ifx{#1\undefined}
}%
\providecommand \@ifnum [1]{%
 \ifnum #1\expandafter \@firstoftwo
 \else \expandafter \@secondoftwo
 \fi
}%
\providecommand \@ifx [1]{%
 \ifx #1\expandafter \@firstoftwo
 \else \expandafter \@secondoftwo
 \fi
}%
\providecommand \natexlab [1]{#1}%
\providecommand \enquote  [1]{``#1''}%
\providecommand \bibnamefont  [1]{#1}%
\providecommand \bibfnamefont [1]{#1}%
\providecommand \citenamefont [1]{#1}%
\providecommand \href@noop [0]{\@secondoftwo}%
\providecommand \href [0]{\begingroup \@sanitize@url \@href}%
\providecommand \@href[1]{\@@startlink{#1}\@@href}%
\providecommand \@@href[1]{\endgroup#1\@@endlink}%
\providecommand \@sanitize@url [0]{\catcode `\\12\catcode `\$12\catcode
  `\&12\catcode `\#12\catcode `\^12\catcode `\_12\catcode `\%12\relax}%
\providecommand \@@startlink[1]{}%
\providecommand \@@endlink[0]{}%
\providecommand \url  [0]{\begingroup\@sanitize@url \@url }%
\providecommand \@url [1]{\endgroup\@href {#1}{\urlprefix }}%
\providecommand \urlprefix  [0]{URL }%
\providecommand \Eprint [0]{\href }%
\providecommand \doibase [0]{https://doi.org/}%
\providecommand \selectlanguage [0]{\@gobble}%
\providecommand \bibinfo  [0]{\@secondoftwo}%
\providecommand \bibfield  [0]{\@secondoftwo}%
\providecommand \translation [1]{[#1]}%
\providecommand \BibitemOpen [0]{}%
\providecommand \bibitemStop [0]{}%
\providecommand \bibitemNoStop [0]{.\EOS\space}%
\providecommand \EOS [0]{\spacefactor3000\relax}%
\providecommand \BibitemShut  [1]{\csname bibitem#1\endcsname}%
\let\auto@bib@innerbib\@empty
%</preamble>
\bibitem [{\citenamefont {{The MOLLER Collaboration}}(2016)}]{MOLLERcd1}%
  \BibitemOpen
  \bibfield  {author} {\bibinfo {author} {\bibnamefont {{The MOLLER
  Collaboration}}},\ }\href {https://arxiv.org/abs/1411.4088} {\emph {\bibinfo
  {title} {The MOLLER Experiment Measurement Of a Lepton Lepton Electroweak
  Reaction: An Ultra-precise Measurement of the Weak Mixing Angle using Moller
  Scattering - Pre-Conceptual Design Report}}} (\bibinfo {year}
  {2016})\BibitemShut {NoStop}%
\bibitem [{\citenamefont {{D. Adhikari et al. (PREX
  Collaboration)}}(2021)}]{PREX2paper}%
  \BibitemOpen
  \bibfield  {author} {\bibinfo {author} {\bibnamefont {{D. Adhikari et al.
  (PREX Collaboration)}}},\ }\bibfield  {title} {\bibinfo {title} {{Accurate
  Determination of the Neutron Skin Thickness of $^{208}Pb$ through
  Parity-Violation in Electron Scattering}},\ }\href@noop {} {\bibfield
  {journal} {\bibinfo  {journal} {PRL}\ }\textbf {\bibinfo {volume} {126}},\
  \bibinfo {pages} {172502} (\bibinfo {year} {2021})}\BibitemShut {NoStop}%
\bibitem [{\citenamefont {Palatchi}\ and\ \citenamefont
  {Paschke}(2019)}]{patent2}%
  \BibitemOpen
  \bibfield  {author} {\bibinfo {author} {\bibfnamefont {C.}~\bibnamefont
  {Palatchi}}\ and\ \bibinfo {author} {\bibfnamefont {K.}~\bibnamefont
  {Paschke}},\ }\bibfield  {title} {\bibinfo {title} {{RTP Electro-optic Switch
  with E-field Gradient Control and Related Method Thereof}},\ }\href@noop {}
  {\bibfield  {journal} {\bibinfo  {journal} {United States Provisional Patent
  Application No. 62/832,414}\ } (\bibinfo {year} {2019})}\BibitemShut
  {NoStop}%
\bibitem [{\citenamefont {Humensky}(2004)}]{NIM}%
  \BibitemOpen
  \bibfield  {author} {\bibinfo {author} {\bibfnamefont {T.}~\bibnamefont
  {Humensky}},\ }\bibfield  {title} {\bibinfo {title} {{SLAC’s Polarized
  Electron Source Laser System and Minimization of Electron Beam Helicity
  Correlations for the E-158 Parity Violation Experiment}},\ }\href@noop {}
  {\bibfield  {journal} {\bibinfo  {journal} {Nucl. Instrum. Meth.}\ }\textbf
  {\bibinfo {volume} {A521}},\ \bibinfo {pages} {261} (\bibinfo {year}
  {2004})}\BibitemShut {NoStop}%
\bibitem [{\citenamefont {Silwal}(2012)}]{SilwalThesis}%
  \BibitemOpen
  \bibfield  {author} {\bibinfo {author} {\bibfnamefont {R.}~\bibnamefont
  {Silwal}},\ }\emph {\bibinfo {title} {Probing the Strangeness Content of the
  Proton and the Neutron Radius of 208Pb using Parity-Violating Electron
  Scattering}},\ \href
  {https://libraetd.lib.virginia.edu/public_view/3j3332545} {\bibinfo {type}
  {{PhD} dissertation}},\ \bibinfo  {school} {University of Virginia}, \bibinfo
  {address} {Department of Physics} (\bibinfo {year} {2012})\BibitemShut
  {NoStop}%
\bibitem [{\citenamefont {Mueller}(2010)}]{ligonotes}%
  \BibitemOpen
  \bibfield  {author} {\bibinfo {author} {\bibfnamefont {C.}~\bibnamefont
  {Mueller}},\ }\bibfield  {title} {\bibinfo {title} {The electro-optic effect:
  Reading notes}} (\bibinfo {year} {2010})\BibitemShut {NoStop}%
\bibitem [{Cri()}]{CristalLaser}%
  \BibitemOpen
  \href
  {http://www.cristal-laser.com/UserFiles/File/brochures-techniques/rtp.pdf}
  {\emph {\bibinfo {title} {Cristal Laser Brochure}}},\ \bibinfo {organization}
  {Cristal Laser}\BibitemShut {NoStop}%
\bibitem [{\citenamefont {Palatchi}(2019)}]{palatchithesis}%
  \BibitemOpen
  \bibfield  {author} {\bibinfo {author} {\bibfnamefont {C.}~\bibnamefont
  {Palatchi}},\ }\emph {\bibinfo {title} {Laser and Electron Beam Technology
  for Parity Violating Electron Scattering Measurements}},\ \href
  {https://doi.org/10.18130/v3-vvcs-m369} {\bibinfo {type} {{PhD}
  dissertation}},\ \bibinfo  {school} {University of Virginia}, \bibinfo
  {address} {Department of Physics} (\bibinfo {year} {2019})\BibitemShut
  {NoStop}%
\bibitem [{\citenamefont {Brachmann}(2002)}]{SLAC2cellsystem}%
  \BibitemOpen
  \bibfield  {author} {\bibinfo {author} {\bibfnamefont {A.}~\bibnamefont
  {Brachmann}},\ }\bibfield  {title} {\bibinfo {title} {{SLAC's polarized
  electron source laser system for the E-158 parity violation experiment}},\
  }\href {http://www.slac.stanford.edu/cgi-wrap/getdoc/slac-pub-9145.pdf}
  {\bibfield  {journal} {\bibinfo  {journal} {SLAC PUB}\ }\textbf {\bibinfo
  {volume} {9145}} (\bibinfo {year} {2002})}\BibitemShut {NoStop}%
\bibitem [{\citenamefont {Roth}\ \emph {et~al.}()\citenamefont {Roth},
  \citenamefont {Samoka}, \citenamefont {Mojaev},\ and\ \citenamefont
  {Tseitlin}}]{RothDraft}%
  \BibitemOpen
  \bibfield  {author} {\bibinfo {author} {\bibfnamefont {M.}~\bibnamefont
  {Roth}}, \bibinfo {author} {\bibfnamefont {E.}~\bibnamefont {Samoka}},
  \bibinfo {author} {\bibfnamefont {E.}~\bibnamefont {Mojaev}},\ and\ \bibinfo
  {author} {\bibfnamefont {M.}~\bibnamefont {Tseitlin}},\ }\bibinfo {title}
  {{RTP Crystals For Electro-Optic Q-Switching}}\BibitemShut {NoStop}%
\bibitem [{\citenamefont {Roth}(2004)}]{Roth2003}%
  \BibitemOpen
\bibfield  {title} {  }\bibfield  {author} {\bibinfo {author} {\bibfnamefont
  {M.}~\bibnamefont {Roth}},\ }\bibfield  {title} {\bibinfo {title}
  {{Ferroelectric phase transition temperatures of self-flux-grown RbTiOPO4
  crystals}},\ }\href@noop {} {\bibfield  {journal} {\bibinfo  {journal}
  {Optical Materials}\ }\textbf {\bibinfo {volume} {26(4)}},\ \bibinfo {pages}
  {465} (\bibinfo {year} {2004})}\BibitemShut {NoStop}%
\bibitem [{Rai()}]{RaicolSpecSheet}%
  \BibitemOpen
  \href {http://raicol.com/wp-content/uploads/catalog.pdf} {\emph {\bibinfo
  {title} {Raicol Catalog}}},\ \bibinfo {organization} {Raicol}\BibitemShut
  {NoStop}%
\bibitem [{\citenamefont {Gobert}(2013)}]{eoprism}%
  \BibitemOpen
  \bibfield  {author} {\bibinfo {author} {\bibfnamefont {O.}~\bibnamefont
  {Gobert}},\ }\bibfield  {title} {\bibinfo {title} {Linear electro optic
  effect for high repetition rate carrier envelope phase control of ultra short
  laser pulses},\ }\href {doi:10.3390/app3010168} {\bibfield  {journal}
  {\bibinfo  {journal} {Appl. Sci.}\ }\textbf {\bibinfo {volume} {3}},\
  \bibinfo {pages} {168} (\bibinfo {year} {2013})}\BibitemShut {NoStop}%
\bibitem [{\citenamefont {Paschke}(2007)}]{SkewPaschkeeffect}%
  \BibitemOpen
  \bibfield  {author} {\bibinfo {author} {\bibfnamefont {K.}~\bibnamefont
  {Paschke}},\ }\bibfield  {title} {\bibinfo {title} {Controlling
  helicity-correlated beam asymmetries in a polarized electron source},\ }\href
  {http://people.virginia.edu/~kdp2c/pubpage/current/EPJA-32-549-2007.pdf}
  {\bibfield  {journal} {\bibinfo  {journal} {Eur. Phys. J. A}\ }\textbf
  {\bibinfo {volume} {32}},\ \bibinfo {pages} {549} (\bibinfo {year}
  {2007})}\BibitemShut {NoStop}%
\bibitem [{\citenamefont {{Matsui et al.}}(2018)}]{Fumihiko2018}%
  \BibitemOpen
  \bibfield  {author} {\bibinfo {author} {\bibnamefont {{Matsui et al.}}},\
  }\bibfield  {title} {\bibinfo {title} {{Parallel and antiparallel angular
  momentum transfer of circularly polarized light to photoelectrons and Auger
  electrons at the NiL3 absorption threshold}},\ }\href@noop {} {\bibfield
  {journal} {\bibinfo  {journal} {Phys. Rev. B}\ }\textbf {\bibinfo {volume}
  {97}},\ \bibinfo {pages} {035424} (\bibinfo {year} {2018})}\BibitemShut
  {NoStop}%
\bibitem [{\citenamefont {Flood}\ \emph {et~al.}(2010)\citenamefont {Flood},
  \citenamefont {Higgins},\ and\ \citenamefont
  {Suleiman}}]{HelicityControlBoard}%
  \BibitemOpen
  \bibfield  {author} {\bibinfo {author} {\bibfnamefont {R.}~\bibnamefont
  {Flood}}, \bibinfo {author} {\bibfnamefont {S.}~\bibnamefont {Higgins}},\
  and\ \bibinfo {author} {\bibfnamefont {R.}~\bibnamefont {Suleiman}},\ }\href
  {http://hallaweb.jlab.org/equipment/daq/HelicityUsersGuideFeb4.pdf} {\emph
  {\bibinfo {title} {Helicity Control Board User's Guide}}},\ \bibinfo
  {organization} {Jefferson Lab} (\bibinfo {year} {2010})\BibitemShut {NoStop}%
\bibitem [{\citenamefont {Yutsis}(2004)}]{Yutsis}%
  \BibitemOpen
  \bibfield  {author} {\bibinfo {author} {\bibnamefont {Yutsis}},\ }\bibfield
  {title} {\bibinfo {title} {{Temperature dependent dispersion relations for
  RTP and RTA}},\ }\href@noop {} {\bibfield  {journal} {\bibinfo  {journal}
  {App Phys B}\ } (\bibinfo {year} {2004})}\BibitemShut {NoStop}%
\end{thebibliography}%

\end{document}